\documentclass[3pt,sort&compress]{elsarticle}

\usepackage[dvipsnames]{xcolor}
\usepackage{graphicx}
\usepackage{amsmath}
\usepackage{amssymb}
\usepackage{amsfonts}
\usepackage{units}
\usepackage{bm}
\usepackage{hyperref}
\usepackage{enumerate}
\usepackage{longtable}
\usepackage{totcount}

\newtotcounter{citnum} 

\def\oldbibitem{} \let\oldbibitem=\bibitem
\def\bibitem{\stepcounter{citnum}\oldbibitem}

\def\btau{{\bm \tau}}

\def\bG{{\bf G}}

\def\bQ{{\bf Q} }

\def\bt{{\bm \tau}}

\def\<{\langle}
\def\>{\rangle}

\begin{document}

\sloppy

\title{Properties and challenges of hot-phonon physics in metals:
MgB$_2$ and other compounds}

\author[ec]{Emmanuele Cappelluti}
\ead{emmanuele.cappelluti@ism.cnr.it}

\author[fc2]{Fabio Caruso}

\author[dn1,dn2]{Dino Novko}

\address[ec]{Istituto di Struttura della Materia, CNR (ISM-CNR), 34149 Trieste, Italy}

\address[fc2]{Institut f\"ur Theoretische Physik und Astrophysik, Christian-Albrechts-Universit\"at zu Kiel,
D-24098 Kiel, Germany}

\address[dn1]{Institute of Physics, 10000 Zagreb, Croatia}

\address[dn2]{Donostia International Physics Center (DIPC), 20018 Donostia-San Sebasti\'an, Spain}

\begin{abstract}
The ultrafast dynamics of electrons and collective modes
in systems out of equilibrium is crucially governed by the
energy transfer from electronic degrees of freedom,
where the energy of the pump source is usually absorbed,
to lattice degrees of freedom.
In conventional metals such process leads to an overall heating
of the lattice, usually described by an effective lattice temperature
$T_{\rm ph}$, until final equilibrium with all the degrees of freedom
is reached.
In specific materials, however, few lattice modes provide a preferential
channel for the energy transfer, leading to a non-thermal
distribution of vibrations and to the onset of {\em hot phonons},
i.e., lattice modes with a much higher population than the other modes.
Hot phonons are usually encountered in semiconductors or semimetal
compounds, like graphene, where the preferential channel
towards hot modes is dictated by the reduced electronic phase space.
Following a different path,
the possibility of obtaining hot-phonon physics also in metals
has been however also recently prompted in literature,
as a result of a strong anisotropy of the electron-phonon (el-ph) coupling.
In the present paper,
taking MgB$_2$ as a representative example,
we review the physical conditions that allow
a hot-phonon scenario in metals with anisotropic el-ph coupling,
and we discuss the observable fingerprints of hot phonons.
Novel perspectives towards the prediction and experimental observation
of hot phonons in other metallic compounds are also discussed.

\end{abstract}

\maketitle

\tableofcontents

\clearpage


\section{Introduction}
\label{s-intro}

After years of steady formidable improvements and refinements,
time-resolved spectroscopy, nowadays scaled down to femtosecond
resolution, has become one of the major tools
for investigating the physical properties and underlying mechanisms
of condensed matter \cite{giannettireview,petek97,king05,mischa11,bauer15,waldecker16}. 
The increasing interest in this line of research
is stimulated by relentlessly expanding possibilities to 
achieve control of quantum phenomena using light pulses.
Typically, a pump-probe setup is employed,
where a laser pump at $t=0$ triggers electronic particle-hole excitations. The pump can also be calibrated to couple directly with lattice, magnon
or other bosonic degrees of freedom.
Alternatively, coherent modes (phonon, collective excitations, etc) can be also triggered
in \cite{kuznetsov94,merlin97,kuznetsov01,stevens02,ishioka06,bovensiepen06,ishioka08,forst11,boschetto13,udina19}.
A later probe (optical, electronic, transport, diffraction)
is thus used to assess the physical properties at a time delay $t$.
Within this context, pump-probe experiments
have proven to be able to induce a variety of physical processes
otherwise not allowed (or very slow) under steady conditions,
as for instance insulator-metal transitions
\cite{cavalleri01,demsar02,iwai03,rini07,perfetti07book,perfetti07tas2,kawakami09},
light-induced superconductivity \cite{fausti11,kaiser14,mitrano16,cantaluppi18,suzuki19,isoyama21},
chemical reactions \cite{cavanagh93,ho96,kuhn07,matsumoto07,petek10},
structural phase transitions \cite{cavalleri01,fausti09},
local heat-transport, non-linear optics \cite{forst11,mankowsky14,subedi14},
exciton dynamics of relevance for light harvesting \cite{cui14,perfetto16,trovatello19}, 
detecting the effects of surface plasmons and/or other collective modes \cite{sundararaman14,kogar20},
manipulation of magnetic degrees of freedom \cite{kim12,johnson15,hudl19,geilhufe21} etc.
In accordance with the different physical processes
under examination, a wide variety of time-resolved techniques
have been developed, encompassing angle-resolved photoemission,
electromagnetic response (including optics, magneto-optics, transport,
Raman), electron and X-ray diffraction and other techniques.
Insomuch laser pumping can be used to excite bound states,
it can provide also a promising tool for encoding quantum information
in selected quantum
states \cite{ultrafastquantumbook}.
The control and the possible tailoring of the decay processes
of such excited states is thus essential for assessing the time scales
over which the bound state persists, and for ruling the interactions
with other degrees of freedom, and hence the possibility
of reading out the information itself.

Among other degrees of freedom present in a real material,
collective lattice vibrations ({\em phonons})
play a pivotal role in ultrafast pump-probe dynamics.
The energy initially induced into the electronic states by laser pumping
is typically transferred to the lattice within few picoseconds, mostly
in the form of excitation of phonon modes \cite{anisimov74,bib:allen87,elsayed-ali87,schoenlein87,snoke92,fann92,fann94,sun94,groeneveld95,petek97,gusev98,hohlfeld00,delfatti00,echenique00,rethfeld02,pietanza07,bib:lin08}.
In most cases, the energy in the lattice sector
is redistributed over all the lattice modes via phonon-phonon interactions,
leading to a overall thermal heating which is usually
detrimental for sustaining coherent states.
As discussed in the next Section, however,
in specific compounds and under specific conditions,
energy transfer can occur between electrons and {\em few}
selected phonon modes, which become {\em hot},
i.e., they acquire a larger (not-thermal) population
than other lattice vibrations.
This scenario is particularly appealing for several reasons.
On the one hand, quenching the energy transfer from electrons
to few phonon modes can contribute to limit the overall heating
of the remnant lattice modes, and hence to sustain
coherent states. On the other hand, hot-phonon modes
are typically strongly coupled with electrons, so that
their presence can increase the electronic damping.
Manipulating hot-phonon physics can help in optimizing
high- and low-$T_c$ superconducting materials.
Furthermore, hot-phonon modes, when associated with
an additional degrees of freedom as phonon chirality
\cite{zhang14,zhang15,zhu18,chen18,komiyama21},
can represent a promising two-level state, useful
for quantum-information purposes \cite{semiao10}.
For these reasons a deep understanding of the processes
and of the required physical conditions leading
to the onset of hot-phonon modes is a crucial goal
for a robust control of mechanism of ultrafast dynamics
in pump-probe experiments.

\begin{figure}[t!]
\centering
\includegraphics[width=12.5cm]{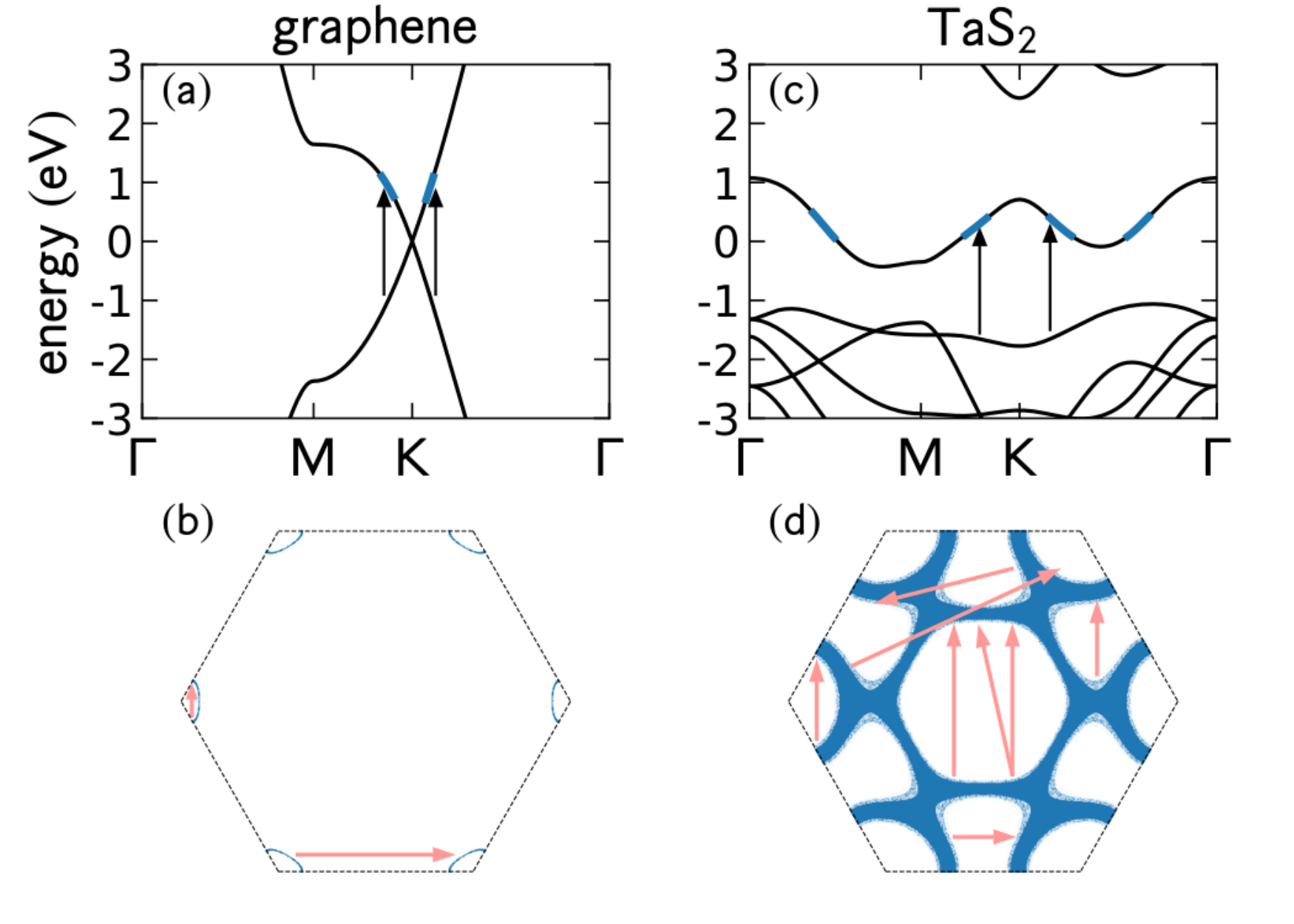}
\caption{
Representative examples of electronic phase space available
in semiconductors or semimetals (panel a) and in metals (panel b) upon
particle-hole pump-driven excitations.
For graphical reasons we show
two paradigmatic
two-dimensional systems: semimetal single-layer graphene, where
the onset of hot phonons has been discussed extensively
in Refs.
\cite{butscher07,ishioka08,richter09,yan09,berciaud10,lui10,wang10,breusing11,scheuch11,huang11,wu2012,golla17,koivi17,novko2019,sidiropoulos21},
and single-layer 1T-TaS$_2$, which is a metal
suitable for pump-probe experiments in both
the 1T and 1H phases
\cite{perfetti07tas2,perfetti08,dean11,petersen11,hellmann12,andreatta19,wang20,simoncig21}
For simplicity,  we show only the conduction, high-energy band where 
electrons are upgraded upon absorption of the photon energy. 
Top panels represent  electronic dispersion and bottom panels
momenta in the Brillouin zone where the electrons are excited.
For a semiconductor (panel a) only electronic states 
with momenta close to the valleys
are populated (blue areas). 
Electron-phonon coupling involves inelastic scattering with low-energy exchange,
so that the initial electron can scatter only into empty states, i.e.,
not all possible phonons are exchanged but only phonons
connecting intravalley and intervalley processes
(red arrows). Only these selected modes can thus absorb
energy from the electronic degrees of freedom. In metals (panel b)
many electrons in the Fermi sea can be excited in different parts
of the Brillouin zone  (blue regions).
Many different electron-phonon driven processes (red arrows)
can occur, scattering an excited electron into available empty states, with very different momenta. These ground state properties were obtained by means of the density-functional-theory package \textsc{quantum espresso}\,\cite{qe1,qe2} (in both cases $24\times24\times1$ momentum grid is used in order to sample the Brillouin zone; unit cell parameters $2.461\,\textrm{\AA}$ and $3.333\,\textrm{\AA}$ are used for graphene and 1H-TaS$_2$, respectively.).
}
\label{f-1sketch}
\end{figure}

For many years, the investigation of hot phonons
has been focused on semiconductors and semimetals
\cite{kocevar72,kocevar85,potz87,joshi89,micke90,kim91,micke96,langot96,lazzeri06,butscher07,ishioka08,richter09,yan09,berciaud10,lui10,wang10,breusing11,scheuch11,huang11,wu2012,golla17,koivi17,novko2019,sidiropoulos21},
whereas standard metals have been considered as
an unfavourable premise for sustaining hot phonons.
In semiconductors and semimetals, indeed,
hot phonons stem from the reduced phase space for electronic decays.
Laser-induced particle-hole excitations are localized
in few spots of the Brillouin zone close to the valleys
(or Dirac points, in case of semimetal graphene).
Electron-phonon scattering is thus operative only for
the ${\bf q}$-momenta connecting available valleys/Dirac cones.
Only such few modes can profit of the energy transfer from
the electronic degrees of freedom, at expenses of
other lattice modes that remain cold (Fig. \ref{f-1sketch}a).

Metals appear at a first glance at odds with such scenario.
Large Fermi surfaces can trigger laser-induced
particle-hole excitations in a vast part of the Brillouin zone.
Electron-phonon scattering, in conventional metals,
is also weakly ${\bf q}$-momentum dependent,
so that a plethora of lattice modes with different
momenta ${\bf q}$ can be effectively excited (Fig. \ref{f-1sketch}b),
leading to global heating of the phonon degrees of freedom
without any preferential lattice mode \cite{anisimov74,bib:allen87,elsayed-ali87,schoenlein87,snoke92,fann92,fann94,sun94,groeneveld95,petek97,gusev98,hohlfeld00,delfatti00,echenique00,rethfeld02,pietanza07,bib:lin08}.

Midway between insulators and metals there are the so-called
``bad metals'', characterized by finite but poor metallic properties.
A particular mention among them goes to high-$T_c$
superconducting cuprates, layered materials that display
strong anisotropy between the in-plane and out-of-plane features.
Even restricting only to the in-plane physics,
which is responsible for the Cooper pairing,
the metallic character is highly non-trivial, having a very complex phase diagram,
including an antiferromagnetic Mott insulator phase at extremely low
doping up to a half-filling, a pseudogap phase in the underdoping region,
where dynamical/short-range correlations appear in the
coupled magnetic-electronic-lattice response,
and a more conventional metallic phase
in the overdoped regime.
Such a rich phase diagram is thought to stem out from the
underlying presence of one (or two) critical points
associated with magnetic and/or lattice instabilities.
Although the electron-phonon coupling was initially disregarded
as a relevant source of scattering \cite{andersonbook,dagotto94},
there are nowadays several
evidences that it might play an important role in the
physical properties of the normal and superconducting states.
Quite crucial is thought to be the electron-phonon coupling
associated with the changes of the distance/angle of the Cu-O-Cu 
bond in the CuO$_2$ layer that represents the relevant unit block \cite{pavarini01}.
The relevance of such electron-phonon interaction
is shown in the observation of a remarkable kink
in the angle-resolved-photoemission spectroscopy (ARPES) \cite{lanzara01,zhou03}.
The energy of such kink matches the energies
of the optical modes associated with the changes
in the Cu-O-Cu bond, providing thus a
fingerprint of the key role of these modes\,\cite{zhenglu21}.
The preferential coupling of the electronic degrees of freedom
with these lattice modes has been as well investigated by means
of time-resolved ultrafast spectroscopy \cite{bib:perfetti07,carbone08,mansart10,bib:dalconte12,johnson15,giannettireview}, that reveals a hot-phonon scenario
where the energy transferred to the electronic sector by the pump laser
is preferentially transferred in a shorter time to such optical modes,
that become hot, and on a much longer time scale to the other lattice
modes.
So far, however,
because of the high sensitivity of the physical properties on
extrinsic conditions (e.g. disorder, substrates, etc.),
the evidence of hot phonons in cuprates
has been used only as a possible way for characterization,
rather than as a suitable tool for manipulating the electronic/lattice
degrees of freedom.

Merging together the advantages of a good metallic character
with suitable conditions for hot phonons,
a novel path for sustaining
hot-phonon physics in metals has been recently pointed out,
as a result of a strong anisotropy of the
electron-phonon coupling \cite{baldini,ncdc}.
The benchmark example of this scenario is magnesium diboride MgB$_2$,
a metal compound that, besides supporting evidence
of hot phonons, displays a significant superconductivity
with critical temperature as high as $T_c\approx 39$ K \cite{bib:nagamatsu01}.
The key ingredient responsible for hot phonons in MgB$_2$
is the peculiar property of having a large part of
the electron-phonon coupling $\lambda$ concentrated in few
phonon modes \cite{kong}.
The possibility of sustaining hot phonons in a metal
opens interesting perspectives in the field of time-resolved
ultrafast dynamics, both on the application side as well
as for understanding the underlying physical mechanisms.
The electron-phonon coupling in metals
plays indeed a crucial role in determining
a variety of physical properties \cite{grimvall},  resulting in a dynamical
renormalization of single-particle excitations,
as detected for instance in angle-resolved photoemission spectroscopy
and in optics; governing the linewidths (and hence lifetimes)
of electrons and of many collective modes (including phonons);
ruling the transport properties, and; least but not last,
in providing the fundamental glue for the
superconducting Cooper pairing \cite{scalapino,allenmitrovic,carbottereview}.

The investigation of possible ways of detecting hot-phonon modes
and of the physical consequences of such scenario
in MgB$_2$ and similar compounds is just in its early stages
and the full potential is still to be explored.
Furthermore, the physical conditions supporting hot-phonon physics
in MgB$_2$ are not unique of this compound but they are expected
to be realized in many other metals and/or charge-doped
semiconductors.
Aim of the present review article is to encompass a useful overview
of the state-of-art knowledge
and to provide a solid basis and guideline for 
future development in this field.
The review is structured as follow: in Sec. \ref{s-processes}
we summarize the physical mechanisms governing
time-resolved dynamics in pump-probe setups,
with a specific focus on the energy-transfer paths 
from electrons to lattice degrees of freedom, and
on the conditions for sustaining hot-phonon physics;
various state-of-the-art studies discussing hot-phonon physics along with the corresponding experimental techniques are reviewed in Sec. \ref{s:detect};
in Sec. \ref{s-mgb2} we discuss the specific properties
of MgB$_2$, governed by a strong anisotropy
of the electron-phonon coupling; the consequences
of such anisotropy of the electron-phonon coupling with respect
to the onset and observation of hot-phonon physics in MgB$_2$
will be examined in Sec. \ref{s-hotmgb2}, while
in Sec. \ref{s-other} we discuss how a similar scenario
can be generalized in other materials.


\section{Energy transfer and relaxation processes in  time-resolved pump-probe experiments}
\label{s-processes}

Condensed-matter physics is in principle a formidably complex topic
where many degrees of freedom (electrons and ions) are coupled
together through different sources of interactions.
In spite of such complexity, we rely on a robust understanding
of most of the physical properties of solids in terms of simplified
but valid modelings.
The Drude model for instance, as well as the Fermi-liquid theory at a more
sophisticate level,
provides a sufficient description of many conventional metals
in terms of non-interacting (or weakly interacting) particles
with few effective parameters, notwithstanding
a large Coulomb-driven electron-electron interaction.
The complex interaction between electron and ions
can be also simplified thanks to the Born-Oppenheimer
principle that permits a useful decoupling of
the quantum nature of electrons and ions.
The further quantum interaction between electrons and lattice vibrations
(phonons) beyond the adiabatic Born-Oppenheimer approximation
can be conveniently treated, thanks to Migdal's theory,
by means of an effective mean-field-like approach,
the Eliashberg theory \cite{grimvall}, which fulfills the Fermi-liquid picture
in the normal state and provides a solid ground for
a quantitative understanding 
of the superconducting properties in low-$T_c$ compounds.
As a general rule, most of our ``textbook'' understanding
of condensed matter applies to systems at thermodynamical
equilibrium, where the different degrees of freedom are
described by common thermodynamical parameters,
more in particular by a unique temperature $T$.

\begin{figure}[t!]
\centering
\includegraphics[width=10.5cm]{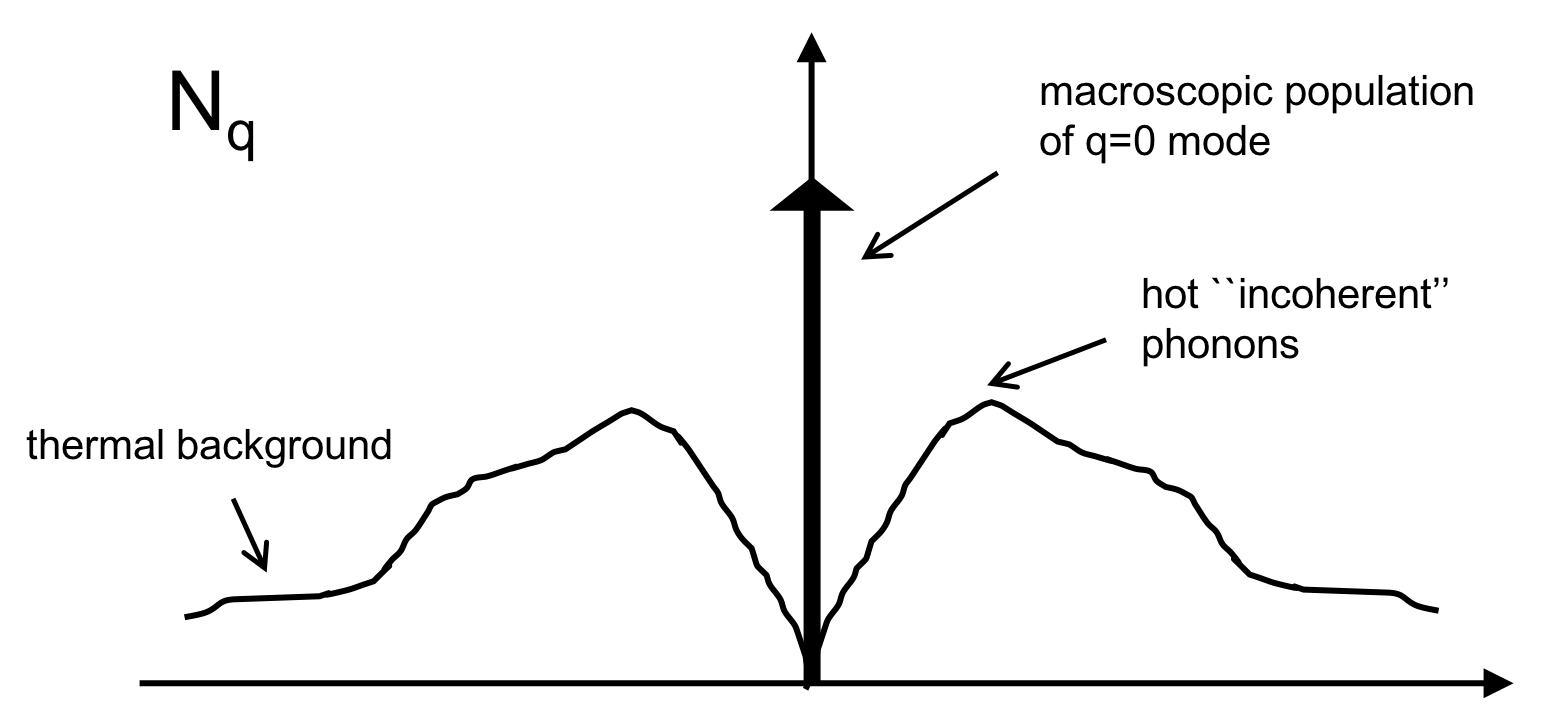}
\caption{
Schematic representation of the nonequilibrium phonon distribution function in GaAs
under femtosecond photoexcitation. The central arrow represents a macroscopic population
of the ${\bf q}=0$ optical mode, giving rise to coherent phonon time dynamics.
This feature is well distinct from the contribution of hot and incoherent phonons,
as well as from the background of cold modes with a thermal population.
Figure inspired by Ref. \cite{kuznetsov01}.
}
\label{f-coh}
\end{figure}

Fundamental aim of pump-probe experiments is to
perturb such initial thermodynamical state
by bringing the systems under non-equilibrium conditions,
and to investigate in real time the physical properties
during the relaxation processes until
the system reaches a new final equilibrium.
Typical pumping lasers (or synchrotron light) are set at energies
much higher than phonons or other collective modes,
so that the incoming pump light mainly induces
particle-hole excitations in the electronic degrees
of freedom.
Interesting refinements of the pump-probe setup
also exploit laser light in a non-linear regime,
as for instance when an impinging photon can trigger
more complex excitations that a single
particle-hole one \cite{petek97}, or when
a sudden sharp pulse can induce coherent (macroscopical) phonon oscillations.
It is worth to remark here that the hot-phonon physics (which in principle might occurs for any generic momentum ${\bf q}$),
as related to {\em decay} processes governed by the electron-phonon interaction,
follows a basically different path than the generation of coherent phonons,
which corresponds to a microscopical population of a ${\bf q}=0$ lattice mode
as a consequence of an appropriate external field (see Fig. \ref{f-coh}).
In the present review we do not survey such scenario and we
focus on hot-phonon physics restricting ourselves
to the most common case
where the external pump triggers linearly particle-hole
excitations (i.e., electronic degrees of freedom)
and where no coherent oscillations are induced.

In a schematic way, the relevant system at equilibrium before
(and in the absence of)
pumping can be described by the standard Hamiltonian:
\begin{eqnarray}
\hat{H}
&=&
\sum_{{\bf k},\sigma,\alpha}
\left(
\epsilon_{{\bf k}\alpha}
-\mu
\right)
\hat{c}_{{\bf k},\sigma,\alpha}^\dagger \hat{c}_{{\bf k},\sigma,\alpha}
+
\sum_{{\bf q},\nu}
\hbar \omega_{{\bf q},\nu}
\hat{a}_{{\bf q},\nu}^\dagger
\hat{a}_{{\bf q},\nu}
\nonumber\\
&&
+
N_{\bf k}^{-\frac{1}{2}}\sum_{{\bf k,q},\sigma,\alpha,\beta}
g_{{\bf k}\alpha,{\bf k+q}\beta}^\nu
\hat{c}_{{\bf k+q},\sigma,\beta}^\dagger \hat{c}_{{\bf k},\sigma,\alpha}
\left(
\hat{a}_{{\bf q},\nu}+\hat{a}_{{\bf -q},\nu}^\dagger
\right)
\nonumber\\
&&
+
H_{\rm p-p}^{\rm anharm}
+
H_{\rm e-e}
,
\label{e:ham}
\end{eqnarray}
where $\mu$ is the electron chemical potential,
$\hat{c}_{{\bf k},\sigma,\alpha}^\dagger$
creates an electron with momentum ${\bf k}$ and spin $\sigma$
in the band $\alpha$
with energy $\epsilon_{{\bf k},\alpha}$,
$\hat{a}_{{\bf q},\nu}^\dagger$
creates a phonon with momentum ${\bf q}$ in the branch $\nu$
with frequency $\omega_{{\bf q},\nu}$,
$g_{{\bf k}\alpha,{\bf k+q}\beta}^\nu$ is the matrix element
of linear electron-phonon coupling scattering
an electron from the state $({\bf k},\beta)$
into the state $({\bf k+q},\alpha)$ with
a phonon emission (adsorption)  with momentum
${\bf q}$ ($-{\bf q}$) in the branch $\nu$.
$N_{\bf k}$ is the number of momenta sampling the Brillouin zone.
The third  line in Eq. (\ref{e:ham})
takes into account
the phonon-phonon scattering due to the lattice anharmonicity
and the Coulomb-driven electron-electron scattering.
In most of the cases, the detailed description of these latter two processes
is not relevant and they do not need to be specified in more detail.
For the sake of simplicity, we consider in the following a model with density-density interaction:
$H_{\rm e-e}=
\sum_{\bf q} V(|{\bf q}|)
\hat{\rho}_{\bf q} \hat{\rho}_{\bf -q}$,
where $\hat{\rho}_{\bf q}=
\sum_{{\bf k},\alpha,\sigma}
\hat{c}_{{\bf k+q},\alpha,\sigma}^\dagger \hat{c}_{{\bf k},\alpha,\sigma}$.
Other interaction terms,
as disorder scattering in the electron-phonon spectra, can be
added when relevant.

For a generic interacting system,
one can define
an electron distribution function \cite{notespin}
$f_{{\bf k},\alpha}=
\langle 
\hat{c}_{{\bf k},\alpha}^\dagger \hat{c}_{{\bf k},\alpha}\rangle$
as well as phonon distribution function
$b_{{\bf q},\nu}=
\langle 
\hat{a}_{{\bf q},\nu}^\dagger \hat{a}_{{\bf q},\nu}\rangle$.
At thermodynamical equilibrium,
both electron and phonon degrees of freedom are 
ruled by a thermal population described 
respectively by the Fermi-Dirac and Bose-Einstein
distributions.
In the non-interacting case, for instance,
we have for the electrons
$f^{(0)}_{{\bf k},\alpha}= f_T[(\epsilon_{{\bf k},\alpha}-\mu)/k_{\rm B}T]$,
and for the phonons
$b^{(0)}_{{\bf q},\nu} = b[\hbar \omega_{{\bf q},\nu}/k_{\rm B}T]$,
where $f_T[x]=1/(\mbox{e}^x+1)$,
$b_T[x]=1/(\mbox{e}^x-1)$, and $k_{\rm B}$ is the Boltzmann constant.
In the interacting case, the energies
$\epsilon_{{\bf k},\alpha}$, $\omega_{{\bf q},\nu}$
at a first qualitative approximation can be
replaced by effective (renormalized) quantities
$\tilde{\epsilon}_{{\bf k},\alpha}$, $\tilde{\omega}_{{\bf q},\nu}$.

\begin{figure}[t!]
 \centering
 \includegraphics[width=12.cm]{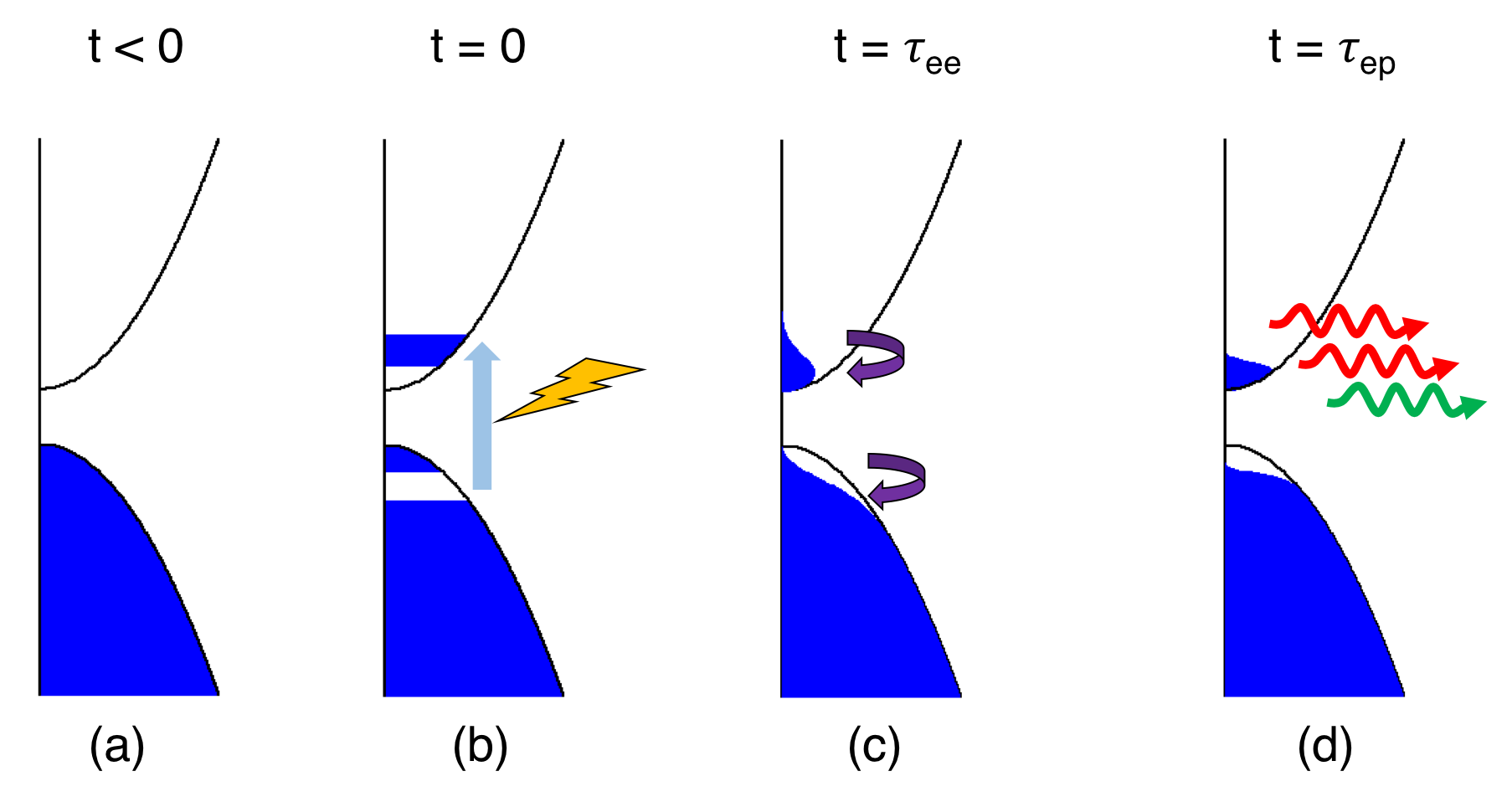}
 \caption{Representative sketch of the time evolution of the electronic distribution
function in a pump-probe experiment. (a) Electronic system at $t < 0$
before the pump excitation. Electrons
and holes obey a thermal distribution, here assumed to be ruled by a
low temperature $T \approx 0$.
(b) Pump laser photons induce at $t=0$ particle-hole excitations,
leading to a highly non-thermal distribution.
(c) In a short time characterized by a time-scale $\tau_{\rm ee}$,
due to the electron-electron interaction,
electrons and holes reach a thermal distribution characterized
by a large electronic temperature $T_{\rm ee}$.
(d) On a time scale $\tau_{\rm ep}$ governed by the
electron-phonon coupling, electrons (and holes) further reduce
their effective temperature by transferring energy to
the lattice modes through emission of hot phonons (red wavy lines)
and to a minor extent of cold phonons (green wavy lines).
For simplicity we depict here the case of a
typical semiconductor where, in the absence of recombination
which occurs on longer times,
electrons and holes obey separate distributions.
For metal the scheme is similar where electrons and holes
excitations stem out from a unique distribution function.
}
\label{f-timeranges}
\end{figure}

Under the assumption of a linear coupling between the pump photon
with energy $h \nu$ with particle-hole excitations,
in a ideal case of a $\delta$-function pump profile
in real time, we could describe at $t=0$
the electronic system with a highly non-thermal distribution
function as sketched in Fig. \ref{f-timeranges},
with a distribution of {\em particle} excitations for
$\epsilon_{{\bf k},\alpha}\ge \mu$ balanced
by a distribution of {\em hole} excitations
for $\epsilon_{{\bf k},\alpha}\le \mu$.

The fundamental information in time-resolved physics concerns
the different time scales within the system under
such initial non-equilibrium state reaches a new final steady state
at equilibrium with the all the degrees of freedom,
and the channels employed to reach the new final equilibrium.
In this regard, it is clear that the time dynamics is crucially
based on many-body processes, and its investigation reveals
the role of the different scattering sources.
From a generic point of view,
neglecting  particle-hole recombination,
which occurs on much larger time scales than
the range of interest here,
all the theoretical approaches
describing the time evolution of the coupled electron/phonon system,
are fundamentally based on the equations of motion
taking into account
the collisional processes among and between the electronic and lattice
degrees of freedom:
\begin{eqnarray}
\frac{
df_{{\bf k},\alpha}(\{f,b\})}{dt}
&=&
\left.
\frac{
df_{{\bf k},\alpha}(\{f,b\})}{dt}
\right|_{\rm ee}
+
\left.
\frac{
df_{{\bf k},\alpha}(\{f,b\})}{dt}
\right|_{\rm ep}
+
I_{\bf k}(t),
\label{e:dftot}
\\
\frac{
db_{{\bf q},\nu}(\{f,b\})
}{dt}
&=&
\left.
\frac{db_{{\bf q},\nu}(\{f,b\})}{dt}
\right|_{\rm ep}
+
\left.
\frac{db_{{\bf q},\nu}(\{f,b\})}{dt}
\right|_{\rm pp}
.
\end{eqnarray}
Here $I_{\bf k}(t)$ accounts for the perturbation induced
by the pump pulse, the terms labeled ``ee'' represent scattering
processes
mediated by the
Coulomb electron-electron scattering, the
terms labeled ``ep'' describe electron-phonon mediated
scattering, while the term labeled as ``pp'' accounts
for phonon-phonon scattering processes driven
by phonon anharmonicity.
The arguments in curly brackets in the electron and phonon
distribution functions, $f_{{\bf k},\alpha}(\{f,b\})$,
$b_{{\bf q},\nu}(\{f,b\})$ reminds that these quantities have
a mutual functional dependence.
Considering, as an illustrative example, collision processes
at the lowest order in a perturbation theory \cite{noteeqdyn},
we can write for instance based on Eq. (\ref{e:ham})
(see \cite{notespin}):
\begin{eqnarray}
\left.
\frac{
df_{{\bf k},\alpha}}{dt}
\right|_{\rm ee}
&=&
\frac{2\pi}{\hbar}
\sum_{{\bf p,q}\beta}
2V(|{\bf q}|)
\left(2-\delta_{\alpha,\beta}\right)
\delta(\epsilon_{{\bf k},\alpha}+\epsilon_{{\bf p},\beta}
-\epsilon_{{\bf k+q},\alpha}-\epsilon_{{\bf p-q},\beta})
\\
&&
\times
\Big\{
f_{{\bf k+q},\alpha}f_{{\bf p-q},\beta}
(1-f_{{\bf k},\alpha})(1-f_{{\bf p},\alpha})
-
f_{{\bf k},\alpha}f_{{\bf p},\beta}
(1-f_{{\bf k+q},\alpha})(1-f_{{\bf p-q},\alpha})
\Big\}
,
\nonumber\\
\left.
\frac{
df_{{\bf k},\alpha}}{dt}
\right|_{\rm ep}
&=&
-
\frac{2\pi}{\hbar}
\sum_{{\bf p}\beta}
\left| g_{{\bf k}\alpha,{\bf p}\beta}^\nu\right|^2
\Big\{
f_{{\bf k},\alpha}(1-f_{{\bf p},\beta})
\\
&&
\times
\left[
(b_{{\bf k-p},\nu}+1)
\delta(\epsilon_{{\bf k},\alpha}-\epsilon_{{\bf p},\beta}-\hbar \omega_{{\bf k-p},\nu})
+
b_{{\bf k-p},\nu}
\delta(\epsilon_{{\bf k},\alpha}-\epsilon_{{\bf p},\beta}+\hbar   \omega_{{\bf k-p},\nu})
\right]
\nonumber\\
&&
-
f_{{\bf p},\beta}(1-f_{{\bf k},\alpha})
\nonumber\\
&&
\nonumber
\times
\left[
(b_{{\bf k-p},\nu}+1)
\delta(\epsilon_{{\bf k},\alpha}-\epsilon_{{\bf p},\beta}+\hbar \omega_{{\bf k-p},\nu})
+
b_{{\bf k-p},\nu}
\delta(\epsilon_{{\bf k},\alpha}-\epsilon_{{\bf  p},\beta}-\hbar \omega_{{\bf k-p},\nu})
\right]
\Big\}
,
\\
\left.
\frac{db_{{\bf q},\nu}}{dt}
\right|_{\rm ep}
&=&
-
\frac{4\pi}{\hbar }
\sum_{{\bf k,q},\alpha,\beta}
\left|
g_{{\bf k}\alpha,{\bf k+q}\beta}^\nu
 \right|^2
f_{{\bf k},\alpha}(1-f_{{\bf  k+q},\beta})
\\
&&\nonumber
\times
\left[
b_{{\bf q},\nu}
\delta(\epsilon_{{\bf k},\alpha}-\epsilon_{{\bf k+q},\beta}+\hbar \omega_{{\bf q},\nu})
-
(b_{{\bf q},\nu}+1)
\delta(\epsilon_{{\bf k},\alpha}-\epsilon_{{\bf k+q},\beta}-\hbar \omega_{{\bf q},\nu})\right]
.
\label{e:bdynep}
\end{eqnarray}
More complex equations can be written down
when the many-body interactions are treated
at a higher order, 
using some kind of approximation,
as for instance in a self-consistent Born scheme
or using random-phase approximation (RPA).
Similar coupled equations can be also equivalently derived within the framework
of {\em ab-initio} approach \cite{perfetto16,molinasanchez17,echenique00}.
Such computation improvements are fundamental when
quantitative predictions are needed, as for instance in
assessing the effective dispersion and linewidths in
time-dependent angle-resolved photoemission spectroscopy (ARPES)
\cite{perfetto16},
two-photon photoemission \cite{petek97}
or in image-potential states \cite{echenique00}.
On the other hand,
the basic physics relevant for the aims of the present paper,
surveying the possible onset of hot phonons in MgB$_2$
and other metals, is already well captured by the
lowest order scheme outlined in Eqs. (\ref{e:dftot})-(\ref{e:bdynep}).

On a macroscopic level, the dynamics
of the electron and phonon distribution functions
drives changes of the electronic and vibrational energies, which can be defined as
\begin{eqnarray}
E_{\rm e}
&=&
N_s
\sum_{{\bf k},\alpha}
\epsilon_{{\bf k},\alpha}
f_{{\bf k},\alpha}
,
\\
E_{\rm p}
&=&
\sum_{{\bf q},\nu}
\hbar \omega_{{\bf q},\nu}
b_{{\bf q},\nu}
,
\end{eqnarray}
where $N_s=2$ is the electron spin degeneracy.

The time evolution of the energies $E_{\rm e}$,
$E_{\rm p}$ can be thus obtained by integrating
Eqs. (\ref{e:dftot})-(\ref{e:bdynep}):
\begin{eqnarray}
\frac{
dE_{\rm e}}{dt}
&=&
\left.
\frac{dE_{\rm e}}{dt}
\right|_{\rm ep}
+
S(t)
,
\label{dEedt}
\\
\frac{
dE_{\rm p}}{dt}
&=&
\left.
\frac{dE_{\rm p}}{dt}
\right|_{\rm ep}
=
-
\frac{dE_{\rm e}}{dt}
,
\label{dEpdt}
\end{eqnarray}
where
\begin{eqnarray}
\left.
\frac{dE_{\rm p}}{dt}
\right|_{\rm ep}
&=&
\frac{4\pi}{\hbar}
\sum_{{\bf k,p},\alpha,\beta,\nu}
\hbar \omega_{{\bf k-p},\nu}
\left|
g_{{\bf k}\alpha,{\bf p}\beta}^\nu
 \right|^2
\delta(\epsilon_{{\bf k},\alpha}-\epsilon_{{\bf p},\beta}
+\hbar
    \omega_{{\bf k-p},\nu})
\nonumber\\
&&
\times
\left[
\left(
f_{{\bf k},\alpha}-f_{{\bf p},\beta}
\right)
b_{{\bf k-p},\nu}
-f_{{\bf p},\beta}
(1-f_{{\bf k},\alpha})
\right]
,
\end{eqnarray}
and
$S(t)=N_s\sum_{{\bf k},\alpha} \epsilon_{{\bf k},\alpha} I_{\bf k}(t)$
accounts for rate of the energy pumping 
in the system \cite{notedepth}.
Equations (\ref{dEedt})-(\ref{dEpdt})
do {\em not} rely on the assumption of a electron/phonon thermal
distributions,
but they are generally valid also for non-thermal distributions,
once the  solution
of Eqs. (\ref{e:dftot})-(\ref{e:bdynep}) is provided.
It is also worth noticing
that the Coulomb electron-electron interaction, as well as
the anharmonicity-driven phonon-phonon term,
are not explicitly appearing in
Eqs. (\ref{dEedt})-(\ref{dEpdt}).
Indeed, 
while electron-electron scattering causes the redistribution of energy and momentum
among electronic degrees of freedom,
the conservation laws constrains
the {\em total} energy and momentum to remain unchanged. 
Correspondingly, electron-electron interaction does not appear explicitly in Eqs. (\ref{dEedt})-(\ref{dEpdt}).
Similar considerations apply for the influence of anharmonic phonon-phonon 
scattering process on vibrational degrees of freedom.
Equations (\ref{dEedt})-(\ref{dEpdt})
underline thus the primary role of the electron-phonon interactions in governing
energy transfer and relaxation among the degrees of freedom of the system.

Equations (\ref{e:dftot})-(\ref{e:bdynep})
[and hence Eqs. (\ref{dEedt})-(\ref{dEpdt})] form a closed set 
of integro-differential equations which can be 
solved explicitly via conventional time-propagation algorithms to investigate the non-equilibrium dynamics of electronic and vibrational degrees of freedom \cite{caruso21,tong21,seiler21}. 
In 3D metallic compounds,
the solution of the Boltzmann equation is 
computationally very demanding due to the
need of extremely dense meshes to sample phonon-assisted  particle-hole transitions
in a sufficiently wide region of reciprocal space
(see Fig. \ref{f-1sketch}b).
Approximate strategies can thus to be employed to reduce the computational complexity of the problem and simplify numerical investigations of the coupled electron-phonon dynamics.

One of the widest scheme employed to this aim is the
two-temperature (2T) model.
Within this approach, electron and lattice degrees of freedom
are assumed to obey a thermal distribution,
$f_{{\bf k},\alpha}[T_{\rm e}]$ and $b_{{\bf q},\nu}[T_{\rm p}]$,
ruled by two district effective temperatures,
an electron $T_{\rm e}$ and a phonon one $T_{\rm p}$,
respectively.
Neglecting particle-hole recombination processes,
which occur on a longer time scale,
for metals the effective temperature $T_{\rm e}$ governs
a normal thermal distribution about the Fermi sea, whereas
in semiconductors/semimetals
$T_{\rm e}$ governs two separate
thermal distributions for electrons in the conduction band
and for holes in the valence band, respectively.

Within the framework of the 2T model,
the electronic and lattice total energies can be parametrized in terms of the electronic and phonon temperature
 $T_{\rm e}$, $T_{\rm p}$, respectively.
In particular, Eqs. (\ref{dEedt})-(\ref{dEpdt})
can be recast in terms of two coupled equations ruling the time
dynamics of the effective temperatures:
\begin{eqnarray}
C_{\rm e}
\frac{d T_{\rm e}}{d t}
&=&
-
G(T_{\rm e}-T_{\rm p})
+
S(t)
,
\label{eq:el2}
\\
C_{\rm p}\frac{d T_{\rm p}}{d t}
&=&
G(T_{\rm e}-T_{\rm p})
,
\label{eq:eph2}
\end{eqnarray}
where we introduced the specific heat capacity of electrons and phonons,
\begin{eqnarray}
C_{\rm e}
&=&
N_s
\sum_{{\bf k},\alpha}
\epsilon_{{\bf k},\alpha}
\frac{\partial f_{{\bf k},\alpha}[T_{\rm e}]}{\partial T_{\rm e}}
,
\label{eq:c_e}
\\
C_{\rm p}
&=&
\sum_{{\bf q},\nu}
\hbar \omega_{{\bf q},\nu}
\frac{\partial b_{{\bf q},\nu}[T_{\rm p}]}{\partial T_{\rm p}}
,
\label{eq:c_e2g}
\end{eqnarray}
and the effective electron-phonon coupling constant $G$, defined by:
\begin{eqnarray}
G
&=&
\frac{2\pi k_{\rm B}}{\hbar}
\sum_{{\bf k},\alpha,\beta}
\sum_{{\bf q},\nu}
\left|
g_{{\bf k}\alpha,{\bf k+q}\beta}^\nu
 \right|^2
\omega_{{\bf q},\nu}
\delta(\epsilon_{{\bf k},\alpha})
\delta(\epsilon_{{\bf k+q},\beta})
.
\label{eq:gph}
\end{eqnarray}

\begin{figure}[t!]
 \centering
 \includegraphics[width=9.5cm]{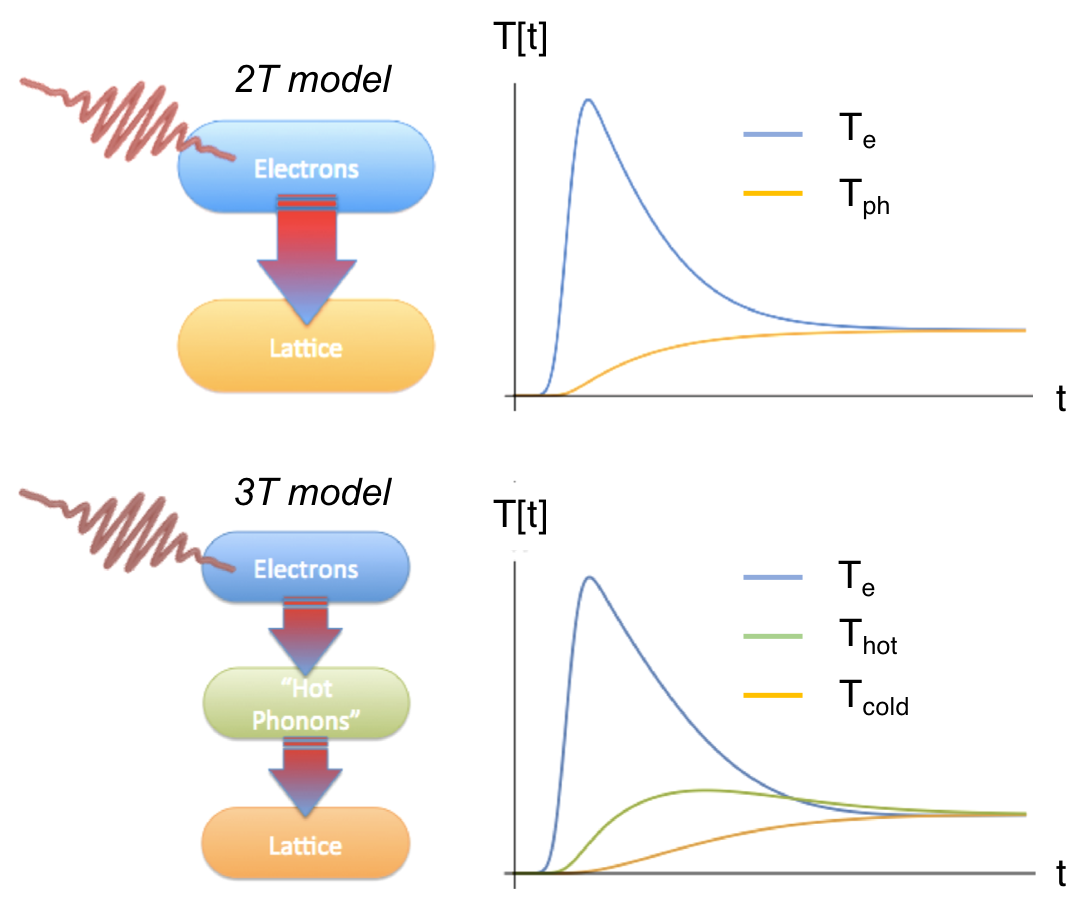}
 \caption{Representative time dynamics of the characteristic
temperatures in the 2T model (top panel) and in the 3T model
(bottom panel). Readaptation from Ref. \cite{johnson17}
}
\label{f-Tmodels}
\end{figure}

The dynamics of the electron and phonon temperatures
is schematically illustrated in the top panel of Fig. \ref{f-Tmodels}
(see also corresponding panels in Fig. \ref{f-timeranges}).
The validity of such approximate description relies
on the analysis of the relevant time scales.
In full generality,
the shortest one is designed to be the time-profile of the pump pulse,
in an ideal case a $\delta(t)$-function.
Due to the Heisenberg's uncertainty principle, however
a quasi-monochromatic laser cannot have a time confinement less
than the inverse of its energy bandwidth, $\Delta t \Delta W \approx
\hbar/2$. Typical pump-pulse have a duration of 
$30-50$ fs, which provides the first intrinsic (not many-body determined)
time scale.

The description in terms of a thermal electronic distributions
further relies on the assumption that electrons reach a their thermal distribution
in a time scale $\tau_{\rm ee}$ that is
much faster than the  characteristic time scales 
of electron-phonon scattering (Fig. \ref{f-timeranges}c).
Such fast electronic thermalization is conventionally attributed
to the  electron-electron Coulomb interaction $V^{\rm C}$
in Eq. (\ref{e:ham}), although the 
effective speed of such internal thermalization
among the electron degrees of freedom can be also very different
in semiconductors and metals.
In the first case, due to the low metallic character
-- and hence due to the poor screening of the Coulomb interaction --
the particle-particle interaction is very strong
and the internal thermalization between
the electronic degrees of freedom is thought to be reached
in few tens of femtoseconds \cite{tan17}.
In conventional metals, on the other hand, since the Coulomb interaction
is effectively screened by the strong metallic character,
the thermalization of electronic degrees of freedom
takes place on longer time scales, although
a detailed description of such processes is 
nowadays still under debate
\cite{groeneveld95,gusev98,delfatti00,hohlfeld00,rethfeld02,pietanza07,cui14,li21,novko21}.

\begin{figure}[t!]
    \centering
    \includegraphics[width=7.5cm]{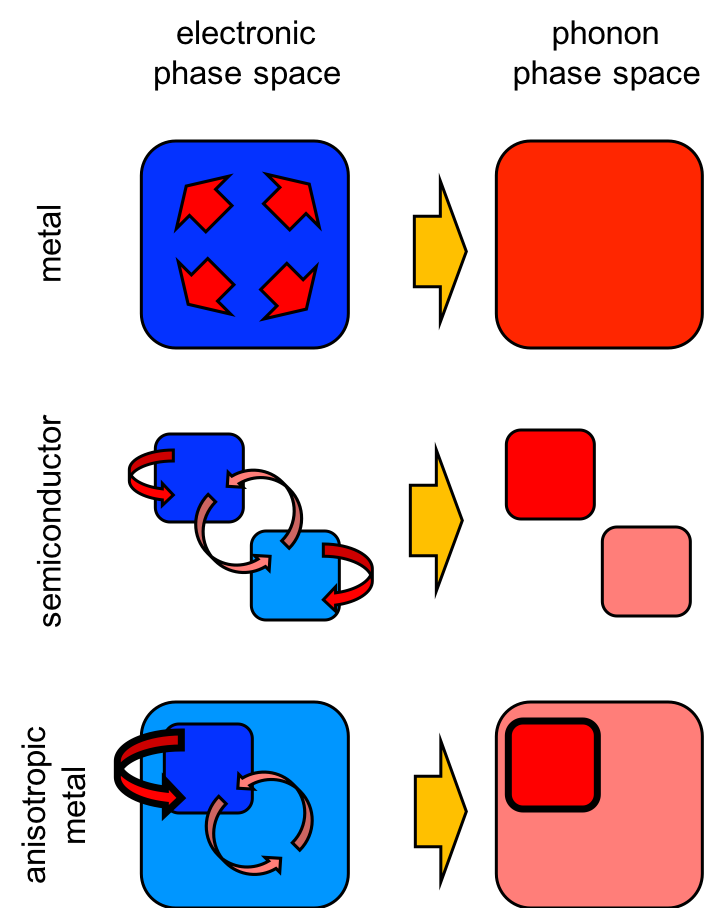}
    \caption{
Sketch of electron and phonon phase spaces
available for scattering for different families of materials.
In metals, the available electronic phase space
(dark blue) is spread in a uniform way all over the Brillouin zone.
Consequently exchanged phonons are scattered in an isotropic way
probing in an equal way the whole phonon phase space.
In semiconductors and semimetals characterized by a valley
degree of freedom (middle panel) pump-driven particle-hole excitations
probe essentially only states close to the valleys (here represented
by dark and light blue regions). Such situation brings to an
intrinsic selection of phonon modes: only intravalley and intervalley
modes, with small (red) and large (pink) momenta respectively, 
can effectively couple to electrons, and become hot.
The available electronic phase space in metals with
anisotropic electron-phonon coupling can be also divided in two
main sector: a core region (dark blue area) characterized
by strong electron-phonon interaction (red), and a large part
(light blue) with weak electron-phonon coupling (pink).
In such anisotropic scenario strongly coupled modes
can profit of energy transfer from the electrons
becoming hot while the remnant modes remain cold.
}
\label{f-phasespace}
\end{figure}

The reliability of the two-temperature model 
relies on the assumption that besides electrons also {\em phonons}
have reached a thermal distribution before the energy transfer
between electron and lattice degrees of freedom, as 
described by Eqs. (\ref{eq:el2})-(\ref{eq:eph2}),
takes place.
This is however a very delicate point.
Phonon thermalization is a natural consequence of the
anharmonic phonon-phonon scattering,
which plays a similar role for the lattice degrees of freedom
as the Coulomb interaction for the electrons.
In most of the compounds of interest, however,
anharmonic phonon-phonon coupling, although 
finite, represents the weakest interaction in Eq. (\ref{e:ham}),
and hence is thought to take place over a time scale $t_{\rm pp}$
longer than the electron-phonon processes.
The idea that phonons can be described by a thermal
distribution  {\em during} the time scale
of the electron-phonon relaxation is thus,
as a matter of fact, not supported by the assumption
of a preliminary anharmonicity-driven phonon thermalization.
In spite of this conundrum, the two-temperature model
has provided a satisfactory description for the time-resolved dynamics
of many known compounds, especially metals.
The reason of such consistency is related to the available
electronic phase space.
As represented in Fig. \ref{f-1sketch}b, in most of the materials
the photoexcited electrons (as well as the photoexcited holes)
are distributed all over the Brillouin zone with large
patches in the momentum space.
The subsequent electron-phonon scattered momenta sample
the whole phonon Brillouin zone to a great extent in an isotropic way,
providing an optimal ground for an easy and efficient thermalization.
The link between the available electron and phonon
phase spaces is sketched in Fig. \ref{f-phasespace}
for standard metals as well as for other compounds.

Quite different can be the scenario for semiconductors and semimetals
characterized by a valley degree of freedom, namely
a band gap (ideally a zero or negative band gap, for semimetals)
with degenerate band edges.
Under such circumstances, the pump-driven excitations
can be concentrated in a few regions nearby band valleys.
A paradigmatic example of this scenario
with a strong topical interest is graphene
(see for instance Fig. \ref{f-1sketch}a),
a two-dimensional semimetal characterized
by a linear Dirac dispersion extending for $\pm 1$ eV
around the K, and K$^\prime$ points of the Brillouin zone.
The reduced phase space for particle-hole excitations
implies a reduced phase space also for 
the electron-phonon scattering,
where only few lattice modes ruling {\em intra}valley processes
(small phonon momenta) or {\em inter}valley processes
(large phonon momenta) are allowed, as schematically depicted
in Fig. \ref{f-phasespace}.
Such selected modes (small-${\bf q}$ momenta;
large-${\bf q}$ momenta) provide thus the main channels
for energy transfer from the electronic degree of freedom to the
lattice sector.
Absorbing energy from the excited electrons,
these modes become {\em hot phonons}, i.e., 
they acquire a non-thermal population  much larger than the
remnant lattice modes that stay {\em cold},
obeying an effective thermal distribution with a moderate temperature.
Considered that anharmonic phonon-phonon mixing occurs
on a much longer time scale than the dynamics relevant here,
hot phonons can survive and be effectively detected
on the scale of hundreds of femtoseconds or even few picoseconds.

In order to describe the time dynamics
of a coupled electron-phonon system
under hot-phonon conditions, a useful 
{\em three-temperature} (3T) model is often employed\,\cite{bib:perfetti07,johannsen13,stange15,novko2019,caruso20,ncdc,maklar21,majchrzak21}.
In such extended version of the 2T model, the parameter
$T_{\rm e}$ still describes the effective temperature
of the electron degrees of freedom which are assumed to be
thermalized before the model applies. The majority
of the lattice modes stay cold, and are thus described
in an efficient way
by a thermal distribution governed by the effective temperature
$T_{\rm cold}$.
On the other hand hot modes, adsorbing energy directly from
the electronic sector, and not redistributing to the other modes,
acquire a phonon population $b_{\rm hot}$ much larger
than the cold modes.
The basic (reasonable) assumption of the 3T model
is that few hot modes are essentially degenerate
(as for instance the optical $E_{2g}$ modes in graphene)
displaying similar energies ($\omega_{\rm hot}$)
and a similar phonon population $b_{\rm hot}$.
The phonon population $b_{\rm hot}$
can be hence conveniently expressed in terms
of an effective temperature $T_{\rm hot}$ for the hot modes:
\begin{eqnarray}
b_{\rm hot}
&=&
\frac{1}{\exp\left[
\frac{\displaystyle \hbar \omega_{\rm hot}
}{
\displaystyle T_{\rm hot}}\right]-1}
.
\end{eqnarray}

Within the validity of the three-temperature model,
Eqs. (\ref{eq:el2})-(\ref{eq:eph2})
can be generalized as:
\begin{eqnarray}
C_{\rm e}
\frac{d T_{\rm e}}{\partial t}
&=&
-
G_{\rm hot}(T_{\rm e}-T_{\rm hot})
- G_{\rm cold}(T_{\rm e}-T_{\rm cold})
+
S(t)
,
\label{eq:el}
\\
C_{\rm hot}\frac{d T_{\rm hot}}{d t}
&=&
G_{\rm hot}(T_{\rm e}-T_{\rm hot})
-
V_{\rm anharm}\left(T_{\rm hot}-T_{\rm cold}\right)
,
\label{eq:e2g}
\\
C_{\rm cold}\frac{d T_{\rm cold}}{d t}
&=&
G_{\rm cold}(T_{\rm e}-T_{\rm cold})
+
V_{\rm anharm}\left(T_{\rm hot}-T_{\rm cold}\right)
,
\label{eq:ph}
\end{eqnarray}
where $C_{\rm hot}$, $C_{\rm cold}$ account for the specific heat
capacities of the hot and cold modes, respectively.
In similar way $G_{\rm hot}$,  $G_{\rm cold}$ describe
the electron-phonon relaxation rates related
to hot and cold modes.
In Eqs. (\ref{eq:el})-(\ref{eq:ph}) we have also explicitly included
a term $\propto V_{\rm anharm}$
due to the anharmonic phonon-phonon coupling restoring
(on a long time scale) the global equilibrium among the
lattice degrees of freedom.
This term is conventionally expressed in term of the
specific heat capacity of the hot modes by introducing
a characteristic time scale $\tau_{\rm anharm}$,
$V_{\rm anharm}=C_{\rm hot}/\tau_{\rm anharm}$ \cite{perfetti07}.
A sketch of the time dynamics of the three representative temperatures
is shown in the bottom panel of Fig. \ref{f-Tmodels}.


\section{Detecting hot phonons}
\label{s:detect}

The concept of hot phonon refers
to a scenario in which one (or few) vibrational modes of the lattice  are characterized by a population $b_{{\bf q},\nu}$ significantly larger than the majority of other modes. These latter (cold) modes obey a standard thermal distribution 
ruled by a background temperature $T_{\rm cold}$.
In spite of such compelling definition, an active role of hot phonons
in a material is often not probed directly and their
presence is inferred in indirect ways.
One of the first experimental evidences
of the presence of hot phonons was prompted by the analysis
of the electric-field dependence of transport properties in
low-carrier-density doped semiconductors 
\cite{shah83,potz87,micke96}. 
Within the framework of a Boltzmann approach,
the charge transport is governed
by an electronic distribution $f_{{\bf k},\vec{E}}$
 {\em in presence} of an electric field $\vec{E}$.
The electric field in non-interacting systems would result
in a constant shift in time of the electronic distribution,
and hence to an infinite conductivity.
A steady state is however obtained as a balance
between such velocity drift and collision
processes that scatter back the electrons, namely \cite{mahanbook,zimanbook}:
\begin{eqnarray}
-e \vec{E} \cdot \vec{\nabla}_{\bf k} f_{{\bf k},\vec{E}}
&=&
\left.
\frac{
df_{\bf k}(\{f,b\})}{dt}
\right|_{\rm ee}
+
\left.
\frac{
df_{\bf k}(\{f,b\})}{dt}
\right|_{\rm ep}
,
\end{eqnarray}
where the scattering terms
$df_{{\bf k}}(\{f,b\})/dt|_{\rm ee}$,
$df_{{\bf k}}(\{f,b\})/dt|_{\rm ep}$
have been discussed in the previous Section.
The net result
can be regarded as an
effective displacement of the Fermi volume along the direction
of $\vec{E}$ \cite{mahanbook,zimanbook}, leading to a net finite electronic
momentum.
While the Boltzmann equations are often solved
assuming an unperturbed phonon distribution function
$b^{(0)}_{{\bf q},\nu}$ (hence with a thermal profile),
such assumption holds true only in the linear regime, valid
for small electric fields and for not too low temperatures where phonon drag effects can occur.
In real systems (as well as in a variety of more refined theoretical
modeling), the coupling between the electron and lattice degrees
of freedom gives rise to a phonon distribution function
$b_{{\bf q},\nu}(\{f,b\})$ which may differ
significantly
from the thermal unperturbed one $b^{(0)}_{{\bf q},\nu}$,
affecting as well the resulting steady electronic function
$f_{{\bf k}}(\{f,b\})$.
The role of a non-thermal $b_{{\bf q},\nu}(\{f,b\})$,
initially regarded as ``phonon disturbance'' have been experimentally detected
and theoretically investigated for a long time,
especially in high electric-field regime,
and it becomes more evident under hot-phonon conditions,
where only few lattice modes are excited by the coupling
with the electrons\,\cite{kocevar72,shah83,kocevar85,price85a,price85b,lugli88,micke96,lazzeri06,weng21}.

It has also been shown that hot phonons and non-equilibrium phonons
play a crucial role
in tailoring the interband properties in narrow gap semiconductors,
with respect to impact ionization processes and quantum laser
cascades \cite{micke90,micke90b,paula97,paula97b,paula98,iotti10,shi14},
which provide in this class of materials another way to detect them.

The presence and an active role of hot phonons
is also commonly revealed in the analysis of the time dependence
of electronic properties upon non-equilibrium conditions.
Given the plethora of ultrafast pump-probe experiments
nowadays accessible, the physical properties under investigation
can vary in a wide range, from optical probes
(transmission
\cite{langot96,wang10,huang11,breusing11,pogna21,chan21},
reflectivity
\cite{ishioka08,gadermaier10,mansart10,bib:dalconte12,bib:dalconte15,price15,golla17,baldini},
absorption \cite{yang16,yang17,fu17,novko2019,novko2021})
to non-linear optics \cite{scheuch11},
time-resolved photoelectron
spectroscopy \cite{bib:perfetti07,bib:johannsen13,rettig13,avigo13,stange15,yang17new,caruso20},
photoluminescence \cite{lui10,sekiguchi21},
time-dependent Raman probes \cite{hannah13}, 
ultrafast diffraction \cite{carbone08,carbone10,mansart13,johnson15,harb16,waldecker16,waldecker17,konstantinova18,karam18,zhan20,seiler21}.

Similarly, hot-phonon effects
have been traced down
in electron properties
and in time-resolved Raman spectroscopy
by means of theoretical simulations \cite{potz87,joshi89,kim91}.

\begin{figure}[t!]
\centering
\includegraphics[width=9cm]{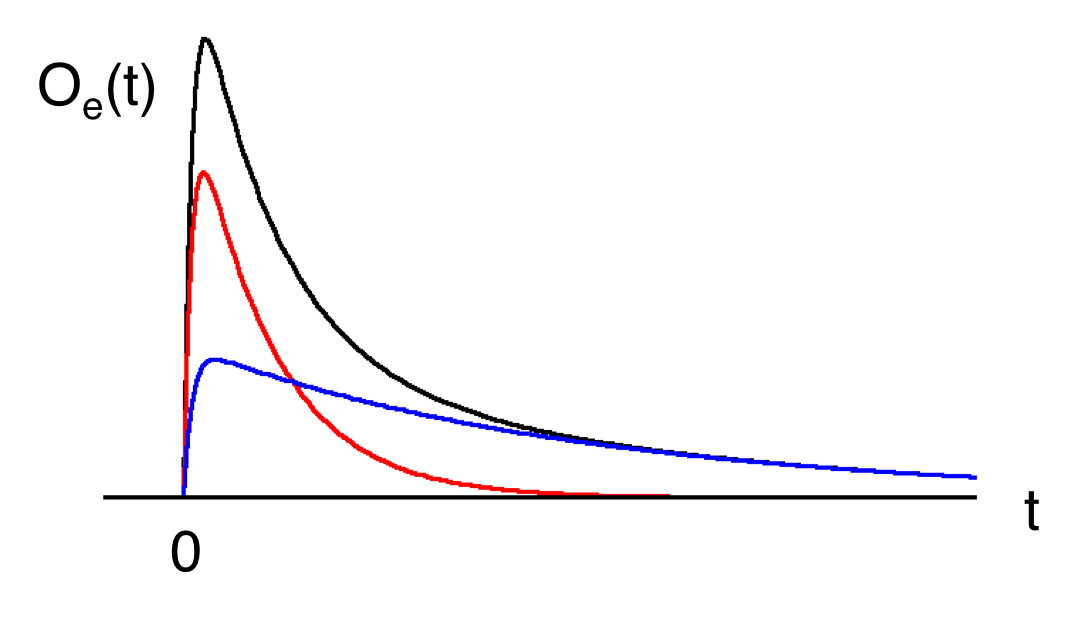}
\caption{
Sketch of the pump-probe time dynamics of a generic electronic observable
$O_{\rm e}(t)$ in the presence of two different channels of
relaxation.
The total time dependence (black line) results from
the overlap of two components: one ruled by a faster decay (red line) associated with scattering with
hot-phonons;
and a slower one (blue line), which is associated with
energy transfer to cold modes. On this scale, the shorter time over
which electrons thermalize among themselves is not depicted.}
\label{f-Ovst}
\end{figure}

Core of these analyses is the evidence of multiple relaxation time scales
in the ultrafast dynamics, which can be roughly rationalized
in terms of two different channels with different characteristic
time scales 
(biexponential decay)
as depicted in Fig. \ref{f-Ovst} : one (faster) channel governed
by the energy transfer between the electron degrees of freedom
and the hot phonons; and a second slower one associated with the
final thermalization of the remnant cold lattice modes.
On the experimental ground these different time scales,
and their relative weights, can be revealed by
an appropriate fitting with two relaxation rates,
\begin{eqnarray}
O_{\rm e}(t)
&\approx&
\alpha_1\exp\left(-\frac{t}{\tau_1}\right)
+
\alpha_2\exp\left(-\frac{t}{\tau_2}\right),
\end{eqnarray}
where $O_{\rm e}$ is a generic physical observable ruled
by the electronic degrees of freedom.

A deeper quantitative insight can be gained if one
is able to translate the time dependence of the electronic
probe $O_{\rm e}$ in the time dependence
of a corresponding effective electronic temperature $T_{\rm e}$.
The multiple time-scale dependence of $T_{\rm e}$
can be compared and fitted with the predictions of a three-temperature
model as described in Eqs. (\ref{eq:el})-(\ref{eq:ph})\,\cite{perfetti07,johannsen13,stange15}.
From this analysis fundamental information are hence extracted,
 as the factors $G_{\rm hot}$,  $G_{\rm cold}$ which are
proportional to the electronic coupling with the hot and cold bosonic
modes, respectively.

It is worth to pointing out that the above described ways of detecting
hot phonons in a material rely on the analysis of hot-phonon effects
on {\em electronic} properties.
On the other hand, since a hot phonon is an object defined
by its bosonic population which is a property logically independent
of the electron-phonon coupling, it looks more natural devising
and employing, when possible, techniques that are based
on the analysis of purely lattice features.
Along this perspective, 
considered that hot phonons are usually located at high-symmetry
points of the Brillouin zone, and in particularly at the  zone-center,
a popular tool is the analysis, when feasible, of the relative
intensity
of the Stokes and anti-Stokes peaks in Raman
spectroscopy \cite{yan09,berciaud10,kang10,wu2012,hannah13,yang17new,ncdc,pellatz21}.
The Stokes and anti-Stokes intensity for a ${\bf q=0}$
Raman active mode $\nu$,
$I_{\rm S}$,
$I_{\rm aS}$, respectively, are indeed known to be governed
by the population of the  Raman mode:
\begin{eqnarray}
I_{\rm S}
&\approx&
1+n(\omega_\nu),
\\
I_{\rm aS}
&\approx&
n(\omega_\nu).
\end{eqnarray}
The population $n(\omega_\nu)$ can be thus simply estimated
from the relation \cite{wu2012,ncdc,pellatz21}:
\begin{eqnarray}
\frac{I_{\rm aS}}{I_{\rm S}}
&=&
\frac{\omega_{\nu,{\rm aS}}^4}{\omega_{\nu,{\rm S}}^4}
\frac{n(\omega_\nu)}{1+n(\omega_\nu)}
\approx
n(\omega_\nu)
,
\end{eqnarray}
where $\omega_{\nu,{\rm S}}$,
$\omega_{\nu,{\rm aS}}$ are the frequencies of the Stokes
and anti-Stokes resonances, respectively, and where
 the last relation holds true
under standard conditions
$(\omega_{\nu,{\rm S}}-\omega_{\nu,{\rm aS}})/
(\omega_{\nu,{\rm S}}+\omega_{\nu,{\rm aS}}) \ll 1$,
$n(\omega_\nu \ll 1$.
An effective temperature for this mode can be also promptly estimated
from\,\cite{berciaud10}
\begin{eqnarray}
n(\omega_\nu)
&=&
\left[\exp\left(\frac{\omega_\nu}{T_\nu}\right)-1
\right]^{-1}
\approx
\exp\left(-\frac{\omega_\nu}{T_\nu}\right)
.
\end{eqnarray}
The comparison between the time evolution of the effective temperature $T_\nu$
of a candidate \emph{hot mode} $\nu$ with the electronic temperature
$T_{\rm e}$
assessed by other means can thus provide a robust evidence
of a hot-phonon scenario.
Such evidence can be furthermore corroborated if
the time evolution of another Raman-active mode,
belonging to the cold-mode sector, can be experimentally resolved,
providing a direct check of the different time dynamics
of the two ( hot and cold) lattice mode sectors.

While the Raman spectroscopy can inspect in a direct experimental way
the population of few (Raman-active) modes at the zone-center of the
Brillouin zone, from a theoretical perspective
the snapshot at given time $t$ of the distribution
of phonon population $b_{{\bf q},\nu}$
for each branch $\nu$ and momentum ${\bf q}$ 
can be obtained by the direct solution of
Eqs. (\ref{e:dftot})-(\ref{e:bdynep}),
permitting to reveal thus
selective high populations for particular momenta.
This approach has been indeed widely employed,
ranging from the seminal work by P\"otz \cite{potz87}
to later works \cite{langot96,lazzeri06,butscher07,richter09} up to more recent refined
analyses \cite{zhan20,seiler21,caruso21}.

Furthermore, direct information about
the time evolution of the lattice dynamics under non-equilibrium
conditions
can be conveniently addressed by means
of ultrafast diffraction
measurements \cite{carbone08,carbone10,trigo10,mansart13,chatelain14,johnson15,chase16,harb16,waldecker16,waldecker17,stern18,konstantinova18,cotret19,zhan20,seiler21,otto21}.
At a first level, useful information can be gained by means of the
time-resolved analysis of the intensity of the Bragg peaks,
which can be related to the magnitude mean square lattice
displacements as:
\begin{eqnarray}
\frac{I({\bf Q},t)}{I({\bf Q},0)}
&=&
\sum_i \mbox{e}^{-2W_i({\bf Q})},
\end{eqnarray}
where ${\bf Q}$ denotes the wave-vector
in the crystallographic reciprocal space,
$I({\bf Q},0)$ is the intensity before the pump excitation
and where
$W_i({\bf Q})$ is the well-known Debye-Waller factor
for a given atom $i$ in the unit cell \cite{zimanbook}:
\begin{eqnarray}
W_i({\bf Q})
&=&
\frac{1}{2}
\left\langle
\left(
{\bf Q} \cdot {\bf u}_i
\right)^2
\right\rangle
.
\end{eqnarray}
A sudden increase of the Debye-Waller factor
portends a sudden energy transfer from the electron
to the lattice mode sector.
Since the Debye-Waller factor results from the contribution
of {\em all} the phonons in the Brillouin zone, it cannot reveal
in a direct way the onset of hot-phonon physics.
Hot modes and, in general, a non-thermal phonon distribution
can however be inferred in an indirect way from
the possible anisotropy of the estimated mean-square lattice
displacements
and from the observation of two-component decay
rate \cite{carbone08,carbone10,mansart13,johnson15,harb16,waldecker16,waldecker17,konstantinova18,zhan20,seiler21}.

\begin{figure}[t!]
\centering
\includegraphics[width=11cm]{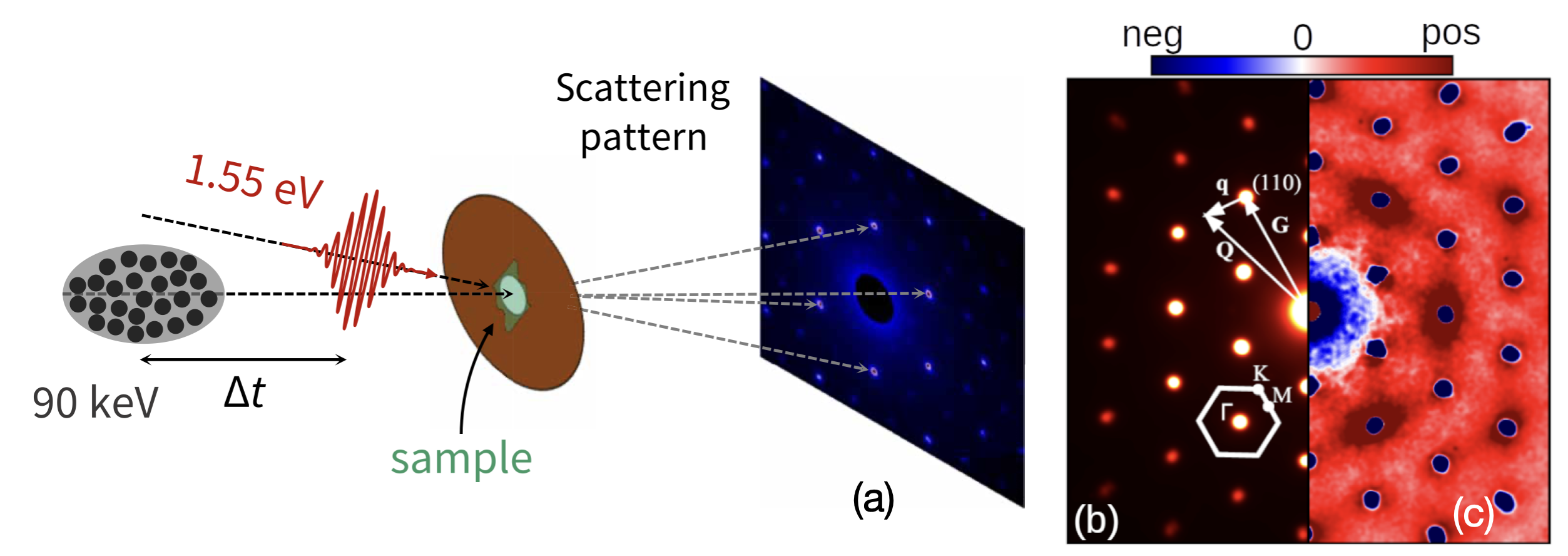}
\caption{(a) Schematic illustration of a UEDS experimental setup. From Ref.~\cite{otto21}. 
(b) Diffuse scattered intensity in bulk MoS$_2$. 
(c) Transient changes of the scattered electron intensity for a time delay $\tau = 100$~ps following photo-excitation. From Ref.~\cite{Zacharias_joint_PRB}.
    }
\label{f-uedsfig}
\end{figure}

Non-thermal phonon populations can be efficiently investigated by means
of ultrafast electron diffuse scattering (UEDS), which provides
a versatile diagnostic tool
for detecting and investigating the emergence of hot phonons in condensed matter 
\cite{trigo10,chase16,harb16,waldecker16,stern18}.
In UEDS the probe is represented by an electron beam hitting the
sample.
The scattered electrons are detected  with momentum resolution via an
electron camera, as schematically depicted in Fig.~\ref{f-uedsfig}a. 
While a fraction of electrons emerges from the sample unaffected (i.e., without undergoing scattering of any sort),
additionally, electrons can undergo elastic and inelastic
scattering processes with the lattice. 
The intensity of the scattered electrons can be Taylor-expanded as a sum of zero-phonon and one-phonon contributions \cite{Zacharias_joint_PRB,notezach}: 
\begin{eqnarray}
I({\bf Q})
&=&
 I_0({\bf Q}) + I_1 ({\bf Q})  + \cdots \quad.
\end{eqnarray}
The first term here accounts for Bragg (elastic) scattering,
which can be expressed in the present formalism as:
\begin{eqnarray}
I_0(\bQ)
&\propto& 
 \sum_{i,j} f_i (\bQ) f^*_j (\bQ)
    \cos\big[ \bQ \cdot (\bt_i - \bt_j)\big]
\nonumber\\ 
 &&
\times \mbox{e}^{-W_i (\bQ)} e^{-W_j (\bQ)}  \delta_{\bQ,\bG}.
\end{eqnarray}
where  $\btau_i$ are the coordinates of the $i$-th atom in the unit
cell, $W_i$  the corresponding Debye-Waller factor, and $f_i$ is the
atomic scattering amplitude. 

Further interesting information about the lattice properties
are conveyed in the one-phonon term which can be expressed as:
\begin{eqnarray}
I_1(\bQ)
&\propto& 
\hbar  \sum_{i,j} f_i (\bQ) f^*_j (\bQ)
\frac{ e^{-W_i (\bQ)}
   \mbox{e}^{-W_j (\bQ)}}{\sqrt{M_i M_j}}
\nonumber  \\
&&
\times  
\sum_{\nu } \text{Re}
\Big[ \bQ \cdot {\bf e}_{i,\nu}(\bQ)
   \bQ \cdot {\bf e}^{*}_{j, \nu} (\bQ)
\mbox{e}^{i\bQ \cdot [\bt_i - \bt_j]} 
\Big] 
\frac{[n_{{\bf Q}  \nu} + 1/2 ]}{\omega_{{\bf Q} \nu}}
,
\end{eqnarray}
where ${\bf e}_{i,\nu}$ is the phonon eigenvectors
for the phonon branch $\nu$, and $M_i$ is the atom mass for atom $i$.
The linear dependence of $I_1$ on the phonon number  $n_{{\bf Q}\nu}$
provides a rationale for the relating transient changes in the scattered
UEDS intensity
[namely,
$\Delta I ({\bf Q}, t) \equiv I({\bf Q}, t) - I ({\bf Q}, 0)$]
to underlying changes in the 
phonon populations, and it provides
thus a suitable experimental framework to 
explore the emergence of hot phonons \cite{waldecker16}. 

UEDS is particularly well suited to detect momentum anisotropies
established in the phonon population following 
photo-excitations \cite{waldecker17,stern18,cotret19,seiler21,otto21}.
On the other hand, 
the determination of a {\it mode-resolved} phonon population
is more challenging  as it would  require the knowledge
of the scattered electron energy with the resolution of few meVs. 
To overcome this limitation, a novel procedure for recovering 
the phonon population dynamics with full mode resolution,
based on the analysis  of the transient scattering intensity,
has recently been proposed by 
de Cotret {\em et al}., with a direct demonstration in graphite \cite{cotret19}.


\section{Anisotropic electron-phonon material: MgB$_2$}
\label{s-mgb2}

\begin{figure}[t!]
    \centering
    \includegraphics[width=11cm]{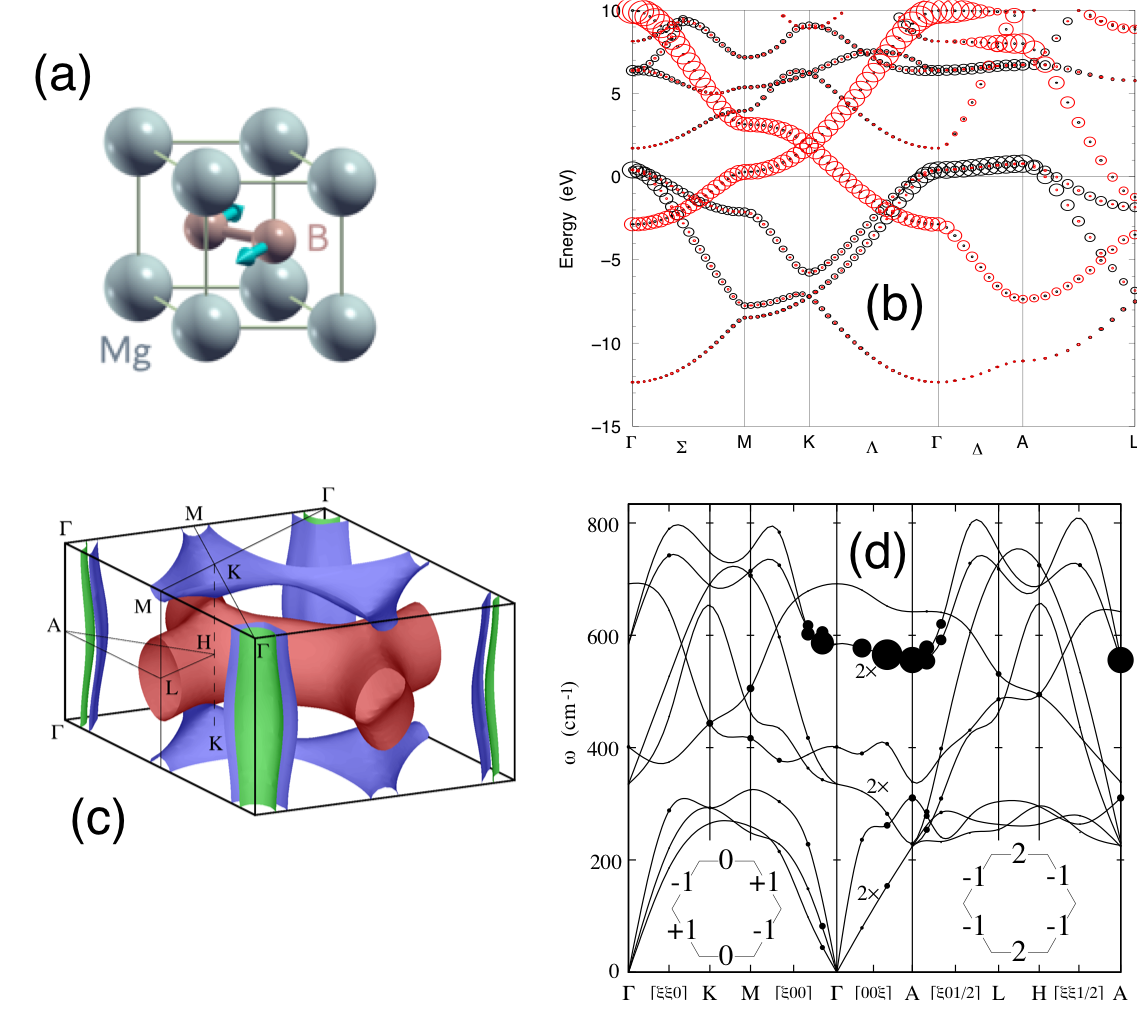}
    \caption{
(a) Crystal structure of MgB$_2$. Also displayed are the lattice displacements
of the in-plane $E_{2g}$ mode at ${\bf q}=0$.
(b) Electronic band dispersion of MgB$_2$. 
The size of the red symbols represents the content
of $p_z$ B orbitals, typical of $\pi$-bands, while
the size of black symbols is related to the content
of $p_x$, $p_y$ B orbitals, representing thus $\sigma$-bands.
From Ref. \cite{bib:kortus01}.
(c) Fermi surface of MgB$_2$. Tubular structures origin from
$\sigma$-bands with a strong two-dimensional character, while
the central complex sheets stem from the $\pi$-bands.
From Ref. \cite{bib:kortus01}.
(d) Phonon dispersion of MgB$_2$.
The size of black dots is proportional to the
magnitude of the electron-phonon coupling $\lambda_{{\bf q},\nu}$.
From Ref. \cite{kong}.
}
\label{f-mgb2}
\end{figure}

The crystal structure of MgB$_2$ is rather simple,
with hexagonal graphene-like planes of B atoms spaced vertically by Mg atoms
located in the center of the hexagons (Fig. \ref{f-mgb2}a).
MgB$_2$
has been on the shelves of chemistry laboratories for decades
without attracting much attention.
A considerable interest about this compound arose however
in 2001 after superconductivity
below $T_c\approx 39$ K was found \cite{bib:nagamatsu01}.
Soon it was clear that the superconducting pairing in MgB$_2$
was driven by the electron-phonon
coupling \cite{an,bib:kortus01,yildirim,bib:liu01,kong,bib:choi02a}.
Discovered during the ``gold rush'' for new high-$T_c$ superconductors triggered by the report of superconductivity with $T_c \lesssim 100$ K
in cuprates
in 1986 and later years \cite{bednorz86,wu87,cava87,hazen88},
MgB$_2$ demonstrated that strong electronic correlations were
not the unique path for high-$T_c$ superconductivity,
but that electron-phonon coupling was not to rule out.
Surely, it was also soon clear that the properties of the electronic structure and of the electron-phonon coupling in MgB$_2$ were quite peculiar with respect to conventional phonon-based low-$T_c$ superconductors.
For instance, in the electronic band structure, two blocks of bands could be identified, as shown in Fig. \ref{f-mgb2}b: a $\sigma$-band sector, with a strong two-dimensional character, built by the $p_{x/y}$ orbitals of boron, similarly as in graphene;
and a $\pi$-band sector, with a strong three-dimensional character,
resulting by the hybridization of Mg atoms with B-$p_z$
orbitals.
It should be further noticed that, while the $\pi$-bands display
large Fermi surfaces, denoting a robust metallic character,
the $\sigma$-bands appear to be only slightly doped \cite{an},
resulting a thin tubular Fermi sheets (Fig. \ref{f-mgb2}c).
Because of the different properties of such two electron-band blocks, in-plane ($x-y$) and out-of-plane ($z$)
electronic and optical properties results to be
anisotropic \cite{masui03,fudamoto03,guritanu06,xi08},
revealing the {\em multiband} character of this material.

Like many metals, MgB$_2$ is not the only compound showing
multiple different Fermi sheets. However, it can be considered
without doubt, along with pnictides, the most evident superconducting material showing
{\em multiple different superconducting gaps}, as assessed by a variety
of experimental and theoretical checks \cite{giubileo,tsuda,chen2001,gonnelli,mou,giustino17}.
The logical shift from anisotropic {\em multiband} properties
to anisotropic {\em multigap} properties should not be overlooked.
Many conventional metals are characterized with multiple Fermi sheets
having different electronic properties but rather homogeneous
superconducting properties over all the bands.
The reason is that the electron-phonon scattering
couples the electrons of one band with all the other bands,
resulting in an effective averaging of the electronic properties
relevant for the superconductivity.
In order to sustain multiple gap features a superconducting material
needs not only to display multiband properties,
but also to possess an {\em anisotropic electron-phonon} coupling
\cite{noteanis} that allows to reflect
the anisotropic multiband character in the superconducting properties.

\begin{figure}[t!]
    \centering
    \includegraphics[width=11cm]{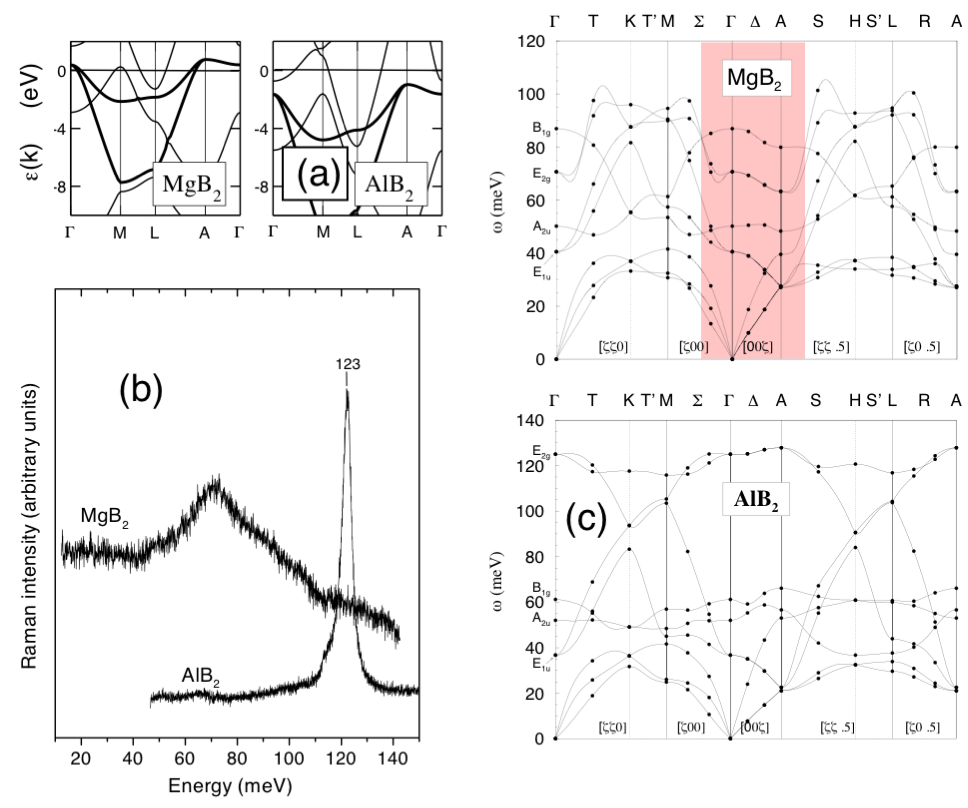}
    \caption{
(a) Electronic band structure of MgB$_2$ and AlB$_2$.
Thick lines denote the $\sigma$-bands which are slightly hole-doped
in MgB$_2$ whereas they are completely filled in AlB$_2$.
From Ref. \cite{boeri02}.
(b) A comparison of the $E_{2g}$ phonon spectrum
in MgB$_2$ and AlB$_2$ as probed
from Raman spectroscopy.
(c) Theoretical phonon dispersion for MgB$_2$ and AlB$_2$.
MgB$_2$ shows a remarkable softening
of the $E_{2g}$ branch for small in-plane ${\bf q}$ momenta
(red area). Such softening results from the Kohn anomaly
due to the metallic screening of the $\sigma$-bands,
which are strongly coupled with the $E_{2g}$ modes.
This feature is not present in AlB$_2$ where the $\sigma$-bands
are completely filled and they do not provide any metallic screening.
Panels (b)-(c) from Ref. \cite{bohnen}.
}
\label{f-compare}
\end{figure}

Although MgB$_2$ has two $\sigma$-bands
and two $\pi$-bands at the Fermi level,
it is a common practice to compact such
electronic space in an unique $\sigma$-block and a unique $\pi$-block.
In such reduced $2 \times 2$ analysis, 
the total electron-phonon coupling described by the Eliashberg
function $\alpha^2F(\omega)$ can be split in four
components $\alpha^2F_{ij}(\omega)$ ($i,j=\sigma,\pi$).
One can as well compute the dimensionless electron-phonon
coupling $\lambda_{ij}=2\int d\omega \alpha^2F_{ij}(\omega)/\omega$
constant that expresses in a compact way
the intraband and interband terms.
The corresponding electron-phonon matrix 
presents a strongly anisotropic structure peaked
in the $\sigma$-$\sigma$ sector
\cite{an,bib:liu01,kong,bib:choi02a,bib:choi02b,golubov},
with typical values \cite{golubov}:
\begin{eqnarray}
\hat{\lambda}
&=&
\left(
\begin{array}{cc}
\lambda_{\sigma\sigma} & \lambda_{\sigma\pi}  \\
\lambda_{\pi\sigma} & \lambda_{\pi\pi} 
\end{array}
\right)
=
\left(
\begin{array}{cc}
1.02 & 0.21 \\
0.16 & 0.45
\end{array}
\right).
\end{eqnarray}
The spectral analysis of the Eliashberg functions $\alpha^2F_{ij}(\omega)$
shows that the predominance of the intraband $\sigma$-$\sigma$
coupling is associated with the appearance
of a remarkable peak in the energy window $[550:650]$ cm$^{-1}$
\cite{yildirim,bib:liu01,kong,bohnen,golubov,bib:choi02b}.
A deeper insight can be gained by investigating the momentum distribution
of the electron-phonon coupling along the different branches of the
phonon dispersion.
A representative plot is shown in Fig. \ref{f-mgb2}d
where the size of the black circles reflects the magnitude
of the electron-phonon coupling.
One can notice that the electron-phonon coupling is strongly
anisotropic also with respect to phonon momentum/branch,
being almost exclusively concentrated 
at small momenta
of the optical $E_{2g}$ branch with in-plane lattice displacements
(Fig. \ref{f-mgb2}d).
Such modes are strongly coupled with the electronic $\sigma$-band and
are essentially the only contributors to $\lambda_{\sigma\sigma}$.
As a consequence of the two-dimensional character of the $\sigma$-bands,
with a negligible dispersion along perpendicular momentum $k_z$, the
electron-phonon coupling of these modes is also weakly dependent
on the vertical component of the phonon momentum $q_z$,
contributing in an equivalent way along the $\Gamma$-A path.
Note that the phonon band structure shows a remarkable
depletion of the $E_{2g}$ dispersion in a close correspondence
of the region with strong electron-phonon coupling.
Such feature can be rationalized as an effect of a
Kohn anomaly driven by the electron-phonon coupling
of the $E_{2g}$ modes with the two-dimensional $\sigma$-bands,
and following qualitatively the behavior of a two-dimensional
Lindhard function.
The role of the electron-phonon coupling
is even more striking when comparing
the phonon dispersion of MgB$_2$ with the similar
isostructural compound AlB$_2$, which does not display
such softening close to the $\Gamma$ point.
AlB$_2$ presents strong similarities with MgB$_2$, with the only
relevant difference of possessing one electron more per cell.
Such additional charge leads to a complete filling of the
$\sigma$-bands that in AlB$_2$ lay 1-1.5 eV below the Fermi level
(see Fig. \ref{f-compare}a),
preventing thus a metallic screening due to the $\sigma$ bands.

The key role of the strongly-coupled zone-center $E_{2g}$ modes
in governing the electron-phonon interaction,
and hence the superconducting properties of MgB$_2$,
has been extensively discussed in literature.
The most striking feature soon after the discovery of
superconductivity in MgB$_2$ was the huge broadening
of the $E_{2g}$ peak observed in Raman spectroscopy,
$\Delta \Omega_{E_{2g}}\approx 25$ meV, that can be compared with
$\Delta \Omega_{E_{2g}}\approx 5$ meV in AlB$_2$ (Fig. \ref{f-compare}b).
Anharmonicity was at the beginning suggested as responsible for such
large broadening in MgB$_2$, supported by first-principle
frozen-phonon calculations
of the energy potential, showing a strong deviations from
a quadratic behavior at large lattice displacements \cite{yildirim}.
On the basis of such anharmonic potential the $\omega_{E_{2g}}$
phonon frequency was estimated to harden from
$\omega_{E_{2g}}=60.3$ meV (harmonic limit)
to $\omega_{E_{2g}}=74.5$ meV (anharmonic calculation \cite{yildirim}. 
The actual relevance of anharmonicity was however
later questioned by further experimental
and theoretical investigations
\cite{rafailov,shukla03,lazzeri03,dastuto07}
showing that the anharmonic contribution due to the
direct phonon-phonon scattering to the phonon linewidths 
and phonon shifts was marginal with respect to the
observed ones \cite{bohnen,goncharov,hlinka,postorino1,quilty02,quilty03,martinho03,renker}.
These controversial findings have been rationalized
by noticing that the strong anharmonicity computed in Ref. \cite{yildirim}
by frozen phonon calculations was not associated with
a quartic term (fingerprint of a direct phonon-phonon coupling) but
it was rather a byproduct of a strong conventional linear electron-phonon
coupling in the presence of a dynamical Lifshitz transition
induced by the effects of the $E_{2g}$ lattice displacement on the electronic
structure \cite{yildirim,boeri02,boeri05,bianconi15}.
The Lifshitz transition appears to be physically accessible in MgB$_2$ due
the small Fermi energy of the $\sigma$-bands ($E_{\rm F}^\sigma\sim
0.4-0.5$ eV) and due to the strong lattice fluctuations
triggered by the strong electron-phonon coupling of the $E_{2g}$ mode.
Once again, such scenario points towards the presence
of a peculiarly strong electron-phonon coupling for the
$E_{2g}$ modes, whereas at the same time the evidence of strong
lattice fluctuations points out the relevance of nonadiabatic
effects \cite{boeri05,cappelluti06,cappelluti06b,novko18}.
A similar situation is encountered in a quantitative analysis
of the Raman $E_{2g}$ phonon frequency and linewidth,
where the experimental data has been successfully reproduced
within the context of  a many-body renormalized nonadiabatic theory
for the phonon self-energy \cite{cappelluti06,novko18}.


\section{Hot-phonon physics in MgB$_2$}
\label{s-hotmgb2}

The  strong anisotropic multiband character of the electron-phonon
coupling in MgB$_2$ has been widely discussed
and assessed since long time, both from the experimental as from the theoretical points
of view.
It was however not recognized until recently that
this scenario could naturally lead in MgB$_2$
also to hot-phonon physics
 \cite{baldini}. Indirect experimental evidences of hot phonons
in time-resolved optical probes were in that work also discussed.

Although prompting the seminal suggestion of hot phonons
in MgB$_2$ that stem from the anisotropic electron-phonon coupling,
the evidence in support of this scenario was in
Ref. \cite{baldini} quite indirect and no quantitative analysis
of the energy transfer processes were given.

\begin{figure}[t!]
    \centering
    \includegraphics[width=7cm]{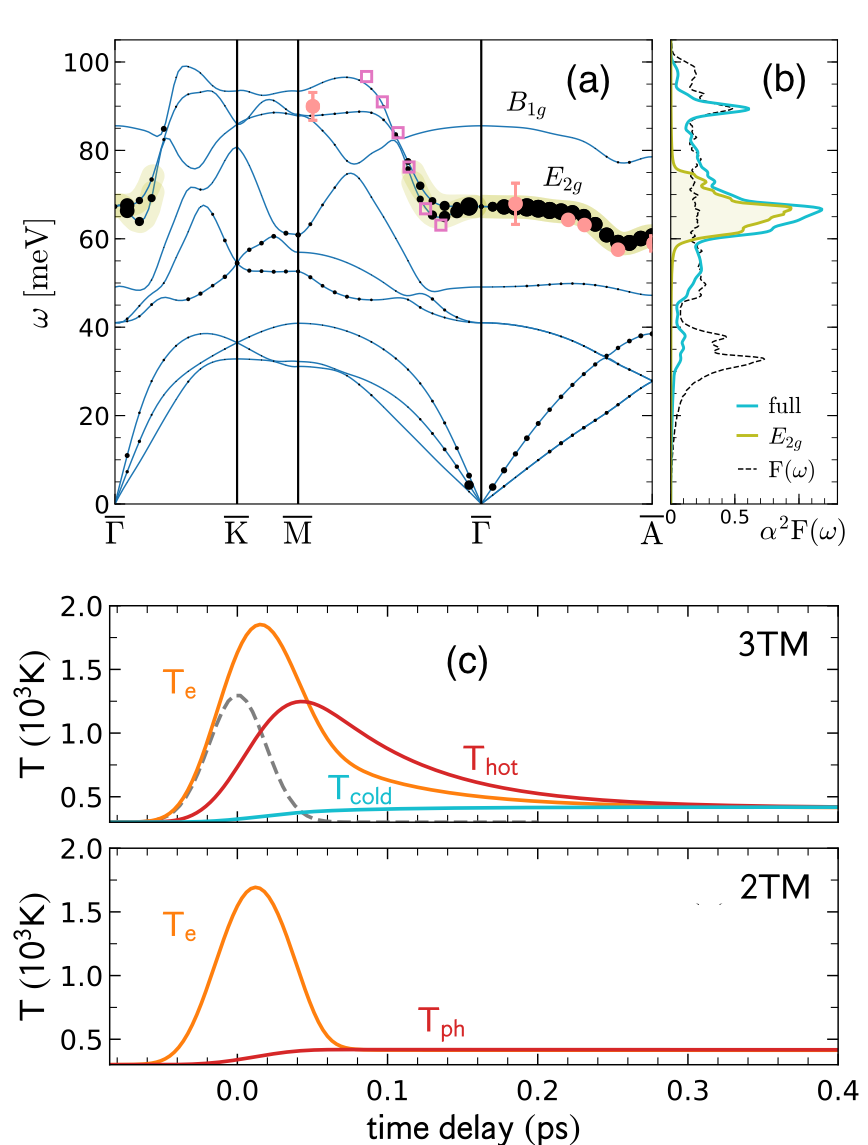}
    \caption{
(a) Plot of the
phonon dispersions of MgB$_2$ (solid lines)
where the size of the black circles represents
the electron-phonon coupling strengths $\lambda_{\mathbf{q}\nu}$.
Also shown are the experimental phonon
energies of the $E_{2g}$ mode close to the $\mathrm{\overline{M}}$ point and
along the $\mathrm{\overline{\Gamma}}-\mathrm{\overline{A}}$ path (red circles)\,\cite{shukla03}, as well as along the $\mathrm{\overline{M}}-\mathrm{\overline{\Gamma}}$ cuts (purple empty squares)\,\cite{bib:baron04} 
as obtained by inelastic X-ray scattering.
The Raman-active $B_{1g}$ mode is also denoted, with the zone-center frequency of
$\omega_{B_{1g}}=86$\,meV and weak el-ph coupling, associated to out-of-plane
lattice vibrations.
(b) Corresponding
phonon density of states $F(\omega)$ (dashed line)
and the total Eliashberg function $\alpha^2F(\omega)$ (blue solid line).
Green color shows the contribution to the Eliashberg function
associated with the hot $E_{2g}$ modes around and along the $\mathrm{\overline{\Gamma}}-\mathrm{\overline{A}}$ path,
$\alpha^2F_{\rm hot}(\omega)$.
Panels (a) and (b) from Ref. \cite{ncdc}.
(c) Time evolution of the characteristic effective 
temperatures $T_{\mathrm{e}}$, $T_{\rm hot}$, $T_{\rm cold}$ in MgB$_2$
for the three-temperature model (top panel)
and for a two-temperature model (bottom panel) where
the electron-phonon coupling in MgB$_2$ is modeled to
be isotropic. 
The grey dashed line shows the pulse profile,
with the pulse duration of 45\, fs
and an absorbed fluence of 12\,J/m$^2$.
From Ref. \cite{cn}.
}
\label{f-ncdc-disp}
\end{figure}

A full microscopical proof of  a key role of hot phonons
in MgB$_2$ was provided in Ref. \cite{ncdc} using {\em ab-initio}
calculations. To this purpose, the electronic and phonon band
structures
were computed from density-functional theory
using the {\sc quantum espresso} package \cite{bib:qe}.
The electron-phonon coupling constants $\lambda_{{\bf q},\nu}$
for each momentum ${\bf q}$ and each branch $\nu$ were also computed
in good agreement with
Fig. \ref{f-mgb2}d.

Equipped with all these theoretical inputs,
the time dynamics of the each single electronic and lattice degree
of freedom can be in principle numerically 
computed, as discussed in Sec. \ref{s-processes}.
As also discussed there, however, in MgB$_2$ as well as in other
metals, due to large Fermi surfaces, this approach is
not computationally affordable.
The {\em key advantage} of MgB$_2$ is that
the strong multiband character and the strong anisotropy
of the electron-phonon coupling, provides a very efficient and compelling way
for simplifying the system, namely for splitting in an unambiguous way
the lattice degrees of freedom (${\bf q},\nu$)  in two subsectors,
the ``hot'' and ``cold'' modes.
As evident from the previous discussion, only the $E_{2g}$ phonon
branch, corresponding to in-plane out-of-phase lattice displacements
of the two B atoms, is strongly coupled.
More delicate is the criterion how to identify the ${\bf q}$ vectors
belonging to the hot-phonon sector. 
In this regard, we are advantaged by the remarkable Kohn anomaly
visible in the phonon spectrum,
that provides a fingerprint of the strong electron-phonon scattering
with these modes.
Such scenario was indeed analyzed in a quantitative way in Ref. \cite{ncdc}.
Obeying to the softening due to the Kohn anomaly,
the $E_{2g}$ branch was there essentially divided
in two blocks  (Fig. \ref{f-ncdc-disp}a):
a strongly-coupled region (hot modes, marked
as green shaded area), for small in-plane ${\bf q}$ momenta
and phonon frequencies $\omega_{{\bf q},E_{2g}} \sim 60-75$ meV,
and a weakly-coupled region (cold modes), for large in-plane ${\bf q}$
momenta and
phonon frequencies $\omega_{{\bf q},E_{2g}} \sim 90$ meV.
Due to the weak dispersion of the $E_{2g}$ modes along $q_z$,
such splitting in the momentum space is reflected
in a marked splitting in the frequency features of the Eliashberg
function $\alpha^2F(\omega)$, as shown
in Fig. \ref{f-ncdc-disp}b,
where the contribution of the strong-coupled modes
results in a pronounced peak
in the window $\omega \in [60:75]$ meV,
which is well detached
by the a higher-frequency smaller peak at $\sim 90$ meV
related to the weakly-coupled $E_{2g}$ modes at
large momenta.
Note that the structure of $\alpha^2F(\omega)$
at $\sim 90$ meV matches with a corresponding peak
in the phonon density of states (dashed line in
Fig. \ref{f-ncdc-disp}b),
whereas the phonon density of states $F(\omega)$
is essentially flat in the frequency window $[60:75]$ meV
pointing out how the peak of the Eliashberg
function at these energies is crucially driven by the
strong electron-phonon coupling, selective for these modes.

Following this observation, the total Eliashberg function 
was divided in Ref. \cite{ncdc} as sum of two terms,
$\alpha^2F(\omega)=\alpha^2F_{\rm hot}(\omega)
+\alpha^2F_{\rm cold}(\omega)$,
where $\alpha^2F_{\rm hot}(\omega)$
contains the contribution of the hot $E_{2g}$ modes along and
around the $\mathrm{\overline{\Gamma}-\overline{A}}$ path
in the relevant energy range $\omega \in [60:75]$ meV,
while $\alpha^2F_{\rm cold}(\omega)$ accounts
for the weakly coupled cold modes in the remnant
parts of the Brillouin zone.
A similar splitting can be performed for the phonon density of states
$F(\omega)=F_{\rm hot}(\omega)+F_{\rm cold}(\omega)$.
The percentage of the $E_{2g}$ hot modes has been estimated to be roughly 5 \%
of the total phonon modes \cite{cn}, whereas
the relative electron-phonon coupling strengths for the hot and cold
modes were found to be
$\lambda_{\rm hot}=0.26$ and $\lambda_{\rm cold}=0.34$,
respectively \cite{ncdc}.

The main relevant parameters ruling Eqs. (\ref{eq:el})-(\ref{eq:ph})
were thus also computed  from the first-principles
calculations in Refs. \cite{ncdc,cn}:
 $C_{\mathrm{e}}=90\,\mathrm{J/m^3K^2}\times T_{\mathrm{e}}$,
$C_{\rm hot}=0.13\,\mathrm{J/m^3K}$, and $C_{\mathrm{cold}}=4.1\,\mathrm{J/m^3K}$,
$G_{\rm hot}=2.8\times 10^{18}\,\mathrm{W/m^3K}$ and 
$G_{\rm cold}=3.6\times 10^{18}\,\mathrm{W/m^3K}$.
The relaxation time $\tau_{\rm anharm}=400$ fs has been estimated
from the linewidth of the $E_{2g}$ modes at $\Gamma$,
due to the direct anharmonic phonon-phonon scattering \cite{shukla03}.
On this ground, within a three-temperature model for a hot-phonon material,
the time evolution of characteristic temperatures $T_{\rm e}$, $T_{\rm hot}$,
$T_{\rm cold}$ was computed upon
a pump pulse with a Gaussian profile of duration 45 fs,
starting from an initial thermal equilibrium at room temperature.
The ultrafast dynamics of $T_{\rm e}$, $T_{\rm hot}$,
$T_{\rm cold}$ (see Fig. \ref{f-ncdc-disp}c)
shows a a sharp increase of the electronic temperature
for these parameters up to $T_{\rm e}\approx 1850$ K.
The energy pumping in the electronic sector is however
efficiently compensated and reversed by the fast energy transfer
from electronic excitations to hot phonons, whose
effective temperature $T_{\rm hot}$
increases up to $T_{\rm hot}^{\rm max}\approx 1250$ K at $t^* \approx 50$\,fs,
with a small delay compared to the time behavior
of the electronic temperature $T_{\rm e}$.
The remnant cold phonon modes follow, on the other hand,
a completely different behavior with a slow monotonic increase
towards the final equilibrium state at $t \ge 0.3-0.4$ ps
where all the degrees of freedom are thermalized with each other.
Such time dynamics is quite different when the fundamental role
of the hot phonons is neglected, as shown in the bottom panel
of Fig. \ref{f-ncdc-disp}c where the electron-phonon coupling
in MgB$_2$ is treated in an {\em isotropic} framework,
i.e. where the energy from the electronic degrees of freedom
is transferred in a equal way to {\em all} the lattice modes, without
a preferential channel. In such case,
where the hot-phonon bottleneck is absent,
employing a two-temperature model
the average phonon temperature $T_{\rm ph}$
is predicted to show just a weak and smooth increase
up to a final thermalization 
at about $T_{\rm ph}=T_{\rm e}=420$ K \cite{cn}.

\subsection{Time-resolved Raman spectroscopy}

The onset of hot-phonon physics,
as captured for instance by the three-temperature model shown in
Fig. \ref{f-ncdc-disp}c, has important observable consequences
on many physical properties.

\begin{figure}[t!]
\centering
\includegraphics[width=10cm]{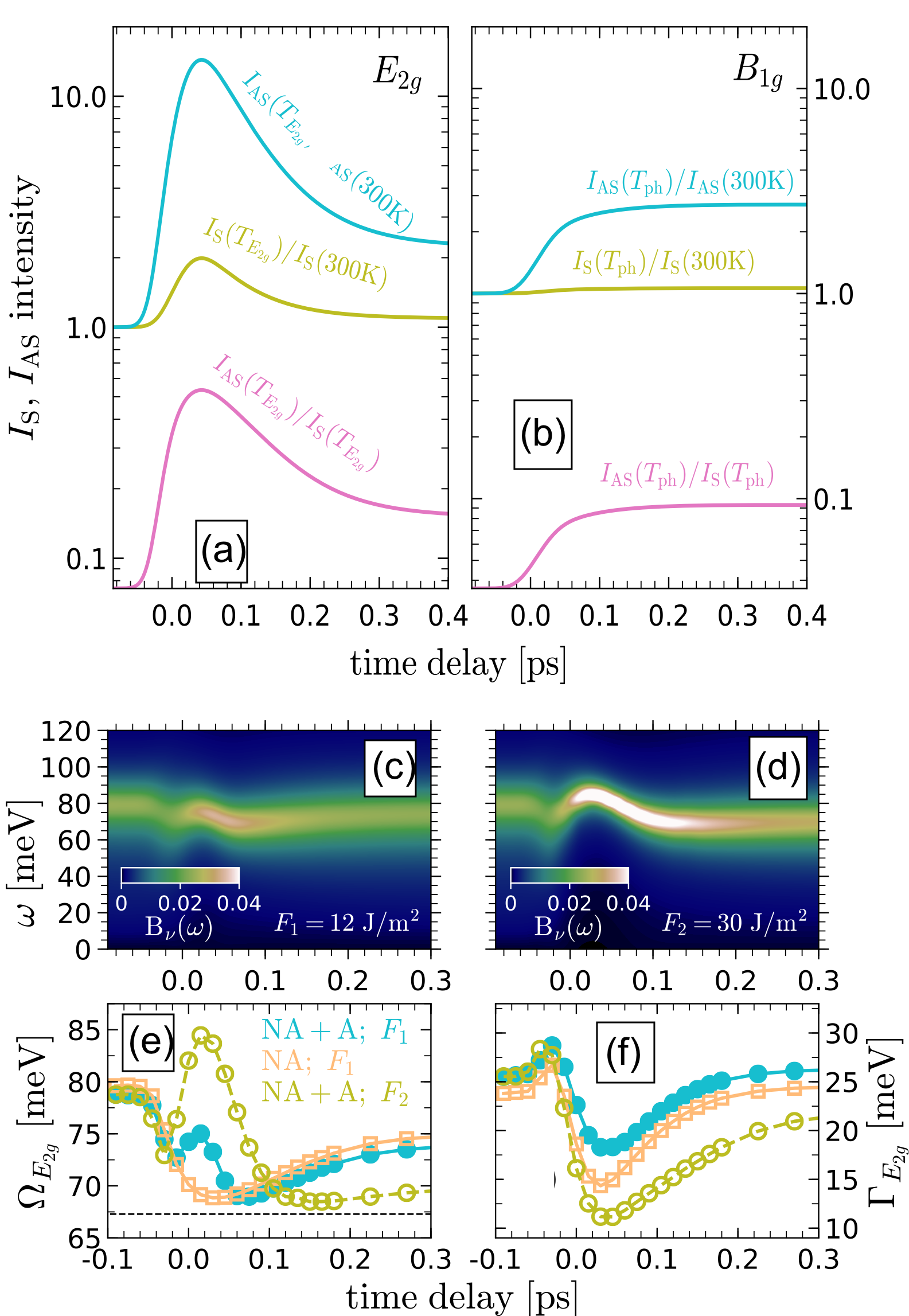}
\caption{
(a)-(b) Predicted time dependence of the Stokes and anti-Stokes
intensities in MgB$_2$, and their ratio,
for the $E_{2g}$ and $B_{1g}$ modes.
(c)-(d)
Computed optical spectral function
$B_{E_{2g}}(\omega;\{T\})$ of the $E_{2g}$ phonon
as function of time delay for two representative fluences,
$F=12$\,J/m$^2$ and $F=30$\,J/m$^2$.
(e)-(f) Time evolution of the Raman peak position
$\Omega_{E_{2g}}$
(panel e) and of the phonon linewidth 
$\Gamma_{E_{2g}}$
(panel f)
using different degrees of approximations
(see text and Ref. \cite{ncdc} for detail).
Figure from Ref. \cite{ncdc}.
}
\label{f-ourRaman}
\end{figure}

One of the most direct fingerprints of hot phonons can be
provided by the analysis of the Raman features \cite{ncdc}.
As discussed in Sec. \ref{s:detect}, for Raman active modes
(such as $E_{2g}$ in MgB$_2$)
the evolution of the hot-phonon temperature $T_{\rm hot}$ can be experimentally
assessed from the time dependence of
the intensities of the Stokes and anti-Stokes peaks,
and their ratio.
The analysis of Ref. \cite{ncdc} is shown
in Fig. \ref{f-ourRaman}a.
Assuming a starting temperature $T=300$ K,
the intensity of the Stoke (anti-Stokes) peak is predict to increase
at $T_{\rm  hot}^{\rm max}$ up to a factor 2 (15 for anti-Stokes),
leading to a corresponding increase of their relative intensity
as large as 
$I_{\rm AS}(T_{\rm hot}^{\rm max})/I_{\rm S}(T_{\rm hot}^{\rm  max})\approx 7$.
Such remarkable anomalies in the intensity
of the $E_{2g}$ Raman resonances can be compared
with predicted time behavior of the Raman active $B_{1g}$ mode
which, as the $E_{2g}$ modes, is characterize by
out-of-phase lattice displacements of the two B atoms,
but along the {\em out-of-plane} direction.
This mode has a frequency $\omega_{B_{1g}}\approx 86$\,meV
not far from the $E_{2g}$ peak but, due to symmetry,
has a negligible linear electron-phonon coupling and
belongs thus to the cold mode sector.
As a consequence its Raman features obey the time dependence
of $T_{\rm cold}$, with negligible effect on the
intensity of the Stokes peaks and a moderate increase
for $I_{\rm AS}$, with the time scales dictated by $T_{\rm cold}$
(Fig. \ref{f-ourRaman}b).
The comparison between the time dynamics
of the Raman properties of the $E_{2g}$ and $B_{1g}$ resonances
provides thus an unambiguous tool to reveal two different temperatures
governing simultaneously different lattice modes.

Peculiar effects of the hot-phonon scenario can be traced
not only by studying frequency-integrated quantities, as
the Raman intensity, but also via {\em spectral} Raman properties.
Theoretically,
these features can be conveniently analyzed 
by computing the many-body phonon self-energy
$\Pi(\omega;\{T\})$
of the $E_{2g}$ mode at ${\bf q}\approx 0$\,\cite{lazzeri06b}.
The phonon spectral function is thus evaluated
as \cite{giustino17}:
\begin{eqnarray}
B(\omega;\{T\})
&=&
-
\frac{1}{\pi}
\mathrm{Im}
\left[
\frac{2\omega_{E_{2g}}}{\omega^2-\omega_{E_{2g}}^2
-2\omega_{E_{2g}}\overline{\Pi}(\omega;\{T\})}
\right],
\label{eq:phononspectral}
\end{eqnarray}
where $\omega_{E_{2g}} =  67$\,meV is the harmonic adiabatic
phonon frequency.
For many conventional systems,
the phonon self-energy can be reasonably computed
by using non-interacting Green's functions,
with the additional inclusion of a phenomenological electronic damping.
Such level of approximation is however 
insufficient in the case of MgB$_2$
where the electronic damping is crucially governed
by the electron-phonon coupling itself \cite{cappelluti06,novko18}.
The interplay between the energy dependence of the
electron and phonon damping gives rise thus to peculiar
nonadiabatic effects that are properly taken into account
in a nonadiabatic approach \cite{novko18}.
It should be stressed that in this framework,
under non-equilibrium conditions,
the phonon self-energy depends on the {\em full}
set of electron and phonon temperatures
$\{T\}=(T_{\rm e}, T_{\rm hot}, T_{\rm cold})$.

The spectral properties of the $E_{2g}$ phonon mode are shown
in Figs. \ref{f-ourRaman}c and \ref{f-ourRaman}d as a function of the time delay,
for two representative values of the pump fluence.
The peculiar features of these spectra
can be parametrized in terms of few quantities,
namely the
renormalized phonon frequency
$\Omega_{E_{2g}}^2=\omega_{E_{2g}}^2
+2\omega_{E_{2g}}\overline{\Pi}(\Omega_{E_{2g}};\{T\})$,
and the many-body phonon linewidth \
$\Gamma_{E_{2g}}=-2\mbox{Im}\overline{\Pi}(\Omega_{E_{2g}};\{T\})$.
The time behavior of these representative parameters
is also shown in Figs. \ref{f-ourRaman}e and \ref{f-ourRaman}f.
Fingerprints of the hot phonons can be detected
in the time dependence of $\Omega_{E_{2g}}$
and $\Gamma_{E_{2g}}$.
The latter one, in particular, appears quite striking since
it reveals a counter-intuitive
{\em reduction} of the phonon linewidth $\Gamma_{E_{2g}}$ 
right after the photo-excitation, followed by a subsequent increase
due to the later thermalization with the cold phonon degrees of freedom.
As discussed in Ref. \cite{ncdc}, such behavior is essentially driven
by the contribution of the nonadiabatic intraband processes,
which can be modeled as:
\begin{eqnarray}
\overline{\Pi}^{\rm intra, NA}(\omega;\{T\})
&=&
\frac{\omega \langle|g_{E_{2g}}|^2\rangle_{T_e}}
{\omega[1+\lambda(\omega;\{T\})]+i\gamma(\omega;\{T\})},
\label{eq:pina}
\end{eqnarray}
where
$\langle|g_{E_{2g}}|^2\rangle_{T_{\rm e}}=
-
\sum_{\mathbf{k}\alpha\sigma}
\left|g_{{\bf k},\alpha,{\bf k},\alpha}^{E_{2g}}\right|^2
\partial f(\epsilon_{\mathbf{k},\alpha};T_{\rm e})/
\partial \epsilon_{\mathbf{k},\alpha}$.
The parameters $\lambda(\omega;\{T\})$,
$\gamma(\omega;\{T\})$
play here the same role as the effective mass function
$m^*(\omega)$ and the optical scattering rate
$\Gamma_{\rm  opt}(\omega)=\tau^{-1}_{\rm opt}(\omega)$
in the extended Drude model and a microscopic derivation
can be provided in a similar way.
A careful analysis has shown that $\gamma(\omega;\{T\})$
can be written as a sum 
of three independent contributions depending
separately on each temperature $T_{\rm e}$, $T_{\rm hot}$,
$T_{\rm cold}$ \cite{ncdc}:
\begin{eqnarray}
\gamma(\omega;\{T\})
&=&
\gamma_{\rm e}(\omega;T_{\rm e})
+
\gamma_{\rm hot}(T_{\rm hot})
+
\gamma_{\rm cold}(T_{\rm cold})
,
\end{eqnarray}
where
\begin{align}
&\gamma_{\rm e}(\omega;T_{\rm e})
=
-
\frac{\pi}{\omega}
\int d\Omega \alpha^2 F(\Omega)
\left[
(\omega+\Omega)\coth\frac{\omega+\Omega}{2k_{\rm B}T_{\rm e}}
-
(\omega-\Omega)\coth\frac{\omega-\Omega}{2k_{\rm B}T_{\rm e}}
\right],
\label{gamma.e}
\\
&\gamma_{\rm hot}(T_{\rm hot})
=
2\pi
\int d\Omega \alpha^2 F_{\rm hot}(\Omega)
\coth\frac{\Omega}{2k_{\rm B}T_{\rm hot}}
,
\\
&\gamma_{\rm cold}(T_{\rm cold})
=
2\pi
\int d\Omega \alpha^2 F_{\rm cold}(\Omega)
\coth\frac{\Omega}{2k_{\rm B}T_{\rm cold}}
.
\label{gamma.ph}
\end{align}
Note that, from this analysis,
employing Kramers-Kronig transformations,
the function $\lambda(\omega;\{T\})$
depends only on the quantity
$\gamma_{\rm e}(\omega;T_{\rm e})$,
and thus only on the electronic temperature, i.e.,
$\lambda(\omega;\{T\})=\lambda_{\rm e}(\omega;T_{\rm e})$.

As intuitive, an increase of the effective temperatures,
expecially $T_{\rm hot}$, leads to an increase of the damping term
$\gamma(\Omega_{E_{2g}};\{T\})$.
What is less intuitive, but easily understandable
on the base of Eq. (\ref{eq:pina}),
is the {\em effect} of the $\gamma(\Omega_{E_{2g}};\{T\})$
on the phonon linewidth $\Gamma_{E_{2g}}$.
Under the room temperature equilibrium conditions,
$\gamma(\Omega_{E_{2g}};T)$ is usually smaller
than the renormalized phonon frequency,
$\gamma(\Omega_{E_{2g}};T)
\ll \Omega_{E_{2g}}[1+\lambda(\Omega_{E_{2g}};T)]$, so that
the phonon damping  scales linearly with
$\gamma(\omega;T)$, i.e.,
$\Gamma_{E_{2g}} \propto \gamma(\Omega_{E_{2g}};T)$.
The strong increase of $T_{\rm hot}$ in the hot-phonon scenario
brings however the material in the unconventional regime
where $\gamma(\Omega_{E_{2g}};\{T\})
\gg \Omega_{E_{2g}}[1+\lambda(\Omega_{E_{2g}};\{T\})]$.
As a consequence, easily noticeable in Eq. (\ref{eq:pina}),
the $E_{2g}$ phonon linewidth scales as
$\Gamma_{E_{2g}} \propto 1/\gamma(\Omega_{E_{2g}};\{T\})$.
The increase of $\gamma(\Omega_{E_{2g}};\{T\})$
in the hot-phonon regime results thus in the {\em decrease}
of $\Gamma_{E_{2g}} $ as observed in Fig. \ref{f-ourRaman}f.
Similar changes of regime occur in other physical contexts,
as the crossover from an Elliott-Yafet to the Dyakonov-Perel
spin-relaxation, or in the case of NMR motional
narrowing \cite{boross13,szolnoki17,mehringbook,rigamonti98}.
If only the nonadiabatic intraband term would be retained,
the same process would lead to a moderate
reduction of the renormalized $E_{2g}$ phonon frequency,
as shown in Fig. \ref{f-ourRaman}e (open orange squares).
The real part of the phonon self-energy,
determining the renormalized phonon frequency, 
is however affected also by (adiabatic) interband transitions which
bring to the additional blueshift (ruled uniquely by $T_{\rm e}$),
visible in Fig. \ref{f-ourRaman}e,
that partially competes with the redshift induced by nonadiabatic
intraband processes.
It is worth stressing out that all
these features have a remarkable dependence not only
on time delay but also on the laser fluence. The
combined analysis of these dependencies can bring
a compelling experimental evidence of the presence
of hot phonons in MgB$_2$ under non-equilibrium pump-probe conditions.

\subsection{Time-resolved correlated lattice dynamics}

As just discussed above, Raman spectroscopy
is a powerful tool for studying the properties of the lattice dynamics,
particularly suited for the ${\bf q}=0$ modes.
For instance, an accurate determination of the intensity of the anti-Stokes
peaks can access directly the population of the
Raman mode under investigation.
Despite these strengths, employing Raman spectroscopy
in pump-probe experiments to completely disentangle the ``hotness''
of electron and lattice degrees of freedom can be challenging.
In its core, Raman spectroscopy is a two-photon
process which involves virtual electronic states,
so that electron and lattice degrees of freedom are unavoidably
mixed together.
A direct detection of the hot-phonon properties that probes
only the {\em lattice} dynamics not entangled with the
electron degrees of freedom can provide thus a compelling
smoking gun.

Along this perspective, 
an alternative  and efficient way for detecting the $E_{2g}$ hot phonons
in MgB$_2$
has been recently theoretically proposed
in Ref. \cite{cn}.
The basic idea is to exploit similar crystallographic information
as the UEDS but, through a Fourier transform,
with {\em real-space} resolution.
Such analysis can determine the interatomic probability distribution
function (PDF) $G(r)$, whose features conveys information
about the crystallographic structure
and about the mean-square displacements, not only
of single-atom motion but also of the
{\em correlated} interatomic motion \cite{campi}.
Within this context $E_{2g}$ hot phonons can feasibly revealed
in MgB$_2$ due to their intrinsic {\em anticorrelated} in-plane motion
of the two B atoms.

\begin{figure}[t!]
\centering
\includegraphics[width=10cm]{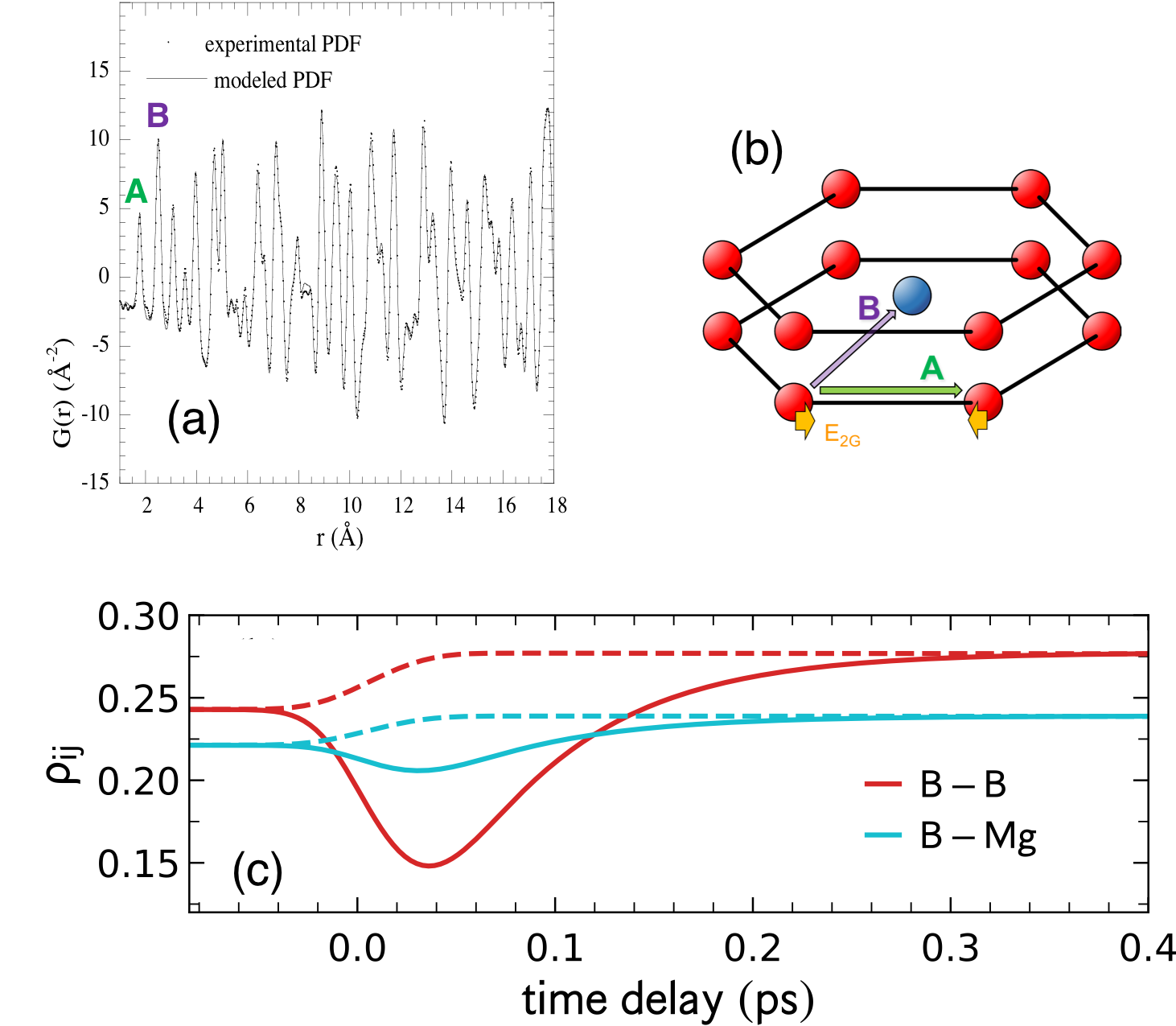}
\caption{
(a) Measured probability distribution function $G(r)$
at room temperature thermal equilibrium. From Ref. \cite{campi}.
Labeled with A and B are the peaks corresponding to the
nearest neighbor B-B and B-Mg pairs.
(b) Lattice structure of MgB$_2$ where the
B-B and the B-Mg bonds are marked (respectively, label A and B).
Also shown is the lattice displacement eigenvector of the $E_{2g}$
mode.
(c) Computed correlation factors $\rho_{\rm B-B}$, $\rho_{\rm B-Mg}$
as a function
of time delay for a typical
pump-probe experiment.
Solid lines correspond to a three-temperature model
accounting for hot-phonon physics, while the dashed lines
represent a two-temperature model where phonons are assumed
to obey a thermal distribution without hot phonons.
From Ref. \cite{cn}.
}
\label{f-corr}
\end{figure}

The experimental assessment of the correlated interatomic
motion in MgB$_2$ was demonstrated in Ref. \cite{campi}
using neutron diffraction. The advantage of this technique
is that, since {\em local} properties are under investigation,
the needed data can be collected in powders.
Although directional information is in this case lost,
the lattice dynamics of a given interatomic bond
can be unambiguously
determined due to the specific distance \cite{jeong99,jeong03,thorpebook}.
An example of the probability distribution
function $G(r)$ in MgB$_2$ under equilibrium conditions is shown in
Fig. \ref{f-corr}a. Different peaks can be related (from the different
interatomic distance) to different atom pairs $i$-$j$
(Fig. \ref{f-corr}b), whereas
the width of each peak provides an estimate
to the mean-square displacement of the relative
distance ${\bf r}_i-{\bf r}_j$.
The basic information of this crystallographic approach
in real space can be summarized in: ($i$) the mean-square
lattice displacement for a given atom $i$ along a given direction
$\alpha=x,y,z$:
\begin{equation}
\sigma^2(i_\alpha) = \langle [{\bf u}_i \cdot \hat{\bf r}_\alpha]^2 \rangle,
\end{equation}
where ${\bf u}_{i}$ is the lattice displacement of atom $i$ from
its average position and $\hat{\bf r}_\alpha$ is
the unit vector pointing along the direction $\alpha=x,y,z$;
 ($ii$)  the mean-square {\em relative} displacement of
atomic pairs projected onto the vector joining the atom
pairs\,\cite{jeong03,campi}:
\begin{equation}
\sigma^2_{ij} = \langle [({\bf u}_i-{\bf u}_j)\cdot
\hat{\bf r}_{ij}]^2 \rangle ,
\end{equation}
where $\hat{\bf r}_{ij}$
is the unit vector connecting atoms $i$ and $j$.

It is also convenient to introduce
the dimensionless correlation factor
$\rho_{ij}$ defined by the implicit relation
\begin{eqnarray}
\sigma^2_{ij}
&=&
 \sigma^2(i_j) + \sigma^2(j_i)
-2\sigma(i_j)\sigma(j_i)\rho_{ij},
\label{e-rho}
\end{eqnarray}
where 
$\sigma^2(i_j)=\langle [({\bf u}_i\cdot\hat{{\bf r}}_{ij}]^2 \rangle$
\cite{notecorr}
.
Positive values of correlation factor $\rho_{ij} > 0$ describe a situation where
the couple of atoms $i$, $j$ move in phase, so that the resulting value of
$\sigma^2_{ij}$ is smaller than for the uncorrelated case. On the
other hand, a predominance of counter-phase atomic vibrations
is expected to result in a negative correlation factor $\rho_{ij} < 0$.
Microscopically, the correlation factor
$\rho_{ij}$ results from the contribution of {\em all}
the phonons of the Brillouin zone, weighted with the appropriate
projection factor.
For systems at thermal equilibrium,
$\rho_{ij}$ is usually slightly positive,
reflecting a slight dominance of the acoustic (in-phase) modes
in contributing to the lattice dynamics.

A key role in detecting hot phonons in MgB$_2$ is played by the nearest
neighbor in-plane boron-boron lattice dynamics.
The $E_{2g}$ modes, characterized by out-of-phase motion (Fig. \ref{f-corr}b),
have a strong anticorrelation character, but at thermal equilibrium
their contribution is balanced by all the other modes.
On the other hand, in the hot-phonon regime the strong increase
of the phonon population of the $E_{2g}$ modes leads
to a striking anomaly in $\rho_{\rm B-B}$ \cite{cn}, which shows
a remarkable dip at $T_{\rm hot}^{\rm max}$ Fig. \ref{f-corr}c.
The experimental determination of the correlation factor
$\rho_{\rm B-B}$ could provide thus a robust fingerprint
of the presence of the hot phonons in MgB$_2$, as well as
a useful estimate of their effective temperature
and of the time scale when {\em all} the lattice degrees of freedom reach
their final thermalization \cite{cn}.

It is worth mentioning that
the real-space approach here described for MgB$_2$
for detecting hot phonons
relies basically on the crucial properties
that hot phonons have a well specific lattice displacement pattern
dictated by the selection rules of the electron-phonon coupling
and by the fact of laying at high-symmetry points.
Hot phonons occurs naturally, because of
the phase space considerations,
at high-symmetry points of the Brillouin zone,
with well-defined and well-known
symmetry and atomic contents.
Such analysis does not apply thus only to MgB$_2$
but it is quite general and it can be extended to a wide
range of materials (semiconductors, metals or semimetals).

\subsection{Time-resolved reflectivity}

Hot phonons occur because of a preferential transfer of energy
from the electronic degrees of freedom to few selected lattice modes,
through the available electron-phonon coupling.
Hot phonons are thus a result of the many-body
electron-phonon interaction, and fingerprints
of a hot-phonon regime have been since long time detected
also in electronic properties, as discussed in Sec. \ref{s:detect}.

\begin{figure}[t!]
\centering
\includegraphics[width=12cm]{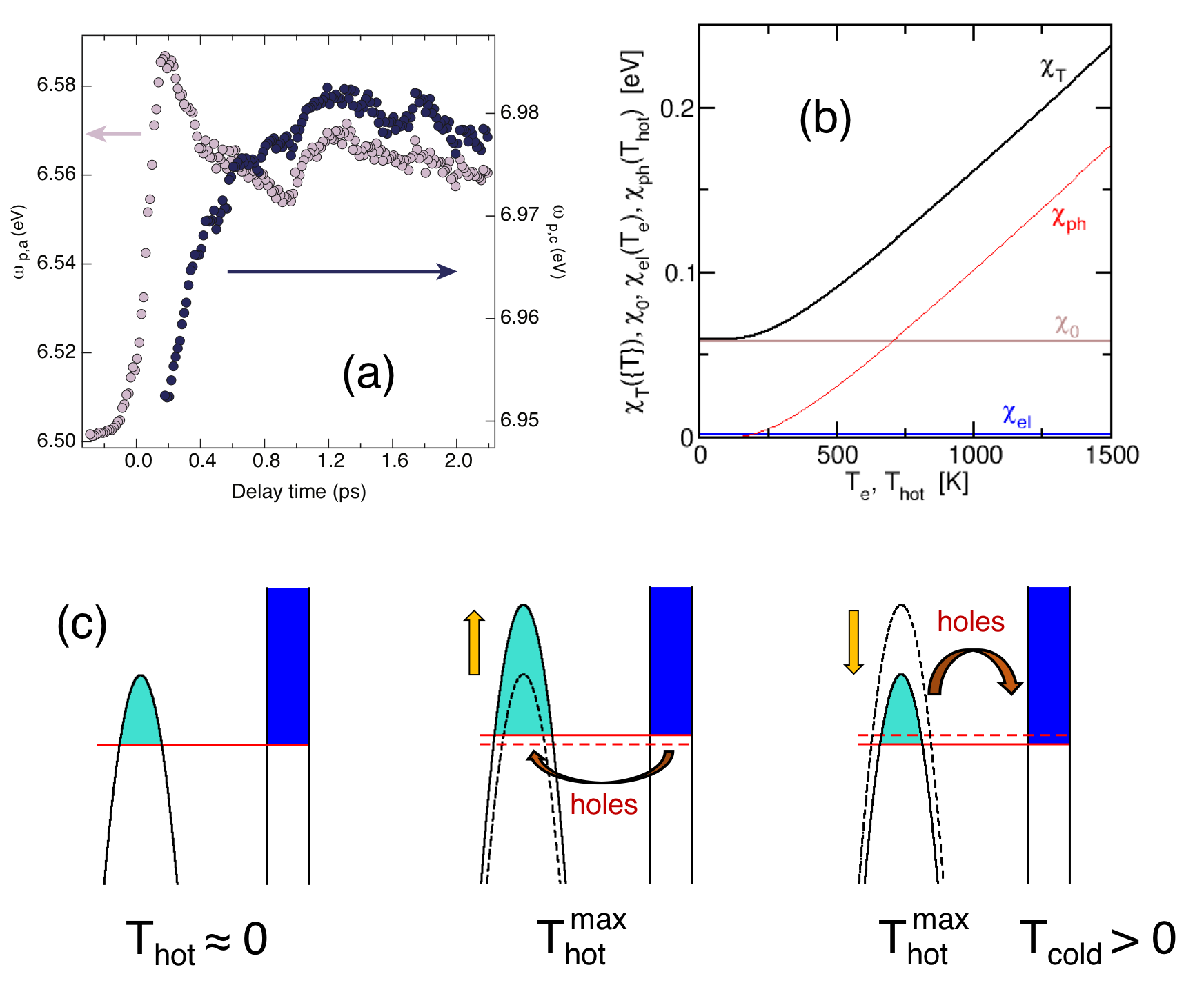}
\caption{
(a) Temporal evolution of the in-plane plasma frequency
$\omega_{p,a}$ (pink circles, left side bar) and
out-of-plane plasma frequency $\omega_{p,c}$
(black dots, right side bar). From Ref. \cite{baldini}.
(b) Plot of the different contributions 
$\chi_0$, $\chi_{\rm el}(T_{\rm e})$, $\chi_{\rm ph}(T_{\rm hot})$
to the bandshift
as a function of the relative ruling temperatures.
(c)
Sketch of the processes governing the time dependence
of the in-plane and out-of-plane plasma frequencies.
Since the evolution of the electronic temperature $T_{\rm e}$
does not play any relevant role in this dynamics,
for graphical representation we use $T_{\rm e}=0$.
Having in mind the peculiarity of MgB$_2$, it is also convenient
to discuss the band structure in terms of {\em hole}
charge (here depicted in green and blue, for $\sigma$
and $\pi$ bands respectively).
Soon the after the pump pulse (left panel) the hot-phonon 
temperature is still low and the carrier density is determined
by the noninteracting bandstructure plus the bandshift contribution
at $T=0$. In the immediate time window
 (middle panel) the preferential energy transfer towards the
 hot phonons leads to a sizable upwards shift of
the $\sigma$ bands and an increase of the $\sigma$-band 
carrier density, reflected in an increase of $\omega_{p,a}$.
At a later stage (right panel) the interband electron-phonon scattering
gives rise to a redistribution of energy towards the cold lattice
modes.
The hot-phonon temperature is consequently reduced 
and therefore the $\sigma$-bands shift back.
This process is thus accompanied by a charge transferring from
the $\sigma$-bands to the three-dimensional $\pi$-bands,
leading to a reduction of the in-plane plasma frequency $\omega_{p,a}$
and to an increase of the out-of-plane plasma frequency $\omega_{p,c}$.
}
\label{f-baldini}
\end{figure}

As a matter of fact, the first experimental report
of hot phonons in MgB$_2$ was deduced from the analysis
of time-resolved optical properties \cite{baldini},  profiting of the specific
peculiarities of this compound.
Unlike previous investigations that have also used reflectivity measurements, and
where hot phonons were inferred from the multi-exponential
decay of the reflectivity, fitted with three- or four-temperature
models including hot-phonon channels,
the core of the analysis in Ref. \cite{baldini}
was the observation
of an anomalous blueshift of the in-plane $\omega_{p,a}$
in the first time window ($\sim 170$ fs) after pumping,
followed by a later blueshift
of the out-of-plane plasma frequency $\omega_{p,c}$
as well, as shown in Fig. \ref{f-baldini}a.

Such blueshift, not observed before in other materials,
is quite anomalous since in conventional materials
the increase of the effective electron and lattice temperatures
would lead to a natural {\em reduction} of the electronic kinetic
energy, and hence of the plasma frequency.
Other possible sources of anomalous shifts of the plasma frequency
under nonequilibrium conditions (electron heating,
bandstructure renormalization due to the non-linear coupling
with the ultrashort laser pulse) 
would predict
as well a redshift \cite{baldini}
rather than the experimentally observed blueshift.

The increase of the in-plane plasma frequency $\omega_{p,a}$
(and of the out-of-plane one $\omega_{p,c}$ after a finite delay)
has been rationalized as a result of the onset of $E_{2g}$
hot modes in the presence of the strong particle-hole asymmetry
of the $\sigma$ bands which, as discussed in Sec. \ref{s-mgb2},
are characterized by a Fermi level very close
to the top edge, and hence with a very low carrier density.
The many-body electron-phonon coupling is known in this situation
to be responsible not only for the usual finite electronic damping term
$\Gamma({\bf k},\omega)=-\mbox{Im}\Sigma({\bf k},\omega)$,
resulting from the imaginary part of the electronic self-energy,
but also for a finite {\em bandshift}, related to
the finite {\em real} part of the self-energy,
$\chi({\bf k},\omega)=\mbox{Re}\Sigma({\bf k},\omega)$
\cite{ortenzi09,benfatto09,benfatto11}.
While the consequences of the first effect in the transient
optical properties have been discussed in detail \cite{echenique00,echenique02,hase03},
the role of the real part of the self-energy
is usually neglected in most materials
where no strong particle-hole asymmetry is present or
when the particle- or hole-character of Fermi multiple sheets
is washed out by an isotropic electron-phonon coupling averaging
over all the electron and hole Fermi pockets.
In this perspective MgB$_2$ is peculiar,
in a similar fashion as iron-based pnictides,
because the strong hole-character
of a subsector of the electronic degree of freedom (the $\sigma$ bands)
is highlighted by the anisotropic electron-phonon coupling,
selecting the relevant scattering for the electron-phonon self-energy
in the $\sigma$ sector, and hence enforcing an effective
hole-character \cite{ortenzi09}.

The role of hot phonons in this context can be captured in
a simple modeling where only the dominant electron-phonon coupling
in the $\sigma$ bands is taken into
account, described in a very good approximation by
two-dimensional  parabolic bands, while the $\pi$ bands
act as a charge reservoir \cite{baldini}.
The in-plane plasma frequency, dominated by the $\sigma$ bands,
can be evaluated thus as:
\begin{eqnarray}
\omega_{p,\sigma}
&=&
\sqrt{\frac{4\pi e^2 \tilde{n}_{\sigma}}{m}}
,
\end{eqnarray}
where $m$ is the effective mass of the parabolic $\sigma$ bands
$k_{\rm F}$ the Fermi momentum for the non-interacting system,
and $\tilde{n}$ the many-body renormalized carrier density
\begin{eqnarray}
\tilde{n}
&=&
\frac{k_{\rm F}^2}{2\pi}
=
\frac{2m[E_{\rm T}+\chi(T)]}{2\pi}
.
\end{eqnarray}

A key role is thus played by the bandshift
which, assuming different temperatures
for the electrons and the hot-phonon modes,
can be computed in a perturbation approach  as \cite{notebaldini}:
\begin{eqnarray}
\chi(\{T\})
&=&
\chi_0
+
\chi_{\rm el}(T_{\rm e})+\chi_{\rm ph}(T_{\rm hot})
,
\end{eqnarray}
where
\begin{eqnarray}
&&\chi_0
=
-\lambda_{\rm hot}
\frac{\omega_{E_{2g}}}{2}
\ln\left|
\frac{E_{\rm T}+\omega_{E_{2g}}}
{E_{\rm B}+\omega_{E_{2g}}}
\right|
,
\\
&&\chi_{\rm el}(T_{\rm e})
 = 
-\lambda_{\rm hot} \frac{T_{\rm e}}{2}
\sum_n \frac{\omega_{E_{2g}}^2}{\omega_n^2+\omega_{E_{2g}}^2}
\ln \left| \frac{E_{\rm T}^2+\omega_n^2}{E_{\rm B}^2+\omega_n^2}
\right|
\nonumber\\
&&
- \lambda_{\rm hot}
\frac{\omega_{E_{2g}}}{2}
f_{T_{\rm e}}(\omega_{E_{2g}})
\ln\left|
\frac{E_{\rm T}^2-\omega_{E_{2g}}^2}
{E_{\rm B}^2-\omega_{E_{2g}}^2}
\right|
+ \lambda_{\rm hot}
\frac{\omega_{E_{2g}}}{2}
\ln\left|
\frac{E_{\rm T}+\omega_{E_{2g}}}
{E_{\rm B}-\omega_{E_{2g}}}
\right|
,
\\
&&\chi_{\rm ph}(T_{\rm hot})
=
-
\lambda_{\rm hot}
\frac{\omega_{E_{2g}}}{2}
b_{T_{\rm hot}}(\omega_{E_{2g}})
\ln\left|
\frac{E_{\rm T}^2-\omega_{E_{2g}}^2}
{E_{\rm B}^2-\omega_{E_{2g}}^2}
\right|
,
\end{eqnarray}
and where $\lambda_{\rm hot}$ is the
dimensionless electron-phonon coupling for the hot phonons,
$E_{\rm T/B}$ represents the band top/bottom edge,
respectively,
estimated with respect to the Fermi level,
$f_{T_{\rm e}}(x)$, $b_{T_{\rm hot}}(x)$ are the Fermi and Bose
distributions ruled by the electronic and hot-phonon temperatures
respectively,
and $\omega_n=\pi T_{\rm e}(2n+1)$ are fermionic Matsubara frequencies
also governed by $T_{\rm e}$.
$\chi_0$ is a temperature-independent contribution,
and it is identical at equilibrium or under non-equilibrium
conditions, whereas $\chi_{\rm el}(T_{\rm e})$,
$\chi_{\rm ph}(T_{\rm hot})$ scale to zero with the corresponding
temperatures.
It is crucial to note here that,
as discussed in Ref. \cite{baldini} and shown
in Fig. \ref{f-baldini}b,
the bandshift $\chi(\{T\})$
is dominated by the phonon temperature $T_{\rm hot}$,
whereas it shows a negligible dependence on the
electronic temperature $T_{\rm e}$.
An effective increase of the phonon temperature leads
thus in MgB$_2$ to a selective upwards bandshift of
the $\sigma$ bands, resulting thus in an increase
of the corresponding in-plane (hole-like)
plasma frequency.

The overall experimental scenario has been
explained in few time steps governed by the
hot-phonon temperature \cite{baldini}, as depicted in Fig. \ref{f-baldini}c.
At $t=0$ the pump pulse induces particle-hole
excitations that soon thermalize towards
an effective electronic temperature \cite{notethermalization}.
The preferential coupling with the $\sigma$-bands
brings to a sudden increase of the hot-phonon temperature
$T_{\rm hot}$ which is accompanied by a many-body-driven
upwards bandshift of the $\sigma$ bands.
Charge transfer from the $\pi$ bands leads thus to
a net increase of the $\sigma$-band carrier density
and hence to a blueshift of the in-plane plasma frequency $\omega_{p,a}$
dominated by the $\sigma$ bands.
At a later stage, the cooling down of the hot-phonon temperature
$T_{\rm hot}$, and the onset of phonon interband scattering
reduces the bandshift of the $\sigma$ sector and
charges flow back to the $\pi$ bands with strong three-dimensional
character leading to the subsequent increase
of the out-of-plane plasma frequency  $\omega_{p,c}$.

One can notice that the present scenario
depends strongly on the intraband character of the
electron-phonon coupling with the hot phonons.
A similar situation might be encountered in
iron-based pnictides or similar materials,
where spin-fluctuations play the role
of the exchanged hot bosons.
In those compounds however spin-fluctuations mediate
{\em interband} exchange between band with different particle/hole
character.
In such context the situation can be reversed and hot-phonon driven
bandshift might result in a net redshift of the plasma frequency
\cite{ortenzi09,benfatto09,benfatto11}.
So far, no experimental check of this scenario
in time-resolved spectroscopy
has been carried out at our knowledge.

\section{Wider view on other compounds and perspectives}
\label{s-other}

As widely discussed above, good metals are usually characterized
by large Fermi surfaces. As a consequence,
electron-phonon scattering is spread over a wide
phonon momentum space, and no hot phonons
are sustained.
At odds with this scenario, MgB$_2$ provides
a bright example that a different path for achieving
hot phonons is possible.
The key element for this alternative context to
semiconductors and semimetals
is a strong anisotropy of the electron-phonon coupling,
where a sizable amount of the electron-phonon
coupling is concentrated in a few phonon modes, along with a weak residual
electron-phonon interaction widely spread over all the other modes.
Such framework leads to an effective separation of the
energy transfer from electrons to lattice into two parallel channels,
where the small specific heat capacity
of the strongly-coupled modes leads to a fast and large increase
of the quantum population of the corresponding phonon states
that get easily hot.

While the above mechanism is fulfilled in the most striking way in MgB$_2$,
which represents a benchmark example,
the physical conditions for this scenario are quite general
and provide a guideline for predicting hot phonons in
other metallic compounds.

A promising candidate along this direction is hole-doped diamond,
which was discovered to become superconductor with $T_c \approx 4$ K
upon 2.8\% of boron doping \cite{ekimov04}.
Baring interesting similarities with MgB$_2$,
boron-doped diamond has indeed regarded as ``three-dimensional''
MgB$_2$ \cite{boeri04}.
From the electronic point of view,
such  high level of boron doping,
achieved thanks to the small size of boron,
leads indeed to a slight hole doping of the
three-dimensional bands of diamond,
in similar way as in the $\sigma$-bands of MgB$_2$.
Hole-doping in diamond
is also accompanied by a downshift and broadening of the
only  Raman-active mode. This scenario can be also explained,
just as in MgB$_2$, as result of a Kohn anomaly
with proper differences in diamond dictated by the three-dimensional
character \cite{boeri04}.
Fano-like asymmetries of the Raman features,
pointing out a relevant electron-phonon coupling \cite{thomsencardona,thomsen},
have been also reported in boron-doped diamond \cite{wang02,szirmai12,mortet19,mortet20},
although Fano asymmetry appears to be absent in N-doped samples \cite{ishioka08}.
Based on such a strong similarity of the physical properties,
we can predict an effective hot-phonon scenario for
boron-doped diamond.

\begin{figure}[t!]
\centering
\includegraphics[width=12cm]{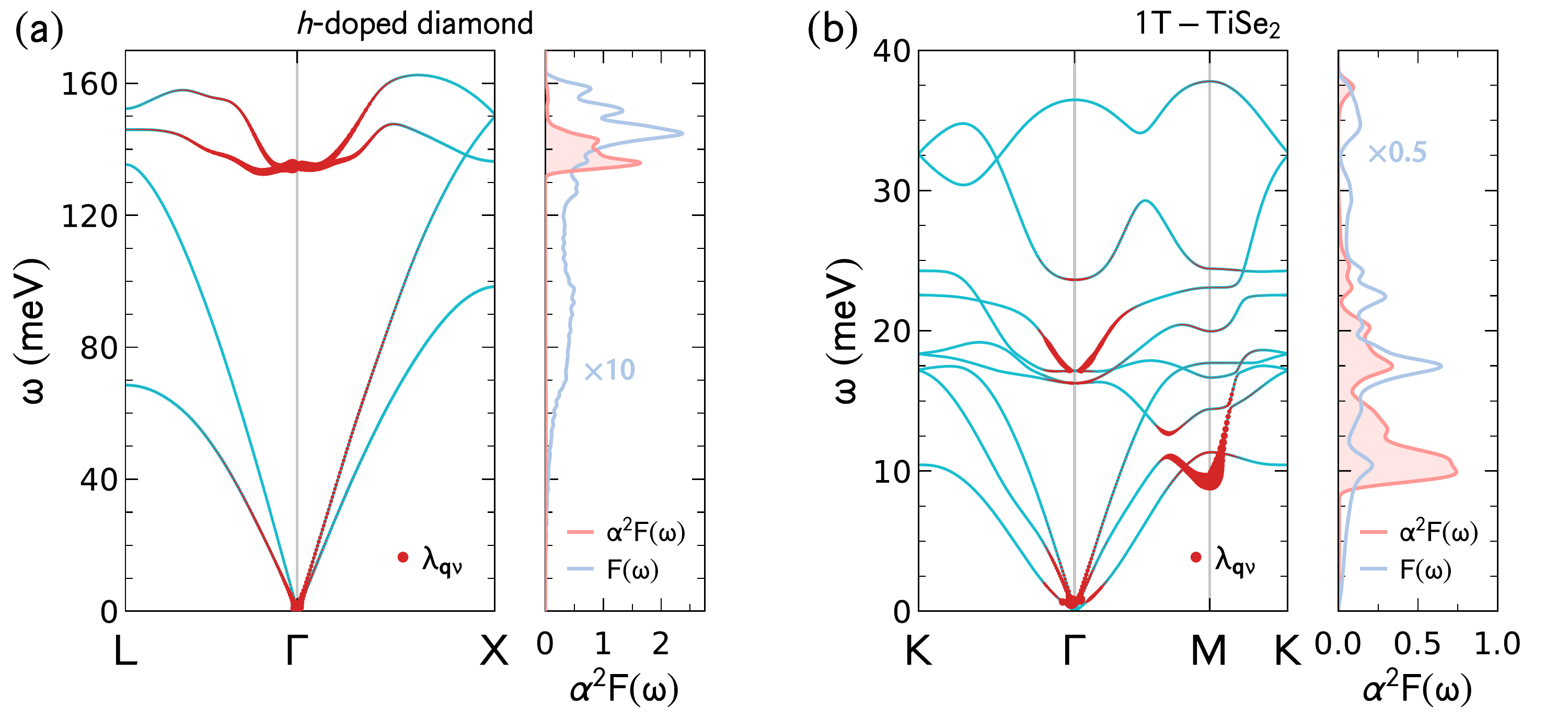}
\caption{Phonon band structure and electron-phonon coupling of (a) 2.6
  \%
hole-doped diamond and (b) 1T-TiSe$_2$ single layer. The strength of the momentum- and band-resolved electron-phonon coupling $\lambda_{{\bf q}\nu}$ is represented with the size of the red dots. The corresponding phonon density of states $\mathrm{F(\omega)}$ and Eliashberg functions $\mathrm{\alpha^2 F(\omega)}$ are depicted with the blue and red lines. The ground state electronic properties were obtained by means of the density-functional-theory package \textsc{quantum espresso}\,\cite{qe1,qe2}. For obtaining phonon properties and electron-phonon coupling, the combination of density functional perturbation theory\,\cite{baroni01} and wannierization procedure\,\cite{wannier,epwan} is used with the help of EPW code\,\cite{epw}. Eliashberg functions $\mathrm{\alpha^2 F(\omega)}$ are obtained from the electron-phonon coupling constants $\lambda_{{\bf q}\nu}$ calculated on the $\mathbf{k}=40\times40\times40$ and $\mathbf{q}=40\times40\times40$ momentum grids in the case of hole-doped diamond and $\mathbf{k}=600\times600\times1$ and $\mathbf{q}=120\times120\times1$ in the case of 1T-TiSe$_2$.
}
\label{diam_tise2}
\end{figure}

Such prediction can be sustained at a quantitative level
by the analysis of the momentum-resolved electron-phonon coupling
$\lambda_{{\bf q},\nu}$ shown in Fig. \ref{diam_tise2}a
for a 2.7 \% of boron doping.
Just as in MgB$_2$, a sizable electron-phonon coupling
(here $\lambda \approx 0.21$)
is concentrated in a few optical modes.
The relevance of the electron-phonon coupling
is shown, like for MgB$_2$, by the comparison
between the frequency dependence
of Eliashberg function $\alpha^2F(\omega)$ and of the phonon
density of states $F(\omega)$:
the peak structure in $\alpha^2F(\omega)$
at $\omega\approx 130-150$\,meV,
together with the knowledge of $\lambda_{{\bf q},\nu}$,
identifies in an unambiguous way
the hot-phonon modes, whereas the frequency structure
of $F(\omega)$, spread over a wide range of frequencies,
with no correspondence in $\alpha^2F(\omega)$,
characterizes the cold modes.
The very weak coupling associated with the cold modes makes
boron-doped diamond
an extreme case of anisotropic electron-phonon system where
the linear coupling of electrons with the cold lattice sector is almost
negligible.
This situation might be even more favourable than MgB$_2$
for detecting hot phonons
since hot phonons are expected to show longer decay rate to the cold modes.
At the same time, these conditions appear appealing also from
the electronic point of view, since the reduced coupling
with the remnant lattice modes
would imply a slower heating of the sample,
which is usually harmful for optoelectronics.

Possible candidates for hot phonons appear also among
related compounds,  like hole-doped Si and Ge,
where for small doping the amount of strongly-coupled modes
is a few percentage \cite{boeri04}, just like in doped diamond.
Although the deformation potential
in these materials has been estimated to be significantly lower
than in diamond, this is not necessarily negative
for sustaining hot phonons.
Indeed, it should be remarked that, as long the coupling with the
cold modes is negligible, the largest hot-phonon temperature
$T_{\rm hot}^{\rm max}$ achievable in the time dynamics
is ruled uniquely by the ratio of the specific heat capacities $C_i$,
regardless of the electron-phonon relaxation parameters $G_i$.
Within this context, the smaller values of $G_i$ in hole-doped Si (or
Ge) with respect to hole-doped diamond would not imply
a smaller $T_{\rm hot}^{\rm max}$, but rather a longer time to reach
it.

Furthermore, several boron-related materials with electronic and vibrational band-structures resembling bulk MgB$_2$ have been theoretically discussed as a promising high-${\rm T_c}$ materials, such as hole-doped LiBC and its different allotropes\,\cite{rosner02,an02,dewhurst03,li18,gao15} as well as
lithium boride (LiB)\,\cite{kolmogorov06,calandra07}. Each of these compounds are characterized by hexagonal layered structure, where boron and carbon atoms occupy hexagonal sites, while lithium is placed between hexagonal layers (with the similar role as Mg in MgB$_2$). Here also the B-C or B-B bond-stretching phonon mode at the center of the BZ contribute predominantly to the overall electron-phonon coupling strength, which makes these materials ideal for exploring hot phonons.
Realization of the efficient hot phonon physics could also possible by extended to MgB$_2$-related monolayers, such as MgB$_2$ monolayer\,\cite{bekaert17}, hydrogenated monolayer MgB$_2$\,\cite{bekaert19}, and LiBC monolayer\,\cite{modak21}, where in each case the significant strength of electron-phonon coupling is concentrated in only a few modes.
Another boron-carbon-based candidate
for hot-phonon is SrRbB$_6$C$_6$, which is predicted
to sustain high-$T_c$ superconductivity
and where a sizable amount of the electron-phonon coupling in concentrated in few selected
lattice modes
in a energy window [40:50] meV \cite{dicataldo21,dicataldoprivate}

Some common trends can be pointed out in the above materials. 
In hole-doped diamond for instance, as in MgB$_2$,
the strong coupling of selected modes candidates for hot-phonon
physics is accompanied by a marked Kohn anomaly which 
is revealed by a large phonon softening
in the phonon dispersion.
Kohn anomalies restricted to few ${\bf q}$ momenta,
on the other hand, are commonly
observed in systems close to a charge-density-wave (CDW) or to a structural
transition.
Within this context, the approaching of a second-order
structural transition is signalized by a softening
of a given phonon mode $\omega_\nu({\bf q})$
which vanishes $\omega_\nu({\bf q}) \to 0$ at the quantum critical
point (Peierls instability). Such softening is accompanied by
a formal divergence of the electron-phonon coupling mediated
by this mode, as a result of the relation $\lambda=
2\int d\omega \alpha^2F(\omega)/\omega$.
According to this picture, one can naively argue that compounds at the verge
of a second-order structural transition would be the best candidates
for hot phonons.
However, although the basic argument is valid,
one should also take into account that the relaxation rates $G_i$
governing the energy transfer from the electrons to the lattice modes
can be qualitatively estimated as $G_i \propto \lambda_i\langle
\omega_i^2 \rangle$ \cite{bib:allen87,bib:lin08}.
This implies that softening modes for $\omega_\nu({\bf q}) \to 0$
do not actually contribute at the energy transfer from the electronic
degrees of freedom.
The most favourable conditions for hot phonons,
balancing all the effects, are thus encountered in systems
{\em approaching} but not too close to a lattice instability.
A representative example of a possible candidate
is 1T-TiSe$_2$, a layered transition-metal dichalcogenide
with a semimetal character at room temperature.
Such compound undergoes a charge-density-wave
transition with a 2 $\times$ 2 $\times$ 2 periodicity
below $T_{\rm CDW}\approx 200$ K \cite{disalvo76}.
This structural transition, which persists also
in few-layer and single-layer samples \cite{watson20},
is preceded by the sizable softening of
a phonon mode at the M and L points of the three-dimensional Brillouin
zone \cite{holt01}, suggesting an electron-phonon driven mechanism,
although electronic models based on an excitonic condensate
have been also proposed \cite{monney11}.
Recent UEDS studies reveal under pump-probe conditions
an unconventional behavior of the intensity of the
diffuse scattering at the M/L points, resulting from
the competing increasing of the phonon population
and the reduced electronic screening \cite{otto21}.
The possibility of hot-phonon physics
is pointed out by an ab-initio investigation
in the normal state of the momentum-resolved
electron-phonon coupling in single-layer TiSe$_2$, as shown
in Fig. \ref{diam_tise2}b,
where the red dots on top of the phonon dispersion
represent the strength of the electron-phonon coupling.
We note that a strong electron-phonon coupling  is concentrated
in few lattice modes at the M point which shows at the same
time a strong Peierls-like softening.
The relevance of these modes with respect to a hot-phonon scenario
is underlined by the comparison in the right
panel of Fig. \ref{diam_tise2}b
between the phonon density of states
$F(\omega)$ and the Eliashberg function $\alpha^2F(\omega)$.
Just like in MgB$_2$, the Kohn anomaly is reflected in a strong peak
($\sim 10$ meV)
in the Eliashberg function $\alpha^2F(\omega)$ which is
not expected from the phonon density of states but it
is uniquely driven by the electron-phonon coupling.
A direct evidence of hot phonons in 1T-TiSe$_2$ is one
of the future experimental challenges in this field\,\cite{karam18}.

The closeness to a charge-density-wave phase
is a promising
path for tailoring the onset
of hot phonons in metals which is not restricted
to TiSe$_2$, but is just a representative example.
Other systems close to a charge-density-wave instability
can be investigated as possible candidates,
not only in the family
of layered transition-metal dichalcogenides $MX_2$
($M=$Nb, Ta; $X=$S, Se), but also in three-dimensional
compounds, as A15 (Nb$_3$Sn, V$_3$Si
or barium-bismuthate perovskites \cite{meregalli98})
Recent experimental studies based on  time-resolved ARPES
and time-resolved X-ray diffraction show
a critical role of
strongly coupled phonons and mode-selective electron-phonon
coupling in 1T-TaSe$_2$\,\cite{shi19,zhang20}, 1T-VSe$_2$\,\cite{majchrzak21}, and several 3D systems
with the CDW order\,\cite{tao13,storeck20,Dolgirev20,maklar21}.
Quite remarkable,
one of these interesting examples shows that a nonthermal phonon distribution
in the form of hot phonons can lead to the CDW melting and to a later CDW recovery which is
markedly different from the equilibrium case\,\cite{maklar21}.

A further promising development
is suggested by the possibility of
exciting hot-phonon physics in surface phonons.
This scenario bares additional interest in MgB$_2$
due to the topological character of phonons,
which is particularly revealed at the
surfaces \cite{jin19,li20}.
Recent experimental techniques
have been shown to be efficient
in investigating large electron-phonon coupling constants and lattice instabilities on the surfaces \cite{benedek20a,benedek20b}.

\begin{figure}[t!]
\centering
\includegraphics[width=12cm]{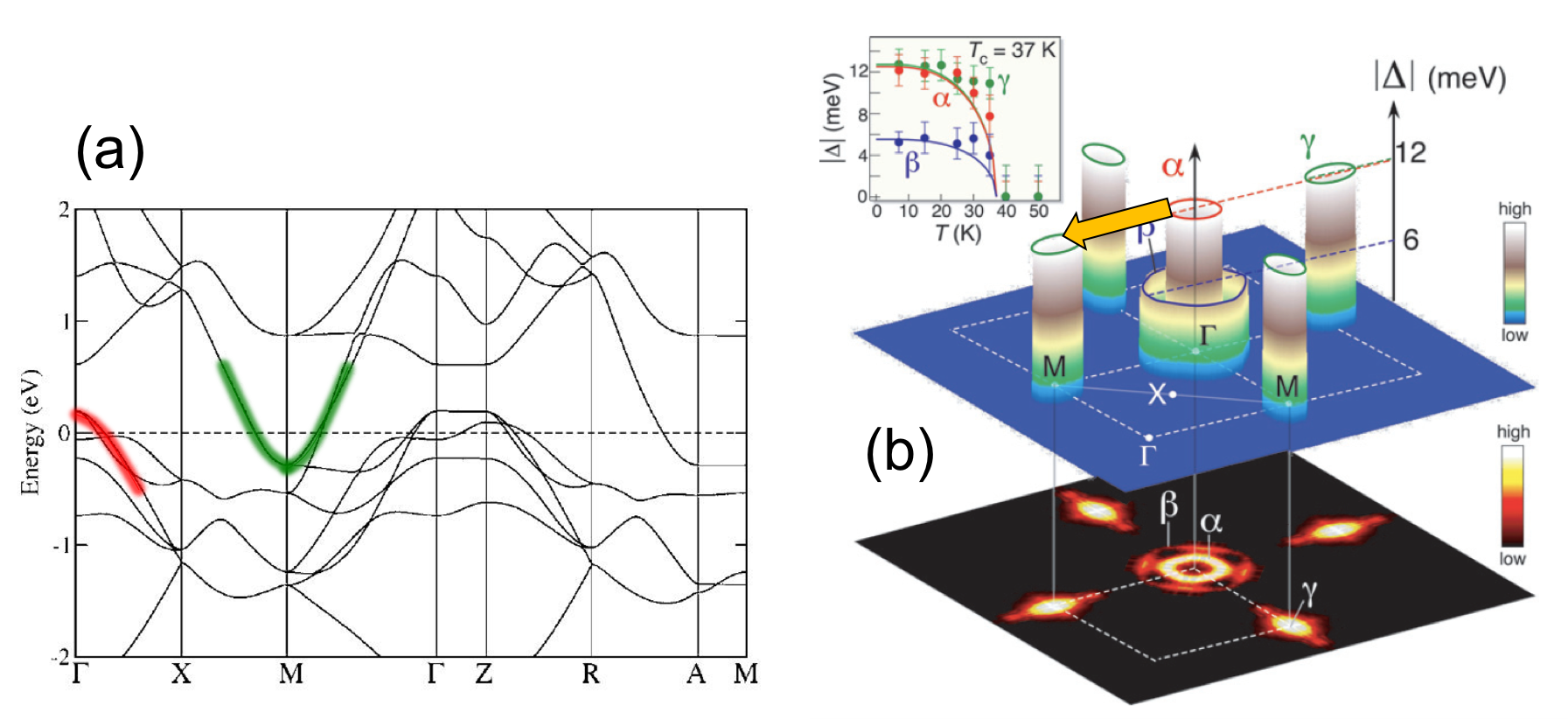}
\caption{Multiband/multigap properties for iron-based pnictides.
(a) Band structure of LaFePO. Hole-like bands at the $\Gamma$ point
are marked in red, while electron-like bands at M point
are marked in green. From Ref. \cite{lebegue07}. (b) Multigap structure of 
Ba$_{0.6}$K$_{0.4}$Fe$_2$As$_2$.
Bottom plane represents ARPES intensity at the Fermi level.
Upper plane represents the value of the superconducting gap
measured by ARPES on different Fermi sheets.
In the inset: temperature evolution of the different superconducting gap
on different Fermi sheets. From Ref. \cite{ding08}.}
\label{f-iron}
\end{figure}

A last mention should be devoted in this framework to iron-based pnictides,
superconducting materials with bulk critical temperatures up to $T_{\rm c} \approx 55$ K \cite{ren08}
and above 100 K in FeSe single-layer on doped SrTiO$_3$ \cite{Ge15}.
Like MgB$_2$, these compounds display a {\em multigap} superconducitity
which, as discussed before, reveals an anisotropic scattering
between the different Fermi sheets
(see Fig. \ref{f-iron} for a typical band structure and
multigap Fermi properties).
However, there is nowadays a wide consensus
that superconductivity in iron-based pnictides, unlike MgB$_2$, is {\em not}
related to the electron-phonon coupling (although present)
but it is driven by spin-fluctuation exchange.
Spin-fluctuations represent a particularly strong channel
for interband scattering, connecting hole-like bands close
to the $\Gamma$ point with electron-like bands at the M point \cite{noteBZiron},
while intraband coupling is relatively weak.
The strong anisotropy of the coupling provides a similar robust platform as MgB$_2$
for hot-mode physics, where spin-fluctuations at $(\pm \pi,\pm \pi)$
represent the preferential bosonic channel for energy transfer
from the electron sector.
Future work in this direction can assess the validity
of this scenario.

\section{Concluding remarks}

In this paper we have reviewed the recent progresses regarding
a novel path for observing hot phonons in metals as
a consequence of a strong anisotropic electron-phonon coupling.
Based on a summary of the fundamental physics involved
in pump-probe out-of-equilibrium experiments,
we have discussed the physical conditions that allow
sustaining non-thermal populations of few selected modes
in metals.

A special focus has been devoted to MgB$_2$ 
that represents maybe the best prototype of metallic compounds
fulfilling such conditions.
Physical observables of the hot-phonon scenario
in MgB$_2$ have been discussed,
both on the experimental and theoretical grounds.
Besides MgB$_2$, other materials have been outlined as promising
candidates for realization of hot phonons.
Among them, low-carrier doped semiconductors,
as for instance boron-doped diamond,
and compounds with soft modes at the verge
of a structural transition.

The feasibility of controlling the onset and the properties
of hot phonons, not only in semiconductors
but also in metals, opens the way to the possibility
of including metal components in ideal optoelectronic devices
where a hot-phonon bottleneck could be desirable
for controlling the electronic relaxation decay rates and
the effective lattice heating.
The possibility of exciting hot phonons with a chiral character
might open also new interesting perspectives in the field
of quantum information where a basic quantum bit could be
engraved in the chiral state of a long-lived chiral phonon.

The continuous and fast development of refined experimental techniques
and theoretical tools provides a promising scenario
where hot-phonon physics in metals could be
a common tool on the advanced research workbench.

\section*{Acknowledgments}

E.C. owes special thanks to
L. Benfatto, F. Carbone, E. Baldini, G. Campi, D. Fausti and A. Tomadin
for many fruitful discussions.
F.C. acknowledges funding by the Deutsche Forschungsgemeinschaft (DFG Projektnummer 443988403) and PRACE for awarding access to Prometheus, CYFRONET, Poland. 
D.N. acknowledges financial support from the Croatian Science Foundation (Grant no. UIP-2019-04-6869) and from the European Regional Development Fund for the ``Center of Excellence for Advanced Materials and Sensing Devices'' (Grant No. KK.01.1.1.01.0001).





\bibliography{myrefs3}

\begin{thebibliography}{100}
\expandafter\ifx\csname url\endcsname\relax
  \def\url#1{\texttt{#1}}\fi
\expandafter\ifx\csname urlprefix\endcsname\relax\def\urlprefix{URL }\fi
\expandafter\ifx\csname href\endcsname\relax
  \def\href#1#2{#2} \def\path#1{#1}\fi

\bibitem{giannettireview}
C.~Giannetti, M.~Capone, D.~Fausti, M.~Fabrizio, F.~Parmigiani, D.~Mihailovic,
  {Ultrafast optical spectroscopy of strongly correlated materials and
  high-temperature superconductors: a non-equilibrium approach}, Adv. Phys. 65
  (2016) 58.

\bibitem{petek97}
H.~Petek, S.~Ogawa, {Femtosecond time-resolved two-photon photoemission studies
  of electron dynamics in metals}, Prog. Surf. Sci. 56 (1997) 239.

\bibitem{king05}
W.~E. King, G.~H. Campbell, A.~Frank, B.~Reed, J.~F. Schmerge, B.~J. Siwick,
  B.~C. Stuart, P.~M. Weber, {Ultrafast electron microscopy in materials
  science, biology, and chemistry}, J. Appl. Phys. 97 (2005) 111101.

\bibitem{mischa11}
R.~Ulbricht, E.~Hendry, J.~Shan, T.~F. Heinz, M.~Bonn, {Carrier dynamics in
  semiconductors studied with time-resolved terahertz spectroscopy}, Rev. Mod.
  Phys. 83 (2011) 543.

\bibitem{bauer15}
M.~Bauer, A.~Marienfeld, M.~Aeschlimann, {Hot electron lifetimes in metals
  probed by time-resolved two-photon photoemission}, Prog. Surf. Sci. 90 (2015)
  319.

\bibitem{waldecker16}
L.~Waldecker, R.~Bertoni, R.~Ernstorfer, J.~Vorberger, {Electron-Phonon
  Coupling and Energy Flow in a Simple Metal beyond the Two-Temperature
  Approximation}, Phys. Rev. X 6 (2016) 021003.

\bibitem{kuznetsov94}
A.~V. Kuznetsov, C.~J. Stanton, {Theory of Coherent Phonon Oscillations in
  Semiconductors}, Phys. Rev. Lett. 73 (1994) 3243.

\bibitem{merlin97}
R.~Merlin, {Generating coherent THz phonons with light pulses}, {Solid State
  Comm.} {102} ({1997}) {207}.

\bibitem{kuznetsov01}
A.~V. Kuznetsov, C.~J. Stanton, {Theory of Coherent Phonon Oscillations in Bulk
  GaAs}, in: K.~T. Tsen (Ed.), {Ultrafast Phenomena in Semiconductors},
  Springer, New York, NY, 2001, p. 353.

\bibitem{stevens02}
T.~E. Stevens, J.~Kuhl, R.~Merlin, {Coherent phonon generation and the two
  stimulated Raman tensors}, Phys. Rev. B 65 (2002) 144304.

\bibitem{ishioka06}
K.~Ishioka, M.~Hase, M.~Kitajima, H.~Petek, {Coherent optical phonons in
  diamond}, Appl. Phys. Lett. 89 (2006) 231916.

\bibitem{bovensiepen06}
U.~Bovensiepen, {Ultra-fast dynamics of coherent lattice and spin excitations
  at the Gd(0001) surface}, {Appl. Phys. A: Mater. Sci. Process.} {82} ({2006})
  {395}.

\bibitem{ishioka08}
K.~Ishioka, M.~Hase, M.~Kitajima, L.~Wirtz, A.~Rubio, H.~Petek, {Ultrafast
  electron-phonon decoupling in graphite}, Phys. Rev. B 77 (2008) 121402.

\bibitem{forst11}
M.~F\"orst, C.~Manzoni, S.~Kaiser, Y.~Tomioka, Y.~Tokura, R.~Merlin,
  A.~Cavalleri, {Nonlinear phononics as an ultrafast route to lattice control},
  {Nat. Phys.} {7}.

\bibitem{boschetto13}
D.~Boschetto, L.~Malard, C.~H. Lui, K.~F. Mak, Z.~Li, H.~Yan, T.~F. Heinz,
  {Real-Time Observation of Interlayer Vibrations in Bilayer and Few-Layer
  Graphene}, Nano Lett. 13 (2013) 4620.

\bibitem{udina19}
M.~Udina, T.~Cea, L.~Benfatto, {Theory of coherent-oscillations generation in
  terahertz pump-probe spectroscopy: From phonons to electronic collective
  modes}, Phys. Rev. B 100 (2019) 165131.

\bibitem{cavalleri01}
A.~Cavalleri, C.~T\'oth, C.~W. Siders, J.~A. Squier, F.~R\'aksi, P.~Forget,
  J.~C. Kieffer, {Femtosecond Structural Dynamics in ${\mathrm{VO}}_{2}$ during
  an Ultrafast Solid-Solid Phase Transition}, Phys. Rev. Lett. 87 (2001)
  237401.

\bibitem{demsar02}
J.~Demsar, L.~Forr\'o, H.~Berger, D.~Mihailovic, {Femtosecond snapshots of
  gap-forming charge-density-wave correlations in quasi-two-dimensional
  dichalcogenides $1T\ensuremath{-}{\mathrm{TaS}}_{2}$ and
  $2H\ensuremath{-}{\mathrm{TaSe}}_{2}$}, Phys. Rev. B 66 (2002) 041101.

\bibitem{iwai03}
S.~Iwai, M.~Ono, A.~Maeda, H.~Matsuzaki, H.~Kishida, H.~Okamoto, Y.~Tokura,
  {Ultrafast Optical Switching to a Metallic State by Photoinduced Mott
  Transition in a Halogen-Bridged Nickel-Chain Compound}, Phys. Rev. Lett. 91
  (2003) 057401.

\bibitem{rini07}
M.~Rini, R.~Tobey, N.~Dean, J.~Itatani, Y.~Tomioka, Y.~Tokura, R.~Schoenlein,
  A.~Cavalleri, {Control of the electronic phase of a manganite by
  mode-selective vibrational excitation}, Nature 449 (2007) 72.

\bibitem{perfetti07book}
L.~Perfetti, P.~Loukakos, M.~Lisowski, U.~Bovensiepen, M.~Wolf, {Time resolved
  photoemission of an insulator-metal transition}, in: {Corkum, P and Jonas, D
  and Miller, RJD and Weiner, AM} (Ed.), {Ultrafast Phenomena XV}, Vol.~{88} of
  {Springer Series in Chemical Physics}, {2007}.

\bibitem{perfetti07tas2}
L.~Perfetti, P.~A. Loukakos, M.~Lisowski, U.~Bovensiepen, H.~Berger,
  S.~Biermann, P.~S. Cornaglia, A.~Georges, M.~Wolf, {Time Evolution of the
  Electronic Structure of $1T\mathrm{\text{\ensuremath{-}}}{\mathrm{TaS}}_{2}$
  through the Insulator-Metal Transition}, Phys. Rev. Lett. 97 (2006) 067402.

\bibitem{kawakami09}
Y.~Kawakami, S.~Iwai, T.~Fukatsu, M.~Miura, N.~Yoneyama, T.~Sasaki,
  N.~Kobayashi, {Optical Modulation of Effective On-Site Coulomb Energy for the
  Mott Transition in an Organic Dimer Insulator}, Phys. Rev. Lett. 103 (2009)
  066403.

\bibitem{fausti11}
D.~Fausti, R.~Tobey, N.~Dean, S.~Kaiser, A.~Dienst, M.~Hoffmann, S.~Pyon,
  T.~Takayama, H.~Takagi, A.~Cavalleri, {Light-Induced Superconductivity in a
  Stripe-Ordered Cuprate}, Science 331 (2011) 189.

\bibitem{kaiser14}
S.~Kaiser, C.~R. Hunt, D.~Nicoletti, W.~Hu, I.~Gierz, H.~Y. Liu, M.~Le~Tacon,
  T.~Loew, D.~Haug, B.~Keimer, A.~Cavalleri, {Optically induced coherent
  transport far above ${T}_{c}$ in underdoped
  ${\mathrm{YBa}}_{2}{\mathrm{Cu}}_{3}{\mathrm{O}}_{6+\ensuremath{\delta}}$},
  Phys. Rev. B 89 (2014) 184516.

\bibitem{mitrano16}
M.~Mitrano, A.~Cantaluppi, D.~Nicoletti, S.~Kaiser, A.~Perucchi, S.~Lupi,
  P.~Di~Pietro, D.~Pontiroli, M.~Ricco\`o, S.~Clark, D.~Jaksch, A.~Cavalleri,
  {Possible light-induced superconductivity in K$_3$C$_{60}$ at high
  temperature}, Nature 530 (2016) 461.

\bibitem{cantaluppi18}
A.~Cantaluppi, M.~Buzzi, G.~Jotzu, D.~Nicoletti, M.~Mitrano, D.~Pontiroli,
  M.~Ricco, A.~Perucchi, P.~Di~Pietro, A.~Cavalleri, {Pressure tuning of
  light-induced superconductivity in K$_3$C$_{60}$}, {Nat. Phys.} {14} ({2018})
  {837}.

\bibitem{suzuki19}
T.~Suzuki, T.~Someya, T.~Hashimoto, S.~Michimae, M.~Watanabe, M.~Fujisawa,
  T.~Kanai, N.~Ishii, J.~Itatani, S.~Kasahara, Y.~Matsuda, T.~Shibauchi,
  K.~Okazaki, S.~Shin, {Photoinduced possible superconducting state with
  long-lived disproportionate band filling in FeSe}, {Commun. Phys.} {2}
  ({2019}) {115}.

\bibitem{isoyama21}
K.~Isoyama, N.~Yoshikawa, K.~Katsumi, J.~Wong, N.~Shikama, Y.~Sakishita,
  F.~Nabeshima, A.~Maeda, R.~Shimano, {Light-induced enhancement of
  superconductivity in iron-based superconductor FeSe$_{0.5}$Te$_{0.5}$},
  {Commun. Phys.} {4} ({2021}) {160}.

\bibitem{cavanagh93}
R.~R. Cavanagh, D.~S. King, J.~C. Stephenson, T.~F. Heinz, {Dynamics of
  nonthermal reactions: femtosecond surface chemistry}, J. Phys. Chem. 97
  (1993) 786.

\bibitem{ho96}
W.~Ho, {Reactions at Metal Surfaces Induced by Femtosecond Lasers, Tunneling
  Electrons, and Heating}, J. Phys. Chem. 100~(31) (1996) 13050.

\bibitem{kuhn07}
{K\"uhn, O. and W\"oste, L. (eds)}, {Analysis and Control of Ultrafast
  Photoinduced Reactions}, Vol.~87 of Chem. Phys., Springer-Verlag Berlin
  Heidelberg, 2007.

\bibitem{matsumoto07}
Y.~Matsumoto, {Photochemistry and Photo-Induced Ultrafast Dynamics at Metal
  Surfaces}, Bull. Chem. Soc. Jpn 80 (2007) 842.

\bibitem{petek10}
H.~Petek, J.~Zhao, {Ultrafast Interfacial Proton-Coupled Electron Transfer},
  {Chem. Rev.} {110} ({2010}) {7082}.

\bibitem{fausti09}
D.~Fausti, O.~V. Misochko, P.~H.~M. van Loosdrecht, {Ultrafast photoinduced
  structure phase transition in antimony single crystals}, Phys. Rev. B 80
  (2009) 161207.

\bibitem{mankowsky14}
R.~Mankowsky, A.~Subedi, M.~Foerst, S.~O. Mariager, M.~Chollet, H.~T. Lemke,
  J.~S. Robinson, J.~M. Glownia, M.~P. Minitti, A.~Frano, M.~Fechner, N.~A.
  Spaldin, T.~Loew, B.~Keimer, A.~Georges, A.~Cavalleri, {Nonlinear lattice
  dynamics as a basis for enhanced superconductivity in YBa2Cu3O6.5}, {Nature}
  {516}.

\bibitem{subedi14}
A.~Subedi, A.~Cavalleri, A.~Georges, Theory of nonlinear phononics for coherent
  light control of solids, Phys. Rev. B 89 (2014) 220301.

\bibitem{cui14}
X.~Cui, C.~Wang, A.~Argondizzo, S.~Garrett-Roe, B.~Gumhalter, H.~Petek,
  {Transient excitons at metal surfaces}, {Nat. Phys.} {10} ({2014}) {505}.

\bibitem{perfetto16}
E.~Perfetto, D.~Sangalli, A.~Marini, G.~Stefanucci, {First-principles approach
  to excitons in time-resolved and angle-resolved photoemission spectra}, Phys.
  Rev. B 94 (2016) 245303.

\bibitem{trovatello19}
C.~Trovatello, F.~Katsch, N.~J. Borys, M.~Selig, K.~Yao, R.~Borrego-Varillas,
  F.~Scotognella, I.~Kriegel, A.~Yan, A.~Zettl, P.~J. Schuck, A.~Knorr,
  G.~Cerullo, S.~Dal~Conte, {The ultrafast onset of exciton formation in 2D
  semiconductors}, {Nat. Comm.} {11} ({2020}) {5277}.

\bibitem{sundararaman14}
R.~Sundararaman, P.~Narang, A.~S. Jermyn, W.~A. Goddard, III, H.~A. Atwater,
  {Theoretical predictions for hot-carrier generation from surface plasmon
  decay}, {Nat. Commun.} {5} ({2014}) {5788}.

\bibitem{kogar20}
A.~Kogar, A.~Zong, P.~E. Dolgirev, X.~Shen, J.~Straquadine, Y.-Q. Bie, X.~Wang,
  T.~Rohwer, I.-C. Tung, Y.~Yang, R.~Li, J.~Yang, S.~Weathersby, S.~Park, M.~E.
  Kozina, E.~J. Sie, H.~Wen, P.~Jarillo-Herrero, I.~R. Fisher, X.~Wang,
  N.~Gedik, {Light-induced charge density wave in LaTe$_3$}, {Nat. Phys.} {16}
  ({2020}) {159}.

\bibitem{kim12}
K.~W. Kim, A.~Pashkin, H.~Schaefer, M.~Beyer, M.~Porer, T.~Wolf, C.~Bernhard,
  J.~Demsar, R.~Huber, A.~Leitenstorfer, {Ultrafast transient generation of
  spin-density-wave order in the normal state of BaFe$_2$As$_2$ driven by
  coherent lattice vibrations}, {Nat. Mater.} {11} ({2012}) {497}.

\bibitem{johnson15}
J.~A. Johnson, T.~Kubacka, M.~C. Hoffmann, C.~Vicario, S.~de~Jong, P.~Beaud,
  S.~Gr\"ubel, S.-W. Huang, L.~Huber, Y.~W. Windsor, E.~M. Bothschafter,
  L.~Rettig, M.~Ramakrishnan, A.~Alberca, L.~Patthey, Y.-D. Chuang, J.~J.
  Turner, G.~L. Dakovski, W.-S. Lee, M.~P. Minitti, W.~Schlotter, R.~G. Moore,
  C.~P. Hauri, S.~M. Koohpayeh, V.~Scagnoli, G.~Ingold, S.~L. Johnson,
  U.~Staub, {Magnetic order dynamics in optically excited multiferroic
  $\mathrm{TbMn}{\mathrm{O}}_{3}$}, Phys. Rev. B 92 (2015) 184429.

\bibitem{hudl19}
M.~Hudl, M.~d'Aquino, M.~Pancaldi, S.-H. Yang, M.~G. Samant, S.~S.~P. Parkin,
  H.~A. D\"urr, C.~Serpico, M.~C. Hoffmann, S.~Bonetti, {Nonlinear
  Magnetization Dynamics Driven by Strong Terahertz Fields}, Phys. Rev. Lett.
  123 (2019) 197204.

\bibitem{geilhufe21}
R.~M. Geilhufe, V.~Juri\ifmmode \check{c}\else
  \v{c}\fi{}i\ifmmode~\acute{c}\else \'{c}\fi{}, S.~Bonetti, J.-X. Zhu, A.~V.
  Balatsky, {Dynamically induced magnetism in ${\mathrm{KTaO}}_{3}$}, Phys.
  Rev. Res. 3 (2021) L022011.

\bibitem{ultrafastquantumbook}
J.~Yuen-Zhou, J.~J. Krich, I.~Kassal, A.~S. Johnson, A.~Aspuru-Guzik,
  {Ultrafast Spectroscopy}, 2053-2563, IOP Publishing, 2014.

\bibitem{anisimov74}
S.~Anisimov, B.~Kapeliovich, T.~Perel'man, {Electron emission from metal
  surfaces exposed to ultrashort laser pulses}, Sov. Phys. JETP 39 (1974) 375.

\bibitem{bib:allen87}
P.~B. Allen, {Theory of thermal relaxation of electrons in metals}, Phys. Rev.
  Lett. 59 (1987) 1460.

\bibitem{elsayed-ali87}
H.~E. Elsayed-Ali, T.~B. Norris, M.~A. Pessot, G.~A. Mourou, {Time-resolved
  observation of electron-phonon relaxation in copper}, Phys. Rev. Lett. 58
  (1987) 1212.

\bibitem{schoenlein87}
R.~W. Schoenlein, W.~Z. Lin, J.~G. Fujimoto, G.~L. Eesley, {Femtosecond studies
  of nonequilibrium electronic processes in metals}, Phys. Rev. Lett. 58 (1987)
  1680.

\bibitem{snoke92}
D.~W. Snoke, W.~W. R\"uhle, Y.-C. Lu, E.~Bauser, {Evolution of a nonthermal
  electron energy distribution in GaAs}, Phys. Rev. B 45 (1992) 10979.

\bibitem{fann92}
W.~S. Fann, R.~Storz, H.~W.~K. Tom, J.~Bokor, {Electron thermalization in
  gold}, Phys. Rev. B 46 (1992) 13592.

\bibitem{fann94}
W.~S. Fann, R.~Storz, H.~W.~K. Tom, J.~Bokor, {Direct measurement of
  nonequilibrium electron-energy distributions in subpicosecond laser-heated
  gold films}, Phys. Rev. Lett. 68 (1992) 2834.

\bibitem{sun94}
C.-K. Sun, F.~Vall\'ee, L.~H. Acioli, E.~P. Ippen, J.~G. Fujimoto,
  {Femtosecond-tunable measurement of electron thermalization in gold}, Phys.
  Rev. B 50 (1994) 15337.

\bibitem{groeneveld95}
R.~H.~M. Groeneveld, R.~Sprik, A.~Lagendijk, {Femtosecond spectroscopy of
  electron-electron and electron-phonon energy relaxation in Ag and Au}, Phys.
  Rev. B 51 (1995) 11433.

\bibitem{gusev98}
V.~E. Gusev, O.~B. Wright, {Ultrafast nonequilibrium dynamics of electrons in
  metals}, Phys. Rev. B 57 (1998) 2878.

\bibitem{hohlfeld00}
J.~Hohlfeld, S.-S. Wellershoff, J.~Güdde, U.~Conrad, V.~Jähnke, E.~Matthias,
  {Electron and lattice dynamics following optical excitation of metals}, Chem.
  Phys. 251 (2000) 237.

\bibitem{delfatti00}
N.~Del~Fatti, C.~Voisin, M.~Achermann, S.~Tzortzakis, D.~Christofilos,
  F.~Vall\'ee, {Nonequilibrium electron dynamics in noble metals}, Phys. Rev. B
  61 (2000) 16956.

\bibitem{echenique00}
P.~Echenique, J.~Pitarke, E.~Chulkov, A.~Rubio, {Theory of inelastic lifetimes
  of low-energy electrons in metals}, Chem. Phys. 251 (2000) 1.

\bibitem{rethfeld02}
B.~Rethfeld, A.~Kaiser, M.~Vicanek, G.~Simon, {Ultrafast dynamics of
  nonequilibrium electrons in metals under femtosecond laser irradiation},
  Phys. Rev. B 65 (2002) 214303.

\bibitem{pietanza07}
L.~Pietanza, G.~Colonna, S.~Longo, M.~Capitelli, {Non-equilibrium electron and
  phonon dynamics in metals under femtosecond laser pulses}, Eur. Phys. J. D 45
  (2007) 369.

\bibitem{bib:lin08}
Z.~Lin, L.~V. Zhigilei, V.~Celli, {Electron-phonon coupling and electron heat
  capacity of metals under conditions of strong electron-phonon
  nonequilibrium}, Phys. Rev. B 77 (2008) 075133.

\bibitem{zhang14}
L.~Zhang, Q.~Niu, {Angular Momentum of Phonons and the Einstein--de Haas
  Effect}, Phys. Rev. Lett. 112 (2014) 085503.

\bibitem{zhang15}
L.~Zhang, Q.~Niu, {Chiral Phonons at High-Symmetry Points in Monolayer
  Hexagonal Lattices}, Phys. Rev. Lett. 115 (2015) 115502.

\bibitem{zhu18}
H.~Zhu, J.~Yi, M.-Y. Li, J.~Xiao, L.~Zhang, C.-W. Yang, R.~A. Kaindl, L.-J. Li,
  Y.~Wang, X.~Zhang, {Observation of chiral phonons}, {Science} {359} ({2018})
  {579}.

\bibitem{chen18}
H.~Chen, W.~Zhang, Q.~Niu, L.~Zhang, {Chiral phonons in two-dimensional
  materials}, {2D Mater.} {6} ({2019}) {012002}.

\bibitem{komiyama21}
H.~Komiyama, S.~Murakami, {Universal features of canonical phonon angular
  momentum without time-reversal symmetry}, Phys. Rev. B 103 (2021) 214302.

\bibitem{semiao10}
F.~L. Semiao, K.~Furuya, G.~J. Milburn, {Vibration-enhanced quantum transport},
  {New J. Phys.} {12} ({2010}) {083033}.

\bibitem{kocevar72}
P.~Kocevar, {Non-ohmic transport and phonon amplification in polar
  semiconductors}, J. Phys. C: Solid State Phys. 5 (1972) 3349.

\bibitem{kocevar85}
P.~Kocevar, {Hot phonon dynamics}, Physica B+C 134 (1985) 155.

\bibitem{potz87}
W.~P\"otz, {Hot-phonon effects in bulk GaAs}, Phys. Rev. B 36 (1987) 5016.

\bibitem{joshi89}
R.~P. Joshi, D.~K. Ferry, {Hot-phonon effects and interband relaxation
  processes in photoexcited GaAs quantum wells}, Phys. Rev. B 39 (1989) 1180.

\bibitem{micke90}
R.~Mickevičius, A.~Reklaitis, {Hot phonon effects on impact ionization
  dynamics in InSb}, Solid State Comm. 73 (1990) 145.

\bibitem{kim91}
D.-s. Kim, P.~Y. Yu, {Hot-electron relaxations and hot phonons in GaAs studied
  by subpicosecond Raman scattering}, Phys. Rev. B 43 (1991) 4158.

\bibitem{micke96}
R.~Mickevičius, V.~Mitin, G.~Paulavičius, V.~Kochelap, M.~A. Stroscio, G.~J.
  Iafrate, {Hot‐phonon effects on electron transport in quantum wires}, J.
  Appl. Phys. 80 (1996) 5145.

\bibitem{langot96}
P.~Langot, N.~Del~Fatti, D.~Christofilos, R.~Tommasi, F.~Vall\'ee, {Femtosecond
  investigation of the hot-phonon effect in GaAs at room temperature}, Phys.
  Rev. B 54 (1996) 14487.

\bibitem{lazzeri06}
M.~Lazzeri, F.~Mauri, {Coupled dynamics of electrons and phonons in metallic
  nanotubes: Current saturation from hot-phonon generation}, Phys. Rev. B 73
  (2006) 165419.

\bibitem{butscher07}
S.~Butscher, F.~Milde, M.~Hirtschulz, E.~Malić, A.~Knorr, {Hot electron
  relaxation and phonon dynamics in graphene}, Appl. Phys. Lett. 91 (2007)
  203103.

\bibitem{richter09}
M.~Richter, A.~Carmele, S.~Butscher, N.~Bücking, F.~Milde, P.~Kratzer,
  M.~Scheffler, A.~Knorr, {Two-dimensional electron gases: Theory of ultrafast
  dynamics of electron-phonon interactions in graphene, surfaces, and quantum
  wells}, J. Appl. Phys. 105 (2009) 122409.

\bibitem{yan09}
H.~Yan, D.~Song, K.~F. Mak, I.~Chatzakis, J.~Maultzsch, T.~F. Heinz,
  {Time-resolved Raman spectroscopy of optical phonons in graphite: Phonon
  anharmonic coupling and anomalous stiffening}, Phys. Rev. B 80 (2009) 121403.

\bibitem{berciaud10}
S.~Berciaud, M.~Y. Han, K.~F. Mak, L.~E. Brus, P.~Kim, T.~F. Heinz, {Electron
  and Optical Phonon Temperatures in Electrically Biased Graphene}, Phys. Rev.
  Lett. 104 (2010) 227401.

\bibitem{lui10}
C.~H. Lui, K.~F. Mak, J.~Shan, T.~F. Heinz, {Ultrafast Photoluminescence from
  Graphene}, Phys. Rev. Lett. 105 (2010) 127404.

\bibitem{wang10}
H.~Wang, J.~H. Strait, P.~A. George, S.~Shivaraman, V.~B. Shields,
  M.~Chandrashekhar, J.~Hwang, F.~Rana, M.~G. Spencer, C.~S. Ruiz-Vargas,
  J.~Park, {Ultrafast relaxation dynamics of hot optical phonons in graphene},
  Appl. Phys. Lett. 96 (2010) 081917.

\bibitem{breusing11}
M.~Breusing, S.~Kuehn, T.~Winzer, E.~Mali\ifmmode~\acute{c}\else \'{c}\fi{},
  F.~Milde, N.~Severin, J.~P. Rabe, C.~Ropers, A.~Knorr, T.~Elsaesser,
  {Ultrafast nonequilibrium carrier dynamics in a single graphene layer}, Phys.
  Rev. B 83 (2011) 153410.

\bibitem{scheuch11}
M.~Scheuch, T.~Kampfrath, M.~Wolf, K.~von Volkmann, C.~Frischkorn, L.~Perfetti,
  {Temperature dependence of ultrafast phonon dynamics in graphite}, Appl.
  Phys. Lett. 99 (2011) 211908.

\bibitem{huang11}
L.~Huang, B.~Gao, G.~Hartland, M.~Kelly, H.~Xing, {Ultrafast relaxation of hot
  optical phonons in monolayer and multilayer graphene on different
  substrates}, Surf. Sci. 605 (2011) 1657.

\bibitem{wu2012}
S.~Wu, W.-T. Liu, X.~Liang, P.~J. Schuck, F.~Wang, Y.~R. Shen, M.~Salmeron,
  {Hot Phonon Dynamics in Graphene}, Nano Letters 12 (2012) 5495.

\bibitem{golla17}
D.~Golla, A.~Brasington, B.~J. LeRoy, A.~Sandhu, {Ultrafast relaxation of hot
  phonons in graphene-hBN heterostructures}, APL Materials 5 (2017) 056101.

\bibitem{koivi17}
J.~Koivistoinen, P.~Myllyperkiö, M.~Pettersson, {Time-Resolved Coherent
  Anti-Stokes Raman Scattering of Graphene: Dephasing Dynamics of Optical
  Phonon}, J. Phys. Chem. Lett. 8 (2017) 4108.

\bibitem{novko2019}
D.~Novko, M.~Kralj, {Phonon-assisted processes in the ultraviolet-transient
  optical response of graphene}, NPJ 2D Mater. Appl. 3 (2019) 48.

\bibitem{sidiropoulos21}
T.~P.~H. Sidiropoulos, N.~Di~Palo, D.~E. Rivas, S.~Severino, M.~Reduzzi,
  B.~Nandy, B.~Bauerhenne, S.~Krylow, T.~Vasileiadis, T.~Danz, P.~Elliott,
  S.~Sharma, K.~Dewhurst, C.~Ropers, Y.~Joly, K.~M.~E. Garcia, M.~Wolf,
  R.~Ernstorfer, J.~Biegert,
  \href{https://link.aps.org/doi/10.1103/PhysRevX.11.041060}{Probing the energy
  conversion pathways between light, carriers, and lattice in real time with
  attosecond core-level spectroscopy}, Phys. Rev. X 11 (2021) 041060.
\newblock \href {http://dx.doi.org/10.1103/PhysRevX.11.041060}
  {\path{doi:10.1103/PhysRevX.11.041060}}.
\newline\urlprefix\url{https://link.aps.org/doi/10.1103/PhysRevX.11.041060}

\bibitem{perfetti08}
L.~Perfetti, P.~A. Loukakos, M.~Lisowski, U.~Bovensiepen, M.~Wolf, H.~Berger,
  S.~Biermann, A.~Georges, {Femtosecond dynamics of electronic states in the
  Mott insulator 1T-TaS2 by time resolved photoelectron spectroscopy}, {New J.
  Phys.} {10} ({2008}) {053019}.

\bibitem{dean11}
N.~Dean, J.~C. Petersen, D.~Fausti, R.~I. Tobey, S.~Kaiser, L.~V. Gasparov,
  H.~Berger, A.~Cavalleri, {Polaronic Conductivity in the Photoinduced Phase of
  $1T\mathrm{\text{\ensuremath{-}}}{\mathrm{TaS}}_{2}$}, Phys. Rev. Lett. 106
  (2011) 016401.

\bibitem{petersen11}
J.~C. Petersen, S.~Kaiser, N.~Dean, A.~Simoncig, H.~Y. Liu, A.~L. Cavalieri,
  C.~Cacho, I.~C.~E. Turcu, E.~Springate, F.~Frassetto, L.~Poletto, S.~S.
  Dhesi, H.~Berger, A.~Cavalleri, {Clocking the Melting Transition of Charge
  and Lattice Order in $1T\mathrm{\text{\ensuremath{-}}}{\mathrm{TaS}}_{2}$
  with Ultrafast Extreme-Ultraviolet Angle-Resolved Photoemission
  Spectroscopy}, Phys. Rev. Lett. 107 (2011) 177402.

\bibitem{hellmann12}
S.~Hellmann, T.~Rohwer, M.~Kallaene, K.~Hanff, C.~Sohrt, A.~Stange, A.~Carr,
  M.~M. Murnane, H.~C. Kapteyn, L.~Kipp, M.~Bauer, K.~Rossnagel, {Time-domain
  classification of charge-density-wave insulators}, {Nat. Comm.} {3} ({2012})
  {1069}.

\bibitem{andreatta19}
F.~Andreatta, H.~Rostami, A.~G. \v{C}abo, M.~Bianchi, C.~E. Sanders, D.~Biswas,
  C.~Cacho, A.~J.~H. Jones, R.~T. Chapman, E.~Springate, P.~D.~C. King, J.~A.
  Miwa, A.~Balatsky, S.~Ulstrup, P.~Hofmann, {Transient hot electron dynamics
  in single-layer ${\mathrm{TaS}}_{2}$}, Phys. Rev. B 99 (2019) 165421.

\bibitem{wang20}
Y.~D. Wang, W.~L. Yao, Z.~M. Xin, T.~T. Han, Z.~G. Wang, L.~Chen, C.~Cai,
  Y.~Li, Y.~Zhang, {Band insulator to Mott insulator transition in 1T-TaS$_2$},
  {Nat. Comm.} {11} ({2020}) {4215}.

\bibitem{simoncig21}
A.~Simoncig, M.~Stupar, B.~Ressel, T.~Saha, P.~Rebernik~Ribic, G.~De~Ninno,
  {Dissecting Mott and charge-density wave dynamics in the photoinduced phase
  of $1T\text{\ensuremath{-}}{\mathrm{TaS}}_{2}$}, Phys. Rev. B 103 (2021)
  155120.

\bibitem{qe1}
P.~Giannozzi, S.~Baroni, N.~Bonini, M.~Calandra, R.~Car, C.~Cavazzoni,
  D.~Ceresoli, G.~L. Chiarotti, M.~Cococcioni, I.~Dabo, A.~Dal~Corso,
  S.~de~Gironcoli, S.~Fabris, G.~Fratesi, R.~Gebauer, U.~Gerstmann,
  C.~Gougoussis, A.~Kokalj, M.~Lazzeri, L.~Martin-Samos, N.~Marzari, F.~Mauri,
  R.~Mazzarello, S.~Paolini, A.~Pasquarello, L.~Paulatto, C.~Sbraccia,
  S.~Scandolo, G.~Sclauzero, A.~P. Seitsonen, A.~Smogunov, P.~Umari, R.~M.
  Wentzcovitch, {QUANTUM ESPRESSO: a modular and open-source software project
  for quantum simulations of materials}, {J. Phys.: Condens. Matter.} {21}
  ({2009}) {395502}.

\bibitem{qe2}
P.~Giannozzi, O.~Andreussi, T.~Brumme, O.~Bunau, M.~B. Nardelli, M.~Calandra,
  R.~Car, C.~Cavazzoni, D.~Ceresoli, M.~Cococcioni, N.~Colonna, I.~Carnimeo,
  A.~Dal~Corso, S.~de~Gironcoli, P.~Delugas, R.~A. DiStasio, Jr., A.~Ferretti,
  A.~Floris, G.~Fratesi, G.~Fugallo, R.~Gebauer, U.~Gerstmann, F.~Giustino,
  T.~Gorni, J.~Jia, M.~Kawamura, H.-Y. Ko, A.~Kokalj, E.~Kucukbenli,
  M.~Lazzeri, M.~Marsili, N.~Marzari, F.~Mauri, N.~L. Nguyen, H.-V. Nguyen,
  A.~Otero-de-la Roza, L.~Paulatto, S.~Ponce, D.~Rocca, R.~Sabatini, B.~Santra,
  M.~Schlipf, A.~P. Seitsonen, A.~Smogunov, I.~Timrov, T.~Thonhauser, P.~Umari,
  N.~Vast, X.~Wu, S.~Baroni, {Advanced capabilities for materials modelling
  with QUANTUM ESPRESSO}, {J. Phys.: Condens. Matter} {29} ({2017}) {465901}.

\bibitem{andersonbook}
P.~Anderson, {The theory of superconductivity in the high-$T_c$ cuprates},
  Princeton, N.J. : Princeton University Press, 1997.

\bibitem{dagotto94}
E.~Dagotto, {Correlated electrons in high-temperature superconductors}, Rev.
  Mod. Phys. 66 (1994) 763.

\bibitem{pavarini01}
E.~Pavarini, I.~Dasgupta, T.~Saha-Dasgupta, O.~Jepsen, O.~K. Andersen,
  {Band-Structure Trend in Hole-Doped Cuprates and Correlation with
  ${\mathit{T}}_{\mathit{c}\mathrm{max}}$}, Phys. Rev. Lett. 87 (2001) 047003.

\bibitem{lanzara01}
A.~Lanzara, P.~Bogdanov, X.~Zhou, S.~Kellar, D.~Feng, E.~Lu, T.~Yoshida,
  H.~Eisaki, A.~Fujimori, K.~Kishio, J.~Shimoyama, T.~Noda, S.~Uchida,
  Z.~Hussain, Z.~Shen, {Evidence for ubiquitous strong electron-phonon coupling
  in high-temperature superconductors}, {Nature} {412} ({2001}) {510}.

\bibitem{zhou03}
X.~Zhou, T.~Yoshida, A.~Lanzara, P.~Bogdanov, S.~Kellar, K.~Shen, W.~Yang,
  F.~Ronning, T.~Sasagawa, T.~Kakeshita, T.~Noda, H.~Eisaki, S.~Uchida, C.~Lin,
  F.~Zhou, J.~Xiong, W.~Ti, Z.~Zhao, A.~Fujimori, Z.~Hussain, Z.~Shen,
  {Universal nodal Fermi velocity}, {Nature} {423} ({2003}) {398}.

\bibitem{zhenglu21}
Z.~Li, M.~Wu, Y.-H. Chan, S.~G. Louie, {Unmasking the Origin of Kinks in the
  Photoemission Spectra of Cuprate Superconductors}, Phys. Rev. Lett. 126
  (2021) 146401.

\bibitem{bib:perfetti07}
L.~Perfetti, P.~A. Loukakos, M.~Lisowski, U.~Bovensiepen, H.~Eisaki, M.~Wolf,
  {Ultrafast Electron Relaxation in Superconducting
  ${\mathrm{Bi}}_{2}{\mathrm{Sr}}_{2}{\mathrm{CaCu}}_{2}{\mathrm{O}}_{8+\ensuremath{\delta}}$
  by Time-Resolved Photoelectron Spectroscopy}, Phys. Rev. Lett. 99 (2007)
  197001.

\bibitem{carbone08}
F.~Carbone, D.-S. Yang, E.~Giannini, A.~H. Zewail, {Direct role of structural
  dynamics in electron-lattice coupling of superconducting cuprates}, Proc.
  Nat. Ac. of Sci. 105 (2008) 20161.

\bibitem{mansart10}
B.~Mansart, D.~Boschetto, A.~Savoia, F.~Rullier-Albenque, F.~Bouquet,
  E.~Papalazarou, A.~Forget, D.~Colson, A.~Rousse, M.~Marsi, {Ultrafast
  transient response and electron-phonon coupling in the iron-pnictide
  superconductor
  $\text{Ba}{({\text{Fe}}_{1\ensuremath{-}x}{\text{Co}}_{x})}_{2}{\text{As}}_{2}$},
  Phys. Rev. B 82 (2010) 024513.

\bibitem{bib:dalconte12}
S.~Dal~Conte, C.~Giannetti, G.~Coslovich, F.~Cilento, D.~Bossini, T.~Abebaw,
  F.~Banfi, G.~Ferrini, H.~Eisaki, M.~Greven, A.~Damascelli, D.~van~der Marel,
  F.~Parmigiani, {Disentangling the Electronic and Phononic Glue in a High-Tc
  Superconductor}, Science 335 (2012) 1600.

\bibitem{baldini}
E.~Baldini, A.~Mann, L.~Benfatto, E.~Cappelluti, A.~Acocella, V.~M. Silkin,
  S.~V. Eremeev, A.~B. Kuzmenko, S.~Borroni, T.~Tan, X.~X. Xi, F.~Zerbetto,
  R.~Merlin, F.~Carbone, {Real-Time Observation of Phonon-Mediated
  $\ensuremath{\sigma}\text{\ensuremath{-}}\ensuremath{\pi}$ Interband
  Scattering in ${\mathrm{MgB}}_{2}$}, Phys. Rev. Lett. 119 (2017) 097002.

\bibitem{ncdc}
D.~Novko, F.~Caruso, C.~Draxl, E.~Cappelluti, {Ultrafast Hot Phonon Dynamics in
  ${\mathrm{MgB}}_{2}$ Driven by Anisotropic Electron-Phonon Coupling}, Phys.
  Rev. Lett. 124 (2020) 077001.

\bibitem{bib:nagamatsu01}
J.~Nagamatsu, N.~Nakagawa, T.~Muranaka, Y.~Zenitani, J.~Akimitsu,
  {Superconductivity at 39 K in magnesium diboride}, Nature 410 (2001) 63.

\bibitem{kong}
Y.~Kong, O.~Dolgov, O.~Jepsen, O.~Andersen, {Electron-phonon interaction in the
  normal and superconducting states of MgB$_2$}, Phys. Rev. B 64 (2001) 020501.

\bibitem{grimvall}
G.~Grimvall, {The electron-phonon interaction in metals}, North-Holland,
  Amsterdam ; New York, 1981.

\bibitem{scalapino}
D.~Scalapino, {The Electron-Phonon Interaction and Strong-Coupling
  Superconductors}, in: P.~R.D. (Ed.), {Superconductivity}, Dekker, New Yprk,
  1969.

\bibitem{allenmitrovic}
P.~B. Allen, B.~Mitrović, {Theory of Superconducting $T_c$}, Vol.~37 of Solid
  State Phys., Academic Press, 1983, p.~1.

\bibitem{carbottereview}
J.~P. Carbotte, {Properties of boson-exchange superconductors}, Rev. Mod. Phys.
  62 (1990) 1027.

\bibitem{notespin}
{For sake of simplicity, we consider here only the case of spin degeneracy. The
  spin index will be thus omitted when not necessary, although it will be
  properly taken into account in the correct counting of physical processes
  (i.e. diagrams). }.

\bibitem{noteeqdyn}
{Such scheme is equivalently named in diffeent contexts as {\em
  Bloch-Boltzmann-Peierls fomulas}, {\em lowest order Born approximation}, or
  {\em Born-Markov approximation}. }.

\bibitem{molinasanchez17}
A.~Molina-Sánchez, D.~Sangalli, L.~Wirtz, A.~Marini, {Ab Initio Calculations
  of Ultrashort Carrier Dynamics in Two-Dimensional Materials: Valley
  Depolarization in Single-Layer WSe$_2$}, Nano Lett. 17 (2017) 4549.

\bibitem{notedepth}
{In more realistic descriptions, the pump-induced energy adsorption can be
  modelled as varying within the sample along the $z$-axis perpendicular with
  the surface. A term taking into account the energy transfer along the
  $z$-direction due to the thermal conductivity is in that case also included.
  }.

\bibitem{caruso21}
F.~Caruso, {Nonequilibrium Lattice Dynamics in Monolayer MoS$_2$}, J. Phys.
  Chem. Lett. 12 (2021) 1734.

\bibitem{tong21}
X.~Tong, M.~Bernardi, Toward precise simulations of the coupled ultrafast
  dynamics of electrons and atomic vibrations in materials, Phys. Rev. Res. 3
  (2021) 023072.

\bibitem{seiler21}
H.~Seiler, D.~Zahn, M.~Zacharias, P.-N. Hildebrandt, T.~Vasileiadis, Y.~W.
  Windsor, Y.~Qi, C.~Carbogno, C.~Draxl, R.~Ernstorfer, F.~Caruso, {Accessing
  the Anisotropic Nonthermal Phonon Populations in Black Phosphorus}, Nano
  Lett. 21 (2021) 6171.

\bibitem{johnson17}
S.~L. Johnson, M.~Savoini, P.~Beaud, G.~Ingold, U.~Staub, F.~Carbone,
  L.~Castiglioni, M.~Hengsberger, J.~Osterwalder, {Watching ultrafast responses
  of structure and magnetism in condensed matter with momentum-resolved
  probes}, Struct. Dyn. 4 (2017) 061506.

\bibitem{tan17}
S.~Tan, A.~Argondizzo, C.~Wang, X.~Cui, H.~Petek, {Ultrafast Multiphoton
  Thermionic Photoemission from Graphite}, Phys. Rev. X 7 (2017) 011004.

\bibitem{li21}
A.~Li, M.~Reutzel, Z.~Wang, D.~Novko, B.~Gumhalter, H.~Petek, {Plasmonic
  Photoemission from Single-Crystalline Silver}, ACS Photon. 8 (2021) 247.

\bibitem{novko21}
D.~Novko, V.~Despoja, M.~Reutzel, A.~Li, H.~Petek, B.~Gumhalter, {Plasmonically
  assisted channels of photoemission from metals}, Phys. Rev. B 103 (2021)
  205401.

\bibitem{johannsen13}
J.~C. Johannsen, S.~Ulstrup, F.~Cilento, A.~Crepaldi, M.~Zacchigna, C.~Cacho,
  I.~C.~E. Turcu, E.~Springate, F.~Fromm, C.~Raidel, T.~Seyller, F.~Parmigiani,
  M.~Grioni, P.~Hofmann, {Direct View of Hot Carrier Dynamics in Graphene},
  Phys. Rev. Lett. 111 (2013) 027403.

\bibitem{stange15}
A.~Stange, C.~Sohrt, L.~X. Yang, G.~Rohde, K.~Janssen, P.~Hein, L.-P. Oloff,
  K.~Hanff, K.~Rossnagel, M.~Bauer, {Hot electron cooling in graphite:
  Supercollision versus hot phonon decay}, Phys. Rev. B 92 (2015) 184303.

\bibitem{caruso20}
F.~Caruso, D.~Novko, C.~Draxl, {Photoemission signatures of nonequilibrium
  carrier dynamics from first principles}, Phys. Rev. B 101 (2020) 035128.

\bibitem{maklar21}
J.~Maklar, Y.~W. Windsor, C.~W. Nicholson, M.~Puppin, P.~Walmsley, V.~Esposito,
  M.~Porer, J.~Rittmann, D.~Leuenberger, M.~Kubli, M.~Savoini, E.~Abreu, S.~L.
  Johnson, P.~Beaud, G.~Ingold, U.~Staub, I.~R. Fisher, R.~Ernstorfer, M.~Wolf,
  L.~Rettig, {Nonequilibrium charge-density-wave order beyond the thermal
  limit}, Nat. Comm. 12 (2021) {2499}.

\bibitem{majchrzak21}
P.~Majchrzak, S.~Pakdel, D.~Biswas, A.~J.~H. Jones, K.~Volckaert,
  I.~Markovi\ifmmode~\acute{c}\else \'{c}\fi{}, F.~Andreatta, R.~Sankar,
  C.~Jozwiak, E.~Rotenberg, A.~Bostwick, C.~E. Sanders, Y.~Zhang, G.~Karras,
  R.~T. Chapman, A.~Wyatt, E.~Springate, J.~A. Miwa, P.~Hofmann, P.~D.~C. King,
  N.~Lanat\`a, Y.~J. Chang, S.~Ulstrup, {Switching of the electron-phonon
  interaction in $1T\text{\ensuremath{-}}{\mathrm{VSe}}_{2}$ assisted by hot
  carriers}, Phys. Rev. B 103 (2021) L241108.

\bibitem{perfetti07}
L.~Perfetti, P.~A. Loukakos, M.~Lisowski, U.~Bovensiepen, H.~Eisaki, M.~Wolf,
  {Ultrafast Electron Relaxation in Superconducting
  ${\mathrm{Bi}}_{2}{\mathrm{Sr}}_{2}{\mathrm{CaCu}}_{2}{\mathrm{O}}_{8+\ensuremath{\delta}}$
  by Time-Resolved Photoelectron Spectroscopy}, Phys. Rev. Lett. 99 (2007)
  197001.

\bibitem{shah83}
J.~Shah, A.~Pinczuk, H.~L. Störmer, A.~C. Gossard, W.~Wiegmann, {Electric
  field induced heating of high mobility electrons in modulation‐doped
  GaAs‐AlGaAs heterostructures}, Appl. Phys. Lett. 42 (1983) 55.

\bibitem{mahanbook}
G.~D. Mahan, {Many Particle Physics, Third Edition}, Plenum, 2000.

\bibitem{zimanbook}
J.~M. Ziman, {Principles of the Theory of Solids}, 2nd Edition, Cambridge
  University Press, 1972.

\bibitem{price85a}
P.~J. Price, {Hot phonon effects in heterolayers}, Superlattices Microstr. 1
  (1985) 255.

\bibitem{price85b}
P.~J.~Price, {Hot phonon effects in heterolayers}, Physica B+C 134 (1985) 164.

\bibitem{lugli88}
P.~Lugli, {Hot phonon dynamics}, Solid-State Electron. 31 (1988) 667.

\bibitem{weng21}
Q.~Weng, L.~Yang, Z.~An, P.~Chen, A.~Tzalenchu, W.~Lu, S.~Komiyama,
  {Quasiadiabatic electron transport in room temperature nanoelectronic devices
  induced by hot-phonon bottleneck}, {Nat. Comm.} {12} ({2021}) {4752}.

\bibitem{micke90b}
R.~Mickevičius, A.~Reklaitis, {Hot intervalley phonons in GaAs}, {J. Phys.:
  Condens. Matter.} {2} ({1990}) {7883}.

\bibitem{paula97}
G.~Paulavičius, V.~V. Mitin, N.~A. Bannov, {Coupled electron and
  nonequilibrium optical phonon transport in a GaAs quantum well}, J. Appl.
  Phys. 82 (1997) 5580.

\bibitem{paula97b}
G.~Paulavičius, R.~Mickevičius, V.~Mitin, M.~A. Stroscio, {Hot-phonon effects
  on electron runaway from GaAs quantum wires}, J. Appl. Phys. 82 (1997) 3392.

\bibitem{paula98}
G.~Paulavičius, V.~Mitin, M.~A. Stroscio, {Hot-optical-phonon effects on
  electron relaxation in an AlGaAs/GaAs quantum cascade laser structure}, J.
  Appl. Phys. 84 (1998) 3459.

\bibitem{iotti10}
R.~C. Iotti, F.~Rossi, M.~S. Vitiello, G.~Scamarcio, L.~Mahler, A.~Tredicucci,
  {Impact of nonequilibrium phonons on the electron dynamics in terahertz
  quantum cascade lasers}, Appl. Phys. Lett. 97 (2010) 033110.

\bibitem{shi14}
Y.~B. Shi, I.~Knezevic, {Nonequilibrium phonon effects in midinfrared quantum
  cascade lasers}, J. Appl. Phys. 116 (2014) 123105.

\bibitem{pogna21}
E.~A.~A. Pogna, X.~Jia, A.~Principi, A.~Block, L.~Banszerus, J.~Zhang, X.~Liu,
  T.~Sohier, S.~Forti, K.~Soundarapandian, B.~Terrés, J.~D. Mehew,
  C.~Trovatello, C.~Coletti, F.~H.~L. Koppens, M.~Bonn, H.~I. Wang, N.~van
  Hulst, M.~J. Verstraete, H.~Peng, Z.~Liu, C.~Stampfer, G.~Cerullo, K.-J.
  Tielrooij, {Hot-Carrier Cooling in High-Quality Graphene Is Intrinsically
  Limited by Optical Phonons}, ACS Nano 15 (2021) 11285.

\bibitem{chan21}
C.~C.~S. Chan, K.~Fan, H.~Wang, Z.~Huang, D.~Novko, K.~Yan, J.~Xu, W.~C.~H.
  Choy, I.~Lon\v{c}ari\'c, K.~S. Wong, {Uncovering the Electron-Phonon
  Interplay and Dynamical Energy-Dissipation Mechanisms of Hot Carriers in
  Hybrid Lead Halide Perovskites}, Adv. Energy Mater. 11 (2021) 2003071.

\bibitem{gadermaier10}
C.~Gadermaier, A.~S. Alexandrov, V.~V. Kabanov, P.~Kusar, T.~Mertelj, X.~Yao,
  C.~Manzoni, D.~Brida, G.~Cerullo, D.~Mihailovic, {Electron-Phonon Coupling in
  High-Temperature Cuprate Superconductors Determined from Electron Relaxation
  Rates}, Phys. Rev. Lett. 105 (2010) 257001.

\bibitem{bib:dalconte15}
S.~Dal~Conte, L.~Vidmar, D.~Golež, M.~Mierzejewski, G.~Soavi, S.~Peli,
  F.~Banfi, G.~Ferrini, R.~Comin, B.~M. Ludbrook, L.~Chauviere, N.~D. Zhigadlo,
  H.~Eisaki, M.~Greven, S.~Lupi, A.~Damascelli, D.~Brida, M.~Capone, J.~Bonča,
  G.~Cerullo, C.~Giannetti, {Snapshots of the retarded interaction of charge
  carriers with ultrafast fluctuations in cuprates}, Nat. Phys. 11 (2015) 421.

\bibitem{price15}
M.~B. Price, J.~Butkus, T.~C. Jellicoe, A.~Sadhanala, A.~Briane, J.~E. Halpert,
  K.~Broch, J.~M. Hodgkiss, R.~H. Friend, F.~Deschler, {Hot-carrier cooling and
  photoinduced refractive index changes in organic-inorganic lead halide
  perovskites}, {Nat. Comm.} {6} ({2015}) {8420}.

\bibitem{yang16}
Y.~Yang, D.~Ostrowski, R.~France, K.~Zhu, J.~van~de Lagemaat, J.~Luther,
  M.~Beard, {Observation of a hot-phonon bottleneck in lead-iodide
  perovskites}, Nat. Photon. 10 (2016) 53.

\bibitem{yang17}
J.~Yang, X.~Wen, H.~Xia, R.~Sheng, Q.~Ma, J.~Kim, P.~Tapping, T.~Harada, T.~W.
  Kee, F.~Huang, Y.-B. Cheng, M.~Green, A.~Ho-Baillie, S.~Huang, S.~Shrestha,
  R.~Patterson, G.~Conibeer, {Acoustic-optical phonon up-conversion and
  hot-phonon bottleneck in lead-halide perovskites}, {Nat. Comm.} {8} ({2017})
  {14120}.

\bibitem{fu17}
J.~Fu, Q.~Xu, G.~Han, B.~Wu, C.~H.~A. Huan, M.~L. Leek, T.~C. Sum, {Hot carrier
  cooling mechanisms in halide perovskites}, {Nat. Comm.} {8} ({2017}) {1300}.

\bibitem{novko2021}
D.~Novko, {First-principles study of ultrafast dynamics of Dirac plasmon in
  graphene}, New Jour. Phys. 23 (2021) 043023.

\bibitem{bib:johannsen13}
J.~C. Johannsen, S.~Ulstrup, F.~Cilento, A.~Crepaldi, M.~Zacchigna, C.~Cacho,
  I.~C.~E. Turcu, E.~Springate, F.~Fromm, C.~Raidel, T.~Seyller, F.~Parmigiani,
  M.~Grioni, P.~Hofmann, {Direct View of Hot Carrier Dynamics in Graphene},
  Phys. Rev. Lett. 111 (2013) 027403.

\bibitem{rettig13}
L.~Rettig, R.~Cort{\'{e}}s, H.~S. Jeevan, P.~Gegenwart, T.~Wolf, J.~Fink,
  U.~Bovensiepen, {Electron{\textendash}phonon coupling in 122 Fe pnictides
  analyzed by femtosecond time-resolved photoemission}, New J. Phys. 15 (2013)
  083023.

\bibitem{avigo13}
I.~Avigo, R.~Cort{\'{e}}s, L.~Rettig, S.~Thirupathaiah, H.~S. Jeevan,
  P.~Gegenwart, T.~Wolf, M.~Ligges, M.~Wolf, J.~Fink, U.~Bovensiepen, {Coherent
  excitations and electron{\textendash}phonon coupling in Ba/{EuFe}$_2$As$_2$
  compounds investigated by femtosecond time- and angle-resolved photoemission
  spectroscopy}, J. Phys.: Condens. Matter 25 (2013) 094003.

\bibitem{yang17new}
J.-A. Yang, S.~Parham, D.~Dessau, D.~Reznik, {Novel Electron-Phonon Relaxation
  Pathway in Graphite Revealed by Time-Resolved Raman Scattering and
  Angle-Resolved Photoemission Spectroscopy}, {Sci. Rep.} {7} ({2017}) {40876}.

\bibitem{sekiguchi21}
F.~Sekiguchi, H.~Hirori, G.~Yumoto, A.~Shimazaki, T.~Nakamura, A.~Wakamiya,
  Y.~Kanemitsu, {Enhancing the Hot-Phonon Bottleneck Effect in a Metal Halide
  Perovskite by Terahertz Phonon Excitation}, Phys. Rev. Lett. 126 (2021)
  077401.

\bibitem{hannah13}
D.~C. Hannah, K.~E. Brown, R.~M. Young, M.~R. Wasielewski, G.~C. Schatz, D.~T.
  Co, R.~D. Schaller, {Direct Measurement of Lattice Dynamics and Optical
  Phonon Excitation in Semiconductor Nanocrystals Using Femtosecond Stimulated
  Raman Spectroscopy}, Phys. Rev. Lett. 111 (2013) 107401.

\bibitem{carbone10}
F.~Carbone, N.~Gedik, J.~Lorenzana, A.~H. Zewail, {Real-Time Observation of
  Cuprates Structural Dynamics by Ultrafast Electron Crystallography}, {Adv.
  Condens. Matter Phys.} {2010} ({2010}) {958618}.

\bibitem{mansart13}
B.~Mansart, M.~J.~G. Cottet, G.~F. Mancini, T.~Jarlborg, S.~B. Dugdale, S.~L.
  Johnson, S.~O. Mariager, C.~J. Milne, P.~Beaud, S.~Gr\"ubel, J.~A. Johnson,
  T.~Kubacka, G.~Ingold, K.~Prsa, H.~M. R\o{}nnow, K.~Conder, E.~Pomjakushina,
  M.~Chergui, F.~Carbone, {Temperature-dependent electron-phonon coupling in
  La${}_{2\ensuremath{-}x}$Sr${}_{x}$CuO${}_{4}$ probed by femtosecond x-ray
  diffraction}, Phys. Rev. B 88 (2013) 054507.

\bibitem{harb16}
M.~Harb, H.~Enquist, A.~Jurgilaitis, F.~T. Tuyakova, A.~N. Obraztsov,
  J.~Larsson, {Phonon-phonon interactions in photoexcited graphite studied by
  ultrafast electron diffraction}, Phys. Rev. B 93 (2016) 104104.

\bibitem{waldecker17}
L.~Waldecker, R.~Bertoni, H.~H\"ubener, T.~Brumme, T.~Vasileiadis, D.~Zahn,
  A.~Rubio, R.~Ernstorfer, {Momentum-Resolved View of Electron-Phonon Coupling
  in Multilayer ${\mathrm{WSe}}_{2}$}, Phys. Rev. Lett. 119 (2017) 036803.

\bibitem{konstantinova18}
T.~Konstantinova, J.~D. Rameau, A.~H. Reid, O.~Abdurazakov, L.~Wu, R.~Li,
  X.~Shen, G.~Gu, Y.~Huang, L.~Rettig, I.~Avigo, M.~Ligges, J.~K. Freericks,
  A.~F. Kemper, H.~A. Duerr, U.~Bovensiepen, P.~D. Johnson, X.~Wang, Y.~Zhu,
  {Nonequilibrium electron and lattice dynamics of strongly correlated
  Bi$_2$Sr$_2$CaCu$_2$O$_{8+\delta}$ single crystals}, {Sci. Adv.} {4} ({2018})
  {eaap7427}.

\bibitem{karam18}
T.~E. Karam, J.~Hu, G.~A. Blake, {Strongly Coupled Electron–Phonon Dynamics
  in Few-Layer TiSe2 Exfoliates}, ACS Photon. 5 (2018) 1228.

\bibitem{zhan20}
D.~Zahn, P.-N. Hildebrandt, T.~Vasileiadis, Y.~W. Windsor, Y.~Qi, H.~Seiler,
  R.~Ernstorfer, {Anisotropic Nonequilibrium Lattice Dynamics of Black
  Phosphorus}, Nano Lett. 20 (2020) 3728.

\bibitem{kang10}
K.~Kang, D.~Abdula, D.~G. Cahill, M.~Shim, {Lifetimes of optical phonons in
  graphene and graphite by time-resolved incoherent anti-Stokes Raman
  scattering}, Phys. Rev. B 81 (2010) 165405.

\bibitem{pellatz21}
N.~Pellatz, S.~Roy, J.-W. Lee, J.~Schad, H.~Kandel, N.~Arndt, C.~Eom,
  D.~Reznik, {Relaxation timescales and electron-phonon coupling in
  optically-pumped YBa$_2$Cu$_3$O$_{6+x}$ revealed by time-resolved Raman
  scattering}, arXiv:2010.15958v3.

\bibitem{trigo10}
M.~Trigo, J.~Chen, V.~H. Vishwanath, Y.~M. Sheu, T.~Graber, R.~Henning, D.~A.
  Reis, {Imaging nonequilibrium atomic vibrations with x-ray diffuse
  scattering}, Phys. Rev. B 82 (2010) 235205.

\bibitem{chatelain14}
R.~P. Chatelain, V.~R. Morrison, B.~L.~M. Klarenaar, B.~J. Siwick, {Coherent
  and Incoherent Electron-Phonon Coupling in Graphite Observed with
  Radio-Frequency Compressed Ultrafast Electron Diffraction}, Phys. Rev. Lett.
  113 (2014) 235502.

\bibitem{chase16}
T.~Chase, M.~Trigo, A.~H. Reid, R.~Li, T.~Vecchione, X.~Shen, S.~Weathersby,
  R.~Coffee, N.~Hartmann, D.~A. Reis, X.~J. Wang, H.~A. Dürr, {Ultrafast
  electron diffraction from non-equilibrium phonons in femtosecond laser heated
  Au films}, Appl. Phys. Lett. 108 (2016) 041909.

\bibitem{stern18}
M.~J. Stern, L.~P. Ren\'e~de Cotret, M.~R. Otto, R.~P. Chatelain, J.-P.
  Boisvert, M.~Sutton, B.~J. Siwick, {Mapping momentum-dependent
  electron-phonon coupling and nonequilibrium phonon dynamics with ultrafast
  electron diffuse scattering}, Phys. Rev. B 97 (2018) 165416.

\bibitem{cotret19}
L.~P. Ren\'e~de Cotret, J.-H. P\"ohls, M.~J. Stern, M.~R. Otto, M.~Sutton,
  B.~J. Siwick, {Time- and momentum-resolved phonon population dynamics with
  ultrafast electron diffuse scattering}, Phys. Rev. B 100 (2019) 214115.

\bibitem{otto21}
M.~R. Otto, J.-H. P\"ohls, L.~P.~R. de~Cotret, M.~J. Stern, M.~Sutton, B.~J.
  Siwick, {Mechanisms of electron-phonon coupling unraveled in momentum and
  time: The case of soft phonons in TiSe$_2$}, Sci. Adv. 7~(20) (2021)
  eabf2810.

\bibitem{Zacharias_joint_PRB}
M.~Zacharias, H.~Seiler, F.~Caruso, D.~Zahn, F.~Giustino, P.~C. Kelires,
  R.~Ernstorfer, {Multiphonon diffuse scattering in solids from first
  principles: Application to layered crystals and two-dimensional materials},
  Phys. Rev. B 104 (2021) 205109.

\bibitem{notezach}
{We neglect here higher-order contributions. For a detailed discussion see Ref.
  \cite{Zacharias_joint_PRL} }.

\bibitem{bib:kortus01}
J.~Kortus, I.~I. Mazin, K.~D. Belashchenko, V.~P. Antropov, L.~L. Boyer,
  {Superconductivity of Metallic Boron in ${\mathrm{MgB}}_{2}$}, Phys. Rev.
  Lett. 86 (2001) 4656.

\bibitem{an}
J.~M. An, W.~E. Pickett, {Superconductivity of MgB$_2$: Covalent Bonds Driven
  Metallic}, Phys. Rev. Lett. 86 (2001) 4366.

\bibitem{yildirim}
T.~Yildirim, O.~G\"ulseren, J.~Lynn, C.~Brown, T.~Udovic, Q.~Huang, N.~Rogado,
  K.~Regan, M.~Hayward, J.~Slusky, T.~He, M.~Haas, P.~Khalifah, K.~Inumaru,
  R.~Cava, {Giant Anharmonicity and Nonlinear Electron-Phonon Coupling in
  MgB$_2$: A Combined First-Principles Calculation and Neutron Scattering
  Study}, Phys. Rev. Lett. 87 (2001) 037001.

\bibitem{bib:liu01}
A.~Y. Liu, I.~I. Mazin, J.~Kortus, {Beyond Eliashberg Superconductivity in
  ${\mathrm{MgB}}_{2}$: Anharmonicity, Two-Phonon Scattering, and Multiple
  Gaps}, Phys. Rev. Lett. 87 (2001) 087005.

\bibitem{bib:choi02a}
H.~J. Choi, D.~Roundy, H.~Sun, M.~L. Cohen, S.~G. Louie, {The origin of the
  anomalous superconducting properties of MgB$_2$}, Nature 418 (2002) 758.

\bibitem{bednorz86}
J.~Bednorz, K.~M\"uller, {Possible high $T_c$ superconductivity in the
  Ba-La-Cu-O system}, Z. Phys. B 64 (1986) 189.

\bibitem{wu87}
M.~K. Wu, J.~R. Ashburn, C.~J. Torng, P.~H. Hor, R.~L. Meng, L.~Gao, Z.~J.
  Huang, Y.~Q. Wang, C.~W. Chu, {Superconductivity at 93 K in a new mixed-phase
  Y-Ba-Cu-O compound system at ambient pressure}, Phys. Rev. Lett. 58 (1987)
  908.

\bibitem{cava87}
R.~J. Cava, B.~Batlogg, R.~B. van Dover, D.~W. Murphy, S.~Sunshine,
  T.~Siegrist, J.~P. Remeika, E.~A. Rietman, S.~Zahurak, G.~P. Espinosa, {Bulk
  superconductivity at 91 K in single-phase oxygen-deficient perovskite
  ${\mathrm{Ba}}_{2}$${\mathrm{YCu}}_{3}$${\mathrm{O}}_{9\mathrm{\ensuremath{-}}\mathrm{\ensuremath{\delta}}}$},
  Phys. Rev. Lett. 58 (1987) 1676.

\bibitem{hazen88}
R.~M. Hazen, C.~T. Prewitt, R.~J. Angel, N.~L. Ross, L.~W. Finger, C.~G.
  Hadidiacos, D.~R. Veblen, P.~J. Heaney, P.~H. Hor, R.~L. Meng, Y.~Y. Sun,
  Y.~Q. Wang, Y.~Y. Xue, Z.~J. Huang, L.~Gao, J.~Bechtold, C.~W. Chu,
  {Superconductivity in the high-${T}_{c}$ Bi-Ca-Sr-Cu-O system: Phase
  identification}, Phys. Rev. Lett. 60 (1988) 1174.

\bibitem{masui03}
T.~Masui, S.~Tajima, {Normal state transport properties of MgB$_2$}, Physica C:
  Superconductivity 385 (2003) 91.

\bibitem{fudamoto03}
Y.~Fudamoto, S.~Lee, {Anisotropic electrodynamics of ${\mathrm{MgB}}_{2}$
  detected by optical reflectance}, Phys. Rev. B 68 (2003) 184514.

\bibitem{guritanu06}
V.~Guritanu, A.~B. Kuzmenko, D.~van~der Marel, S.~M. Kazakov, N.~D. Zhigadlo,
  J.~Karpinski, {Anisotropic optical conductivity and two colors of
  ${\mathrm{MgB}}_{2}$}, Phys. Rev. B 73 (2006) 104509.

\bibitem{xi08}
X.~X. Xi, {Two-band superconductor magnesium diboride}, Rep. Progr. Phys. 71
  (2008) 116501.

\bibitem{giubileo}
F.~Giubileo, D.~Roditchev, W.~Sacks, R.~Lamy, D.~Thanh, J.~Klein, S.~Miraglia,
  D.~Fruchart, J.~Marcus, P.~Monod, {Two-Gap State Density in MgB$_2$: A True
  Bulk Property Or A Proximity Effect?}, Phys. Rev. Lett. 87 (2001) 177008.

\bibitem{tsuda}
S.~Tsuda, T.~Yokoya, T.~Kiss, Y.~Takano, K.~Togano, H.~Kito, H.~Ihara, S.~Shin,
  {Evidence for a Multiple Superconducting Gap in MgB$_2$ from High-Resolution
  Photoemission Spectroscopy}, Phys. Rev. Lett. 87 (2001) 17706.

\bibitem{chen2001}
X.~Chen, M.~Konstantinovi\'c, J.~Irwin, D.~Lawrie, J.~P. Franck, {Evidence for
  Two Superconducting Gaps in MgB$_2$}, Phys. Rev. Lett. 87 (2001) 157002.

\bibitem{gonnelli}
R.~Gonnelli, D.~Daghero, G.~Ummarino, V.~Stepanov, J.~Jun, S.~Kazakov,
  J.~Karpinski, {Direct Evidence for Two-Band Superconductivity in MgB$_2$
  Single Crystals from Directional Point-Contact Spectroscopy in Magnetic
  Fields}, Phys. Rev. Lett. 89 (2002) 247004.

\bibitem{mou}
D.~Mou, R.~Jiang, V.~Taufour, S.~Bud'ko, P.~Canfield, A.~Kaminski, {Momentum
  dependence of the superconducting gap and in-gap states in MgB$_2$ multiband
  superconductor}, Phys. Rev. B 91 (2015) 214519.

\bibitem{giustino17}
F.~Giustino, {Electron-phonon interactions from first principles}, Rev. Mod.
  Phys. 89 (2017) 015003.

\bibitem{noteanis}
{The concept of anisotropy is meant here relatively to the band index space
  $\alpha$, so that electronic/superconducting anisotropic properties mean that
  they are strongly varying upon $\alpha$. }.

\bibitem{bib:choi02b}
H.~J. Choi, D.~Roundy, H.~Sun, M.~L. Cohen, S.~G. Louie, {First-principles
  calculation of the superconducting transition in ${\mathrm{MgB}}_{2}$ within
  the anisotropic Eliashberg formalism}, Phys. Rev. B 66 (2002) 020513.

\bibitem{golubov}
A.~Golubov, J.~Kortus, O.~Dolgov, O.~Jepsen, Y.~Kong, O.~Andersen, B.~Gibson,
  K.~Ahn, R.~Kremer, {Specific heat of MgB$_2$ in a one- and a two-band model
  from first-principles calculations}, J. Phys.: Condens. Matter 14 (2002)
  1353.

\bibitem{bohnen}
K.-P. Bohnen, R.~Heid, B.~Renker, {Phonon Dispersion and Electron-Phonon
  Coupling in MgB$_2$ and AlB$_2$}, Phys. Rev. Lett. 86 (2001) 5771.

\bibitem{boeri02}
L.~Boeri, G.~Bachelet, E.~Cappelluti, L.~Pietronero, {Small Fermi energy and
  phonon anharmonicity in MgB$_2$ and related compounds}, Phys. Rev. B 65
  (2002) 214501.

\bibitem{rafailov}
P.~Rafailov, M.~Dworzak, C.~Thomsen, {Luminescence and Raman spectroscopy on
  MgB$_2$}, Solid State Comm. 122 (2002) 455.

\bibitem{shukla03}
A.~Shukla, M.~Calandra, M.~d'Astuto, M.~Lazzeri, F.~Mauri, C.~Bellin,
  M.~Krisch, J.~Karpinski, S.~M. Kazakov, J.~Jun, D.~Daghero, K.~Parlinski,
  {Phonon Dispersion and Lifetimes in ${\mathrm{M}\mathrm{g}\mathrm{B}}_{2}$},
  Phys. Rev. Lett. 90 (2003) 095506.

\bibitem{lazzeri03}
M.~Lazzeri, M.~Calandra, F.~Mauri, {Anharmonic phonon frequency shift in
  ${\mathrm{MgB}}_{2}$}, Phys. Rev. B 68 (2003) 220509.

\bibitem{dastuto07}
M.~d'Astuto, M.~Calandra, S.~Reich, A.~Shukla, M.~Lazzeri, F.~Mauri,
  J.~Karpinski, N.~D. Zhigadlo, A.~Bossak, M.~Krisch, {Weak anharmonic effects
  in $\mathrm{Mg}{\mathrm{B}}_{2}$: A comparative inelastic x-ray scattering
  and Raman study}, Phys. Rev. B 75 (2007) 174508.

\bibitem{goncharov}
A.~Goncharov, V.~Struzhkin, E.Gregoryanz, J.~Hu, R.~Hemley, H.~k.~Mao,
  G.~Lapertot, S.~Bud’ko, P.~Canfield, {Raman spectrum and lattice parameters
  of MgB$_2$ as a function of pressure}, Phys. Rev. B 64 (2001) 100509.

\bibitem{hlinka}
J.~Hlinka, I.~Gregora, J.~Pokorn\'y, A.~Plecenik, P.~K\'u\v{s}, L.~Satrapinsky,
  \v{S}. Be\v{n}a\v{c}ka, {Phonons in MgB$_2$ by polarized Raman scattering on
  single crystals}, Phys. Rev. B 64 (2001) 140503.

\bibitem{postorino1}
P.~Postorino, A.~Congeduti, P.~Dore, A.~Nucara, A.~Bianconi, D.~Di~Castro,
  S.~De~Negri, A.~Saccone, {Effect of the Al content on the optical phonon
  spectrum in
  ${\mathrm{Mg}}_{1\ensuremath{-}x}{\mathrm{Al}}_{x}{\mathrm{B}}_{2}$}, Phys.
  Rev. B 65 (2001) 020507.

\bibitem{quilty02}
J.~W. Quilty, S.~Lee, A.~Yamamoto, S.~Tajima, {Superconducting Gap in
  ${\mathrm{MgB}}_{2}$: Electronic Raman Scattering Measurements of Single
  Crystals}, Phys. Rev. Lett. 88 (2002) 087001.

\bibitem{quilty03}
J.~W. Quilty, S.~Lee, S.~Tajima, A.~Yamanaka, {$c$-Axis Raman Scattering
  Spectra of ${\mathrm{M}\mathrm{g}\mathrm{B}}_{2}$: Observation of a
  Dirty-Limit Gap in the $\ensuremath{\pi}$ Bands}, Phys. Rev. Lett. 90 (2003)
  207006.

\bibitem{martinho03}
H.~Martinho, C.~Rettori, P.~Pagliuso, A.~Martin, N.~Moreno, J.~Sarrao, {Role of
  the $E_{2g}$ phonon in the superconductivity of MgB$_2$: a Raman scattering
  study}, {Solid State Comm.} {125} ({2003}) {499}.

\bibitem{renker}
B.~Renker, K.~Bohnen, R.~Heid, D.~Ernst, H.~Schober, M.~Koza, P.~Adelmann,
  P.~Schweiss, T.~Wolf, {Strong Renormalization of Phonon Frequencies in
  Mg$_{1-x}$Al$_x$B$_2$ }, Phys. Rev. Lett. 88 (2002) 067001.

\bibitem{boeri05}
L.~Boeri, E.~Cappelluti, L.~Pietronero, {Small Fermi energy, zero-point
  fluctuations, and nonadiabaticity in MgB$_2$}, Phys. Rev. B 71 (2005) 012501.

\bibitem{bianconi15}
A.~Bianconi, T.~Jarlborg, {Lifshitz transitions and zero point lattice
  fluctuations in sulfur hydride showing near room temperature
  superconductivity}, Nov. Supercond. Mater. 1 (2015) {37}.

\bibitem{cappelluti06}
E.~Cappelluti, {Electron-phonon effects on the Raman spectrum in
  $\mathrm{Mg}{\mathrm{B}}_{2}$}, Phys. Rev. B 73 (2006) 140505.

\bibitem{cappelluti06b}
E.~Cappelluti, L.~Pietronero, {Electron–phonon interaction and breakdown of
  the adiabatic principle in fullerides and MgB$_2$}, J. Phys. Chem. Solids 67
  (2006) 1941.

\bibitem{novko18}
D.~Novko, {Nonadiabatic coupling effects in MgB$_2$ reexamined}, Phys. Rev. B
  98 (2018) 041112.

\bibitem{bib:baron04}
A.~Q.~R. Baron, H.~Uchiyama, Y.~Tanaka, S.~Tsutsui, D.~Ishikawa, S.~Lee,
  R.~Heid, K.-P. Bohnen, S.~Tajima, T.~Ishikawa, {Kohn Anomaly in
  ${\mathrm{MgB}}_{2}$ by Inelastic X-Ray Scattering}, Phys. Rev. Lett. 92
  (2004) 197004.

\bibitem{cn}
E.~Cappelluti, D.~Novko, {Fingerprints of hot-phonon physics in time-resolved
  correlated quantum lattice dynamics}, arXiv:2110.13274.

\bibitem{bib:qe}
P.~Giannozzi, S.~Baroni, N.~Bonini, M.~Calandra, R.~Car, C.~Cavazzoni,
  D.~Ceresoli, G.~L. Chiarotti, M.~Cococcioni, I.~Dabo, \textit{et al.},
  {QUANTUM ESPRESSO: a modular and open-source software project for quantum
  simulations of materials}, J. Phys: Condens. Matter 21 (2009) 395502.

\bibitem{lazzeri06b}
M.~Lazzeri, F.~Mauri, {Nonadiabatic Kohn Anomaly in a Doped Graphene
  Monolayer}, Phys. Rev. Lett. 97 (2006) 266407.

\bibitem{boross13}
P.~Boross, B.~Dora, A.~Kiss, F.~Simon, {A unified theory of spin-relaxation due
  to spin-orbit coupling in metals and semiconductors}, {Sci. Rep.} {3}
  ({2013}) {3233}.

\bibitem{szolnoki17}
L.~Szolnoki, A.~Kiss, B.~Dora, F.~Simon, {Spin-relaxation time in materials
  with broken inversion symmetry and large spin-orbit coupling}, {Sci. Rep.}
  {7} ({2017}) {9949}.

\bibitem{mehringbook}
{M. Mehring}, {High Resolution NMR Spectroscopy in Solids}, Vol.~11 of NMR
  Basic Principles and Progress, Springer-Verlag Berlin Heidelberg, 2076.

\bibitem{rigamonti98}
A.~Rigamonti, F.~Borsa, P.~Carretta, {Basic aspects and main results of
  {NMR}-{NQR} spectroscopies in high-temperature superconductors}, Rep. Prog.
  Phys. 61 (1998) 1367.

\bibitem{campi}
G.~Campi, E.~Cappelluti, T.~Proffen, X.~Qiu, E.~Bozin, S.~Billinge,
  S.~Agrestini, N.~Saini, A.~Bianconi, {Study of temperature dependent atomic
  correlations in MgB$_2$}, Eur. Phys. J. B 52 (2006) 15.

\bibitem{jeong99}
I.-K. Jeong, T.~Proffen, F.~Mohiuddin-Jacobs, S.~J.~L. Billinge, {Measuring
  Correlated Atomic Motion Using X-ray Diffraction}, J. Phys. Chem. A 103
  (1999) 921.

\bibitem{jeong03}
I.-K. Jeong, R.~H. Heffner, M.~J. Graf, S.~J.~L. Billinge, {Lattice dynamics
  and correlated atomic motion from the atomic pair distribution function},
  Phys. Rev. B 67 (2003) 104301.

\bibitem{thorpebook}
M.~Thorpe, V.~Levashov, M.~Lei, S.~Billinge, in: S.~Billinge, M.~Thorpe (Eds.),
  {From semiconductors to proteins: beyond the average structure},
  Kluwer/Plenum, New York, 2002, 2002, p. 105.

\bibitem{notecorr}
{These relations hold true in the present form in the ideal clean case. In the
  analysis of experimenta data of real materials, the mean-square displacements
  should be considered at the net of disorder contribution \cite{campi}. }.

\bibitem{ortenzi09}
L.~Ortenzi, E.~Cappelluti, L.~Benfatto, L.~Pietronero, {Fermi-Surface Shrinking
  and Interband Coupling in Iron-Based Pnictides}, Phys. Rev. Lett. 103 (2009)
  046404.

\bibitem{benfatto09}
L.~Benfatto, E.~Cappelluti, C.~Castellani, {Spectroscopic and thermodynamic
  properties in a four-band model for pnictides}, Phys. Rev. B 80 (2009)
  214522.

\bibitem{benfatto11}
L.~Benfatto, E.~Cappelluti, {Effects of the Fermi-surface shrinking on the
  optical sum rule in pnictides}, Phys. Rev. B 83 (2011) 104516.

\bibitem{echenique02}
P.~Echenique, J.~Pitarke, E.~Chulkov, V.~Silkin, {Image-potential-induced
  states at metal surfaces}, {J. Electron. Spectros. Relat. Phenomena} {126}
  ({2002}) {163}.

\bibitem{hase03}
M.~Hase, M.~Kitajima, A.~Constantinescu, H.~Petek, {The birth of a
  quasiparticle in silicon observed in time-frequency space}, {Nature} {426}
  ({2003}) {51}.

\bibitem{notebaldini}
{The different contributions to the total bandshift are here expressed in a
  slightly different but more intuitive way than in Ref. \cite{baldini}. }.

\bibitem{notethermalization}
{Assumption of fast electronic thermalization guided by the Coulomb
  electron-electron interaction is not actually required in the present
  scenario, as shown in Ref. \cite{ncdc}. Its employment here is just for sake
  of simplicity in the later discussion. }.

\bibitem{ekimov04}
E.~Ekimov, V.~Sidorov, E.~Bauer, N.~Mel'nik, N.~Curro, J.~Thompson, S.~Stishov,
  {Superconductivity in diamond}, {Nature} {428} ({2004}) {542}.

\bibitem{boeri04}
L.~Boeri, J.~Kortus, O.~K. Andersen, {Three-Dimensional
  ${\mathrm{M}\mathrm{g}\mathrm{B}}_{2}$-Type Superconductivity in Hole-Doped
  Diamond}, Phys. Rev. Lett. 93 (2004) 237002.

\bibitem{thomsencardona}
C.~Thomsen, M.~Cardona, {Physical Properties of High-Temperature
  Superconductors}, in: T.~Ginsberg (Ed.), {Raman Scattering in High-$T_c$
  Superconductors}, World Scientific, Singapore, 1989, p. 409.

\bibitem{thomsen}
C.~Thomsen, {Light scattering in high-$T_c$ superconductors}, in: M.~Cardona,
  G.~Guntherodt (Eds.), {Light Scattering in Solids VI}, Springer, Berlin,
  1991, p. 285.

\bibitem{wang02}
Y.~G. Wang, S.~P. Lau, B.~K. Tay, X.~H. Zhang, {Resonant Raman scattering
  studies of Fano-type interference in boron doped diamond}, J. Appl. Phys. 92
  (2002) 7253.

\bibitem{szirmai12}
P.~Szirmai, T.~Pichler, O.~A. Williams, S.~Mandal, C.~Bäuerle, F.~Simon, {A
  detailed analysis of the Raman spectra in superconducting boron doped
  nanocrystalline diamond}, Phys. Status Solidi B 249 (2012) 2656.

\bibitem{mortet19}
V.~Mortet, Z.~V. Živcová, A.~Taylor, M.~Davydová, O.~Frank, P.~Hubík,
  J.~Lorincik, M.~Aleshin, {Determination of atomic boron concentration in
  heavily boron-doped diamond by Raman spectroscopy}, Diam. Relat. Mater. 93
  (2019) 54.

\bibitem{mortet20}
V.~Mortet, I.~Gregora, A.~Taylor, N.~Lambert, P.~Ashcheulov, Z.~Gedeonova,
  P.~Hubik, {New perspectives for heavily boron-doped diamond Raman spectrum
  analysis}, Carbon 168 (2020) 319.

\bibitem{baroni01}
S.~Baroni, S.~de~Gironcoli, A.~Dal~Corso, P.~Giannozzi, {Phonons and related
  crystal properties from density-functional perturbation theory}, Rev. Mod.
  Phys. 73 (2001) 515.

\bibitem{wannier}
N.~Marzari, A.~A. Mostofi, J.~R. Yates, I.~Souza, D.~Vanderbilt, {Maximally
  localized Wannier functions: Theory and applications}, Rev. Mod. Phys. 84
  (2012) 1419.

\bibitem{epwan}
F.~Giustino, M.~L. Cohen, S.~G. Louie, {Electron-phonon interaction using
  Wannier functions}, Phys. Rev. B 76 (2007) 165108.

\bibitem{epw}
S.~Poncé, E.~Margine, C.~Verdi, F.~Giustino, {EPW: Electron–phonon coupling,
  transport and superconducting properties using maximally localized Wannier
  functions}, Comput. Phys. Commun. 209 (2016) 116.

\bibitem{rosner02}
H.~Rosner, A.~Kitaigorodsky, W.~E. Pickett, {Prediction of High ${T}_{c}$
  Superconductivity in Hole-Doped LiBC}, Phys. Rev. Lett. 88 (2002) 127001.

\bibitem{an02}
J.~M. An, S.~Y. Savrasov, H.~Rosner, W.~E. Pickett, {Extreme electron-phonon
  coupling in boron-based layered superconductors}, Phys. Rev. B 66 (2002)
  220502.

\bibitem{dewhurst03}
J.~K. Dewhurst, S.~Sharma, C.~Ambrosch-Draxl, B.~Johansson, {First-principles
  calculation of superconductivity in hole-doped LiBC: ${T}_{c}=65\mathrm{K}$},
  Phys. Rev. B 68 (2003) 020504.

\bibitem{li18}
Q.-Z. Li, X.-W. Yan, M.~Gao, J.~Wang, {Electron-phonon coupling and
  superconductivity in LiB1+xC1-x}, {Europhys. Lett.} {122} ({2018}) {47001}.

\bibitem{gao15}
M.~Gao, Z.-Y. Lu, T.~Xiang, {Prediction of phonon-mediated high-temperature
  superconductivity in ${\mathrm{Li}}_{3}{\mathrm{B}}_{4}{\mathrm{C}}_{2}$},
  Phys. Rev. B 91 (2015) 045132.

\bibitem{kolmogorov06}
A.~N. Kolmogorov, S.~Curtarolo, {Prediction of different crystal structure
  phases in metal borides: A lithium monoboride analog to
  $\mathrm{Mg}{\mathrm{B}}_{2}$}, Phys. Rev. B 73 (2006) 180501.

\bibitem{calandra07}
M.~Calandra, A.~N. Kolmogorov, S.~Curtarolo, {Search for high ${T}_{c}$ in
  layered structures: The case of LiB}, Phys. Rev. B 75 (2007) 144506.

\bibitem{bekaert17}
J.~Bekaert, A.~Aperis, B.~Partoens, P.~M. Oppeneer, M.~V. Milo\ifmmode
  \check{s}\else \v{s}\fi{}evi\ifmmode~\acute{c}\else \'{c}\fi{}, {Evolution of
  multigap superconductivity in the atomically thin limit: Strain-enhanced
  three-gap superconductivity in monolayer ${\mathrm{MgB}}_{2}$}, Phys. Rev. B
  96 (2017) 094510.

\bibitem{bekaert19}
J.~Bekaert, M.~Petrov, A.~Aperis, P.~M. Oppeneer, M.~V. Milo\ifmmode
  \check{s}\else \v{s}\fi{}evi\ifmmode~\acute{c}\else \'{c}\fi{},
  {Hydrogen-Induced High-Temperature Superconductivity in Two-Dimensional
  Materials: The Example of Hydrogenated Monolayer ${\mathrm{MgB}}_{2}$}, Phys.
  Rev. Lett. 123 (2019) 077001.

\bibitem{modak21}
P.~Modak, A.~K. Verma, A.~K. Mishra, {Prediction of superconductivity at 70 K
  in a pristine monolayer of LiBC}, Phys. Rev. B 104 (2021) 054504.

\bibitem{dicataldo21}
S.~di~Cataldo, S.~Qulaghasi, G.~Bachelet, L.~Boeri, {High-$T_c$
  Superconductivity in doped boron-carbon clathrates}, arXiv:2110.05333v1.

\bibitem{dicataldoprivate}
{S. di Cataldo and L. Boeri, private communication (2021). }.

\bibitem{disalvo76}
F.~J. Di~Salvo, D.~E. Moncton, J.~V. Waszczak, {Electronic properties and
  superlattice formation in the semimetal ${\mathrm{TiSe}}_{2}$}, Phys. Rev. B
  14 (1976) 4321.

\bibitem{watson20}
M.~D. Watson, A.~Rajan, T.~Antonelli, K.~Underwood, I.~Markovic, F.~Mazzola,
  O.~J. Clark, G.-R. Siemann, D.~Biswas, A.~Hunter, S.~Jandura,
  J.~Reichstetter, M.~McLaren, P.~Le~Fevre, G.~Vinai, P.~D.~C. King,
  {Strong-coupling charge density wave in monolayer TiSe2}, {2D Mater.} {8}
  ({2021}) {015004}.

\bibitem{holt01}
M.~Holt, P.~Zschack, H.~Hong, M.~Y. Chou, T.-C. Chiang, {X-Ray Studies of
  Phonon Softening in ${\mathrm{TiSe}}_{2}$}, Phys. Rev. Lett. 86 (2001) 3799.

\bibitem{monney11}
C.~Monney, C.~Battaglia, H.~Cercellier, P.~Aebi, H.~Beck, Exciton condensation
  driving the periodic lattice distortion of
  $1t\mathrm{\text{\ensuremath{-}}}{\mathrm{tise}}_{2}$, Phys. Rev. Lett. 106
  (2011) 106404.

\bibitem{meregalli98}
V.~Meregalli, S.~Y. Savrasov, {Electron-phonon coupling and properties of doped
  ${\mathrm{BaBiO}}_{3}$}, Phys. Rev. B 57 (1998) 14453.

\bibitem{shi19}
X.~Shi, W.~You, Y.~Zhang, Z.~Tao, P.~M. Oppeneer, X.~Wu, R.~Thomale,
  K.~Rossnagel, M.~Bauer, H.~Kapteyn, M.~Murnane, {Ultrafast electron
  calorimetry uncovers a new long-lived metastable state in 1$T$-TaSe$_2$
  mediated by mode-selective electron-phonon coupling}, Sci. Adv. 5 (2019)
  eaav4449.

\bibitem{zhang20}
Y.~Zhang, X.~Shi, W.~You, Z.~Tao, Y.~Zhong, F.~Cheenicode~Kabeer, P.~Maldonado,
  P.~M. Oppeneer, M.~Bauer, K.~Rossnagel, H.~Kapteyn, M.~Murnane, {Coherent
  modulation of the electron temperature and electron{\textendash}phonon
  couplings in a 2D material}, Proc. Nat. Ac. Sci. 117 (2020) 8788.

\bibitem{tao13}
Z.~Tao, T.-R.~T. Han, C.-Y. Ruan, {Anisotropic electron-phonon coupling
  investigated by ultrafast electron crystallography: Three-temperature model},
  Phys. Rev. B 87 (2013) 235124.

\bibitem{storeck20}
G.~Storeck, J.~G. Horstmann, T.~Diekmann, S.~Vogelgesang, G.~von Witte, S.~V.
  Yalunin, K.~Rossnagel, C.~Ropers, {Structural dynamics of incommensurate
  charge-density waves tracked by ultrafast low-energy electron diffraction},
  Struct. Dyn. 7~(3) (2020) 034304.

\bibitem{Dolgirev20}
P.~E. Dolgirev, A.~V. Rozhkov, A.~Zong, A.~Kogar, N.~Gedik, B.~V. Fine,
  {Amplitude dynamics of the charge density wave in ${\mathrm{LaTe}}_{3}$:
  Theoretical description of pump-probe experiments}, Phys. Rev. B 101 (2020)
  054203.

\bibitem{jin19}
K.-H. Jin, H.~Huang, J.-W. Mei, Z.~Liu, L.-K. Lim, F.~Liu, {Topological
  superconducting phase in high-Tc superconductor MgB$_2$ with
  Dirac{\textendash}nodal-line fermions}, NPJ Comput. Mater. 5~(1) (2019) 57.

\bibitem{li20}
J.~Li, Q.~Xie, J.~Liu, R.~Li, M.~Liu, L.~Wang, D.~Li, Y.~Li, X.-Q. Chen,
  {Phononic Weyl nodal straight lines in ${\mathrm{MgB}}_{2}$}, Phys. Rev. B
  101 (2020) 024301.

\bibitem{benedek20a}
G.~Benedek, S.~Miret-Artés, J.~R. Manson, A.~Ruckhofer, W.~E. Ernst,
  A.~Tamtögl, {Origin of the Electron–Phonon Interaction of Topological
  Semimetal Surfaces Measured with Helium Atom Scattering}, J. Phys. Chem.
  Lett. 11 (2020) 1927.

\bibitem{benedek20b}
G.~Benedek, J.~R. Manson, S.~Miret-Artés, {The Electron–Phonon Interaction
  of Low-Dimensional and Multi-Dimensional Materials from He Atom Scattering},
  Adv. Mater. 32~(25) (2020) 2002072.

\bibitem{lebegue07}
S.~Leb\`egue, {Electronic structure and properties of the Fermi surface of the
  superconductor LaOFeP}, Phys. Rev. B 75 (2007) 035110.

\bibitem{ding08}
H.~Ding, P.~Richard, K.~Nakayama, K.~Sugawara, T.~Arakane, Y.~Sekiba,
  A.~Takayama, S.~Souma, T.~Sato, T.~Takahashi, Z.~Wang, X.~Dai, Z.~Fang, G.~F.
  Chen, J.~L. Luo, N.~L. Wang, {Observation of Fermi-surface-dependent nodeless
  superconducting gaps in Ba$_{0.6}$K$_{0.4}$Fe$_2$As$_2$}, {Europhys. Lett.}
  83 (2008) 47001.

\bibitem{ren08}
R.~Zhi-An, L.~Wei, Y.~Jie, Y.~Wei, S.~Xiao-Li, L.~Zheng-Cai, C.~Guang-Can,
  D.~Xiao-Li, S.~Li-Ling, Z.~Fang, Z.~Zhong-Xian, {Superconductivity at 55 K in
  Iron-Based F-Doped Layered Quaternary Compound Sm[O$_{1-x}$F$_x$]FeAs}, Chin.
  Phys. Lett. 25 (2008) 2215.

\bibitem{Ge15}
J.-F. Ge, Z.-L. Liu, C.~Liu, C.-L. Gao, D.~Qian, Q.-K. Xue, Y.~Liu, J.-F. Jia,
  {Superconductivity above 100 K in single-layer FeSe films on doped
  SrTiO$_3$}, {Nat. Mater.} {14} ({2015}) {285}.

\bibitem{noteBZiron}
{Labelling of the high-symmetry points of the Brillouin zone can be different
  in literature according the choice of the unit cell, including one or two
  iron atoms. In the present notation the electron-like bands are located at
  the M points at $(\pm \pi,\pm \pi)$. }.

\bibitem{Zacharias_joint_PRL}
M.~Zacharias, H.~Seiler, F.~Caruso, D.~Zahn, F.~Giustino, P.~C. Kelires,
  R.~Ernstorfer, {Efficient First-Principles Methodology for the Calculation of
  the All-Phonon Inelastic Scattering in Solids}, Phys. Rev. Lett. 127 (2021)
  207401.

\end{thebibliography}

\end{document}